\documentclass[prx,superscriptaddress,reprint,amsmath,amssymb,aps,longbibliography]{revtex4-2}

\usepackage{xcolor}
\usepackage{ulem}
\usepackage{graphicx}
\usepackage{float}
\usepackage{diagbox}
\usepackage[
pdfstartview=FitH,
bookmarks=true,
breaklinks=true,
colorlinks=true,
linkcolor=blue,anchorcolor=blue,
citecolor=blue,filecolor=blue,
menucolor=blue,urlcolor=blue]{hyperref}

\begin{document}

\title{Entanglement generation via single-qubit rotations in a torn Hilbert space}

\author{Tao Zhang}
\thanks{These authors contribute equally to this work.}
\affiliation{Department of Physics and State Key Laboratory of Low Dimensional Quantum Physics, Tsinghua University, Beijing, 100084, China}

\author{Zhihao Chi}
\thanks{These authors contribute equally to this work.}
\affiliation{Department of Physics and State Key Laboratory of Low Dimensional Quantum Physics, Tsinghua University, Beijing, 100084, China}

\author{Jiazhong Hu}
\email{hujiazhong01@ultracold.cn}
\affiliation{Department of Physics and State Key Laboratory of Low Dimensional Quantum Physics, Tsinghua University, Beijing, 100084, China}
\affiliation{Frontier Science Center for Quantum Information and Collaborative Innovation Center of Quantum Matter, Beijing, 100084, China}

\begin{abstract}
    
    We propose an efficient yet simple protocol to generate arbitrary symmetric entangled states with only global single-qubit rotations in a torn Hilbert space.
    The system is based on spin-1/2 qubits in a resonator 
    such as atoms in an optical cavity or superconducting qubits coupled to a main bus.
    By sending light or microwave into the resonator, it induces
    AC Stark shifts on particular angular-momentum eigenstates (Dicke states) of qubits.
    Then we are able to generate barriers that 
    hinder transitions between adjacent Dicke states and 
    tear the original Hilbert space into pieces.
    Therefore, a simple global single-qubit rotation becomes highly non-trivial, and thus generates entanglement among the many-body system.
    By optimal control of energy shifts on Dicke states, we are able to generate arbitrary symmetric entangled states. We also exemplify that we can create
    varieties of useful states with near-unity fidelities in only one or very few steps, including W states, spin-squeezed states (SSS), and Greenberger-Horne-Zeilinger (GHZ) states.
    Particularly, the SSS can be created by only one step with a squeezing parameter $\xi_R^2\sim1/N^{0.843}$ approaching the Heisenberg limit (HL).
    Our finding establishes a way for universal entanglement generations with only single-qubit drivings where all the multiple-qubit controls are integrated into simply switching on/off microwave. It has direct applications in the variational quantum optimizer which is available with existing technology.

\end{abstract}

\maketitle

\section{Introduction}
Entanglement is one of the most essential ingredients in quantum technology, including quantum metrology \cite{giovannetti2011advances,pezze2018quantum,marciniak2022optimal, giovannetti2006quantum, anisimov2010quantum,lawrie2019quantum,kaubruegger2019variational, kaubruegger2021quantum,kessler2014heisenberg, pezze2020heisenberg,hu2017vacuum,chen2015carving,hosten2016quantum,wang2019heisenberg}, quantum information processing \cite{bennett2000quantum, horodecki2022five, slussarenko2019photonic, xavier2020quantum}, and quantum computations \cite{ladd2010quantum, divincenzo1995quantum,albash2018adiabatic}.
It also plays a central role in many practical applications such as quantum dense coding \cite{guo2019advances, mattle1996dense, braunstein2000dense,wang201818, ouyang_permutation-invariant_2014,ouyang_permutation-invariant_2016}, quantum teleportation \cite{pirandola2015advances, bouwmeester1997experimental, xia2017long,hu2023progress}, and quantum cryptography \cite{gisin2002quantum, bennett1992experimental,yin2020entanglement}. Meanwhile, it is enabling more and more quantum algorithms and leading to the quantum advantages \cite{ daley2022practical, riste2017demonstration}.
Therefore, finding an efficient way to generate entanglement \cite{ horodecki2009quantum, erhard2020advances,mcconnell2015entanglement,haas2014entangled,barontini2015deterministic,kuzmich1997spin,fernholz2008spin,oblak2005quantum,saffman2009spin,polzik2016entanglement,chen2014cavity,de2005conditions,vitagliano2018entanglement,wang2017two,pedrozo2020entanglement,colombo2022time,yao2012observation,bao2022experimental,wang2022flying,cao2023generation,gong2019genuine,lu2014push,zhao2021creation, shankar2021subradiant,dalla2013dissipative,krauter2011entanglement, kastoryano2011dissipative, reiter2016scalable, shen2011steady} is one particularly important subject in quantum science attracting both theoretical and experimental investigations.

Recently, there are pioneer experimental demonstrations generating varieties of entangled states including SSS \cite{malia2022distributed, zhao2020near, eckner2023realizing,chen2011conditional,braverman2019near}, W states \cite{tashima2009local, mikami2005new, fang2019three, eibl2004experimental,sewell2012magnetic}, and GHZ states \cite{omran2019generation,huang2011experimental, su2007experimental,takeda2021quantum, zhang2022synthesizing,wang2019synthesis,song2019generation,ren2020simultaneous}. 
In order to apply these helpful entangled states into practical applications, one important issue is how to scale up these entangled states with more particles, which is a highly non-trivial problem. 
Usually, generation of entanglement requires many-body interaction between particles, and extremely high control precision of interaction is needed when the particle number increases.
Sometimes, people can bypass the highly-precise control and utilize dissipation for creating robust entangled states in many-body systems \cite{shankar2021subradiant,dalla2013dissipative,krauter2011entanglement, kastoryano2011dissipative, reiter2016scalable, shen2011steady}. 
These designs are usually for some particular entangled states.
Thus, a universal protocol to create a wide range of or almost arbitrary entangled states is still essential and important in this area.

\begin{figure*}[htbp]
    \centering
    \includegraphics[width=0.9\textwidth]{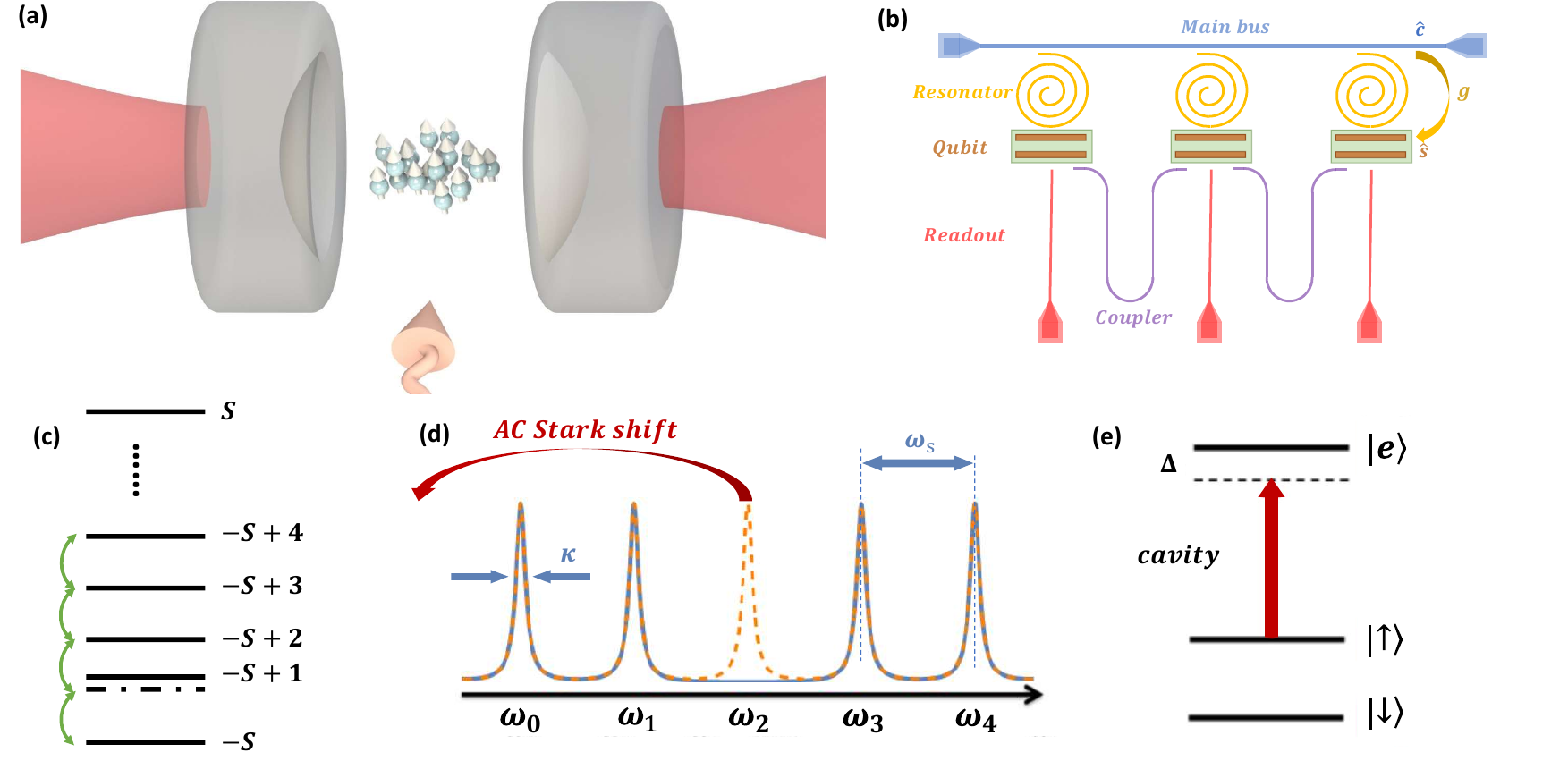}
    \vspace{+0.5cm}
    \hspace{-0.2cm}
    \caption{Setups for entanglement generation via single-qubit rotations in a torn Hilbert space. (a) corresponds to atoms in a cavity. A laser beam incident onto the cavity induces AC stark shift on atoms, and microwave is applied from the side of the cavity to drive the transition between two ground states of atoms. (b) corresponds to superconducting qubits. All transmon qubits are coupled to a main bus by private resonators. The private resonators are used to adjust the coupling strength and the frequency detuning of each qubit coupled to the main bus. The main bus is actually the shared resonator which provides quantized microwave field experienced by all qubits with coupling strength $g$. In our scenario, we need to adjust so that all qubits have the same transition frequency and coupling strength. Readout lines are used for measuring the state of each single qubit, and couplers provide direct interaction between two connected qubits. (c) Dicke state energy levels where one particular Dicke state such as $\left|-S+1\right\rangle$ is shifted away and becomes off-resonant. (d) shows the resonant frequency of the resonator is strongly depending on the qubits' populations.
    Each qubit in $\left|\!\uparrow\right\rangle$ shifts the cavity resonance by an amount of $\omega_s=g^2/\Delta$. When $\omega_s$ is larger than the resonator's linewidth $\kappa$, each spectrum will be resolved. By sending light or microwave to the resonator with one particular frequency, we can shift only one particular Dicke level away from its original location in the energy spectrum.
    (e) The energy level of one qubit. The quantized field couples $| \!\uparrow\rangle$ to the excited state $|e\rangle$ with a detuning $\Delta$.}
    \label{fig1}
\end{figure*}

Inspired by recent works \cite{bretheau_quantum_2015,chakram_multimode_2022} on engineering photons with resolved energy shifts, here we propose a scheme to generate arbitrary symmetric entangled states by only global single-qubit rotations in a torn Hilbert space. 
The description of symmetric state corresponds to a state with the maximal total angular momentum which is on the surface of a Bloch sphere.
The physical system is based on spin-1/2 qubits in a high cooperativity (such as 200) resonator.
Due to the high cooperativity and nonlinearity of resonators, the resonant frequency is highly influenced by the angular momentum of qubits. Each eigenstate of the angular momentum (Dicke state \cite{dicke1954coherence}) corresponds to one particular resonant frequency. By sending light or microwave with different frequencies, the Hilbert space of qubits is torn into pieces.
Under this condition, even a global single-qubit rotation to all qubits can induce strong entanglement among the whole ensemble. In this manuscript, we will mathematically prove the capability to generate arbitrary symmetric states and show a few examples on creating famous entangled states with our scheme, such as Dicke states, GHZ states, and SSS.
All these states only require very few steps to reach a high fidelity above 99\%. Meanwhile, the SSS only need one sequence to achieve a squeezing of $\xi_R^2\sim1/N^{0.843}$ which is close to the HL and useful in metrology.
Particularly, the high cooperativity system has been achieved in varieties of platforms, including the atom-optical cavity systems and the superconducting qubits.
Currently, the cooperativity $\eta$ in the atom-optical cavity system is close to 200 \cite{kroeze2023high}.
In superconducting-qubit system, the cooperativity $\eta$ is even higher. 
A transmission line connecting to gatemon qubits can have an $\eta$ larger than 400 \cite{huo2023gatemon}, and a 3D microwave superconducting cavity containing qubits has an $\eta$ large than $10^6$ \cite{rigetti2012superconducting,chakram2021seamless,milul2023superconducting}. These recent demonstrations suggest our scheme is actually practical and can be experimentally realized.

\section{The system design}
We consider $N$ three-level qubits with two ground states $\left|\downarrow\right\rangle$, $\left|\uparrow\right\rangle$, and an excited state $\left|e\right\rangle$, trapped in a resonator.
This configuration can be applied to either atomic qubits in an optical cavity, or transmon qubits coupled to a cavity bus, as shown in Fig.~\ref{fig1}.
The two ground states of each qubit correspond to a spin-1/2 system  with $s_i=1/2$, and we define the collective spin operator as $\hat{\textbf{S}} \equiv \sum_i \hat{\textbf{s}}_i$.
The resonator mode couples one of the ground states $\left|\uparrow\right\rangle$ to the excited state $\left|e\right\rangle$ with a detuning $\Delta$ and a single-photon Rabi frequency $2g$. The system Hamiltonian is written as
\begin{equation}
    \frac{H'}{\hbar}=\sum_i\left[-\omega_0\left|\downarrow\right\rangle_i\langle \downarrow|_i-\Delta|e\rangle_{i}\langle e|_i+(g\hat{c}^\dagger|\!\uparrow\rangle_i\langle e|_i+h.c.)\right],\label{originalEq}
\end{equation}
where $\omega_0$ is the energy splitting between $\left|\downarrow\right\rangle$ and $\left|\uparrow\right\rangle$. 
$\hat{c}^\dagger$ and $\hat{c}$ are the creation and annihilation operators of the quantized resonator field. $\omega_c$ is the single excitation energy.
By adiabatically eliminating the excited state $\left|e\right\rangle$ and converting into the interaction picture (see Appendix~\ref{effHamiltonian} for a detailed derivation of the adiabatic elimination, and Appendix~\ref{Superconducting} for the Hamiltonian in a superconducting-qubit system), the effective Hamiltonian describing the interaction between the resonator field and $N$ spin-1/2 qubits is written as \cite{schleier2010squeezing, chen2015carving, norcia2018cavity, borregaard2017one, leroux2010implementation}:
\begin{equation}
\frac{H}{\hbar}= \omega_s\left(\hat{S}_z+S\right) \hat{c}^{\dagger} \hat{c},\label{simplifiedEq}
\end{equation}
where $\omega_s = g^2/\Delta$ is the coupling strength, and $S = N/2$ is the total angular momentum number.
This Hamiltonian exactly describes the many-body interaction in the cavity quantum electrodynamics regime. In the following texts, we may use the term of \textit{cavity} to refer the resonator and the term of \textit{light} to refer the incident light, microwave, or other-frequency eletromagnetic fields of the resonator for the words simplicity.

This interaction term can be interpreted as each qubit in the state $\left|\uparrow\right\rangle$ shifts the cavity resonance $\omega_c$ by an amount of $\omega_s$, or alternatively the intra-cavity light shifts the energy of each Dicke state by $(S_z+S)\hbar\omega_s\langle\hat{c}^{\dagger} \hat{c}\rangle$.
When the cavity is illuminated by a monochromatic light beam at the frequency $\omega_n=\omega_c+n \omega_s$, the cavity will be resonant with the incident light only when there are $n$ qubits in the state $\left|\uparrow\right\rangle$. For other atomic states, such as $m$ qubits in $\left|\uparrow\right\rangle$, the cavity becomes off-resonant with the incident beam, and the intensity transmission coefficient is given by
\begin{equation}
T(n,m)=\frac{1}{1+\left[\frac{2(m-n)\omega_s}{\kappa}\right]^2},
\end{equation}
which exhibits a Lorentzian profile.
In particular, when $\kappa \ll \omega_s$, the cavity is strongly off-resonant with the incident light, leading to a significant suppression of the intra-cavity light intensity. Consequently, the transmission equation above tends towards a delta function of $n$ and $m$.
Then, only the state with $n$ qubits in $\left|\uparrow\right\rangle$ experiences a significant intra-cavity intensity.
The Dicke state $\left|S,S_z=-S+n\right\rangle$ will be shifted by the AC-Stark shift with an amount of $n \hbar \omega_s\left\langle\hat{c}^{\dagger} \hat{c}\right\rangle_{n, n}$. Because there is no light sent into the cavity for the other Dickes states, these states experience almost no AC-Stark shift. Therefore, by utilizing this nonlinear property of the cavity system, we can create a Hamiltonian for the qubit ensemble with a form $H_{\textrm{AC}}(n)=\textrm{Diag}\left\{0,\ldots,0,C_n,0,\ldots\right\}$ where there is only one non-zero element on the diagonal $(n+1)$-th term.
Here the matrix is labeled by the Dicke state
$|S,S_z = -S+m\rangle$ with an index $m$. 

\begin{figure}[htbp]
    \includegraphics[width=0.4\textwidth]{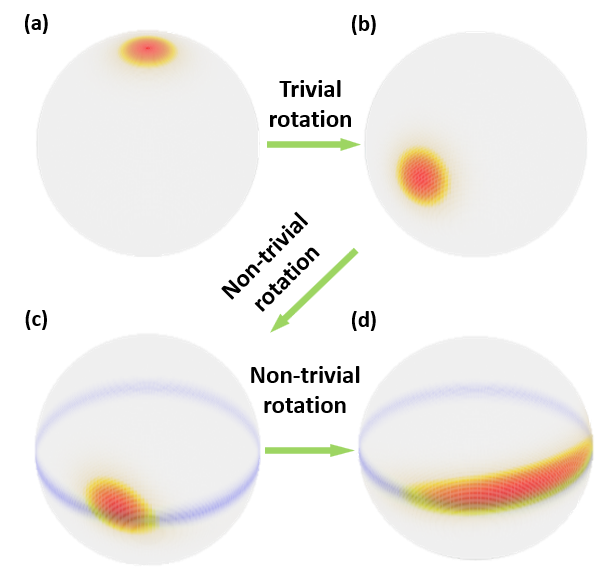}
    \caption{The illustration of how wave functions evolve represented by the Husimi-Q functions on the Bloch sphere with $N=50$.
    (a) A CSS is initialized in $\left|\downarrow\right\rangle ^{\otimes N}$. 
    (b) A trivial rotation rotates the CSS approaching the boundary shown as a blue stripe in (c) and (d).
    (c) Applying the non-trivial rotation, the wave function cannot penetrate the blue stripe and it has to be distorted.
    (d) Continuing the rotation, the wave function is highly squeezed around the blue stripe and the entanglement is created.}
    \label{fig:SSS_demo}
\end{figure}

Now we consider the case of applying a global single-qubit rotation such as $\Omega \hat{S}_x$ and this can be simply achieved by applying a microwave coupling two energy levels of either atoms or superconducting qubits. If there is no cavity incident light, the rotation term $e^{-i\Omega \hat{S}_x t}$ will rotate the wave function on the surface of the Bloch sphere and this is a trivial scenario where no entanglement will be introduced (Fig.~\ref{fig:SSS_demo}a and b). If there is a light with the frequency $\omega_n$, there will be a barrier at the location of $\left|S, -S+n\right\rangle$ on the Bloch sphere because this particular Dicke state is shifted away. 
When the wave function is rotated approaching the state $\left|S,-S+n\right\rangle$, it cannot fully penetrate into this barrier and has to be distorted and turns around on the Bloch sphere (Fig.~\ref{fig:SSS_demo}c and d).
In fact, this light is tearing the original Hilbert space into pieces and then the single-qubit rotation under the light on $e^{-i(H_\textrm{AC}(n)+\Omega \hat{S}_x)t}$ becomes highly non-trivial near the torn boundary. 
Thus, the entanglement emerges near this boundary.

\section{Creating entanglement via non-trivial rotations}

In this section, we describe the details how we create arbitrary symmetric entangled states, and also the considerations of real experimental situations with finite cooperativity, cavity linewidth, and qubit decoherence from the spontaneous decay.
Here, each qubit has a spontaneous decay rate $\Gamma$ in the excited state $|e\rangle$ and the cavity also has a finite linewidth $\kappa$. For a monochromatic incident light with a frequency $\omega_l=\omega_c+\delta$, the cavity amplitude transmission function for $n$ qubits in $\left|\uparrow\right\rangle$ is given by \cite{zhao2021creation,tanji2011advances}
\begin{equation}
\mathcal{T}(\delta, n)=\frac{1}{1+\frac{n \eta}{1+4(\Delta+\delta)^2 / \Gamma^2}-2 i\left[\frac{\delta}{\kappa}-n \eta \frac{(\Delta+\delta) / \Gamma}{1+4(\Delta+\delta)^2 / \Gamma^2}\right]}.
\label{trans}
\end{equation}
Here, $\eta=4 g^2 /(\Gamma \kappa)$ is the cooperativity while the angular momentum is $S_z=-S+n$. This formula has included the spontaneous emission into the free space which will also lead to a broadening of the cavity linewidth.
Note that we have assumed $|\Delta|\gg\Gamma$,  $|\Delta|\gg|\delta|$, and $\left|\downarrow\right\rangle$ state is off-resonant with the cavity mode. Generally, qubits on $\left|\downarrow\right\rangle$ state also contribute to scattering events during the evolution, which will be taken into consideration in the followings during the numerical calculations.

Here we derive our model in the strong coupling regime of the Tavis-Cummings (TC) model \cite{tavis1968exact,tavis1969approximate}, and the TC model may break down according to a recent research \cite{blaha_beyond_2022}. The breakdown of the TC model is due to a large optical depth (OD) of the qubit ensemble. Such a large OD absorbs the cavity light and strongly attenuates its propagation inside a cavity. Therefore, all qubits cannot experience the same quantized field.
For our scheme, the TC model is still valid under the condition of $|\Delta|\gg \Gamma$ according to the results in Ref.~\cite{blaha_beyond_2022}, where a direct quantitative comparison has already been provided. Here we summarize that the major reason is when $|\Delta|\gg \Gamma$, the spontaneous decay into the free space is strongly suppressed due to off-resonance and then the absorption of the qubit ensemble becomes negligible.

When $\omega_s=g^2/\Delta$ is larger than $\kappa$, the shifted cavity lineshape under each Dicke state $|S,-S+n\rangle$ is resolved where transmission spectrums do not overlap with each other. 
Therefore, the intra-cavity light intensity will be negligibly small if we choose $\delta=m\omega_s$ with $m\neq n$. 
In a practical cavity system with a finite cavity linewidth, a modified non-Hermitian Hamiltonian describes the energy shift of the Dicke states under spontaneous decay of excited states:
\begin{eqnarray}
    H_{\rm{AC}}(\delta,C)&=&C\times\rm{Diag}\{0\times|\mathcal{T}(\delta,0)|^2,\nonumber\\
    & &1\times|\mathcal{T}(\delta,1)|^2,2\times|\mathcal{T}(\delta,2)|^2,\ldots\}\label{realH}
\end{eqnarray}
Here $C=\hbar \delta_{\rm AC}\left(1-\mathrm{i} \frac{\Gamma}{2 \Delta}\right)/\left|\mathcal{T}\left(\delta, \delta/\omega_s\right)\right|^2$ is a coefficient which is proportional to the incident light intensity. The real part of Eq.~\ref{realH} characterizes AC Stark shifts for different Dicke states and the imaginary part characterizes the spontaneous-decay-induced decoherence.
Besides sending one monochromatic light, we can also use a light with multiple frequencies $\omega_c+\delta_i$.
Therefore, the overall Hamiltonian of sending light to the resonator becomes $H_{\textrm{AC}}=\sum_i  H_{\rm{AC}}(\delta_i,C_i)$.
Here $\delta_i=n_i\omega_s$ can be chosen from $n_i\in\{0,1,\ldots,N\}$ and then $H_{\textrm{AC}}$ can be tuned as an arbitrary diagonal Hamiltonian by linear combinations of $ H_{\rm{AC}}(\delta_i,C_i)$. 
Here we use a non-Hermitian Hamiltonian which is enough for calculating the fidelity, and we provide a more detailed explanation in Appendix \ref{nonHermitian}. For the performance of spin squeezing, we also provide detailed discussions in Appendix~\ref{SSSIdeal}.

What we did next is just alternatively applying the trivial and non-trivial single-qubit rotations $\Omega \hat{S}_\phi$ of all qubits, corresponding to global single qubit rotation with or without the incident light. Here $\hat{S}_\phi=\hat{S}_x\cos\phi+\hat{S}_y\sin\phi$.
First, all $N$ qubits are initialized in $\left|\downarrow\right\rangle^{\otimes N}$. Then we use the trivial rotation, and the state is rotated into another coherent spin states (CSS) with a form of 
$|\textrm{CSS}\rangle=e^{-i\Omega \hat{S}_{\phi_1} t_1}\left|\downarrow\right\rangle^{\otimes N}$. Next we apply the non-trivial rotation. This leads to a new state 
\begin{eqnarray}
    |\psi_1\rangle &=&e^{-i\left[H_{\textrm{AC,1}}+\tilde{\Omega}_1 \hat{S}_{\phi'_1}\right]\tilde{t}_1}|\textrm{CSS}\rangle\nonumber
    \\
    &=&e^{-i\left[H_{\textrm{AC,1}}+\tilde{\Omega}_1 \hat{S}_{\phi'_1}\right]\tilde{t}_1}e^{-i\Omega \hat{S}_{\phi_1} t_1}\left|\downarrow\right\rangle^{\otimes N}. 
\end{eqnarray}

After one round of the operations, we label this as one sequence with two steps and then we repeat this process again. Each time, we pick up different barrier positions and set different rotation angles or phases. After $j$-th sequences, the state $|\psi_j\rangle$ can be written by a recursive form as
\begin{equation}
    |\psi_j\rangle=e^{-i\left[H_{\textrm{AC,j}}+\tilde{\Omega}_j \hat{S}_{\phi'_j}\right]\tilde{t}_j}e^{-i\Omega \hat{S}_{\phi_j} t_j}|\psi_{j-1}\rangle.
\end{equation}
By repeatedly applying the above sequence with two steps, we are able to approach any target entangled state of total angular momentum $N/2$ with arbitrarily high precision, and we give a mathematical proof in Appendix \ref{proof} based on the representation of the Lie algebra in SU$(N+1)$.

\section{Examples of entangled states}

While by repeatedly applying trivial and non-trivial rotations, we can create any entangled states, most of metrologically useful entangled states can actually be created simply within one or very few steps, and the convergence of fidelity is very fast, which we will exemplify in the following.
Within one sequence, the SSS are created with a squeezing parameter close to the results of two-axis twisting squeezing which is the HL. Meanwhile, we can also generate many useful entangled states within two or three steps and the fidelity can be higher than 99\%.

\subsection{Spin squeezed state}
 
\begin{figure}[htbp]
    \centering
    \includegraphics[width=0.5\textwidth]{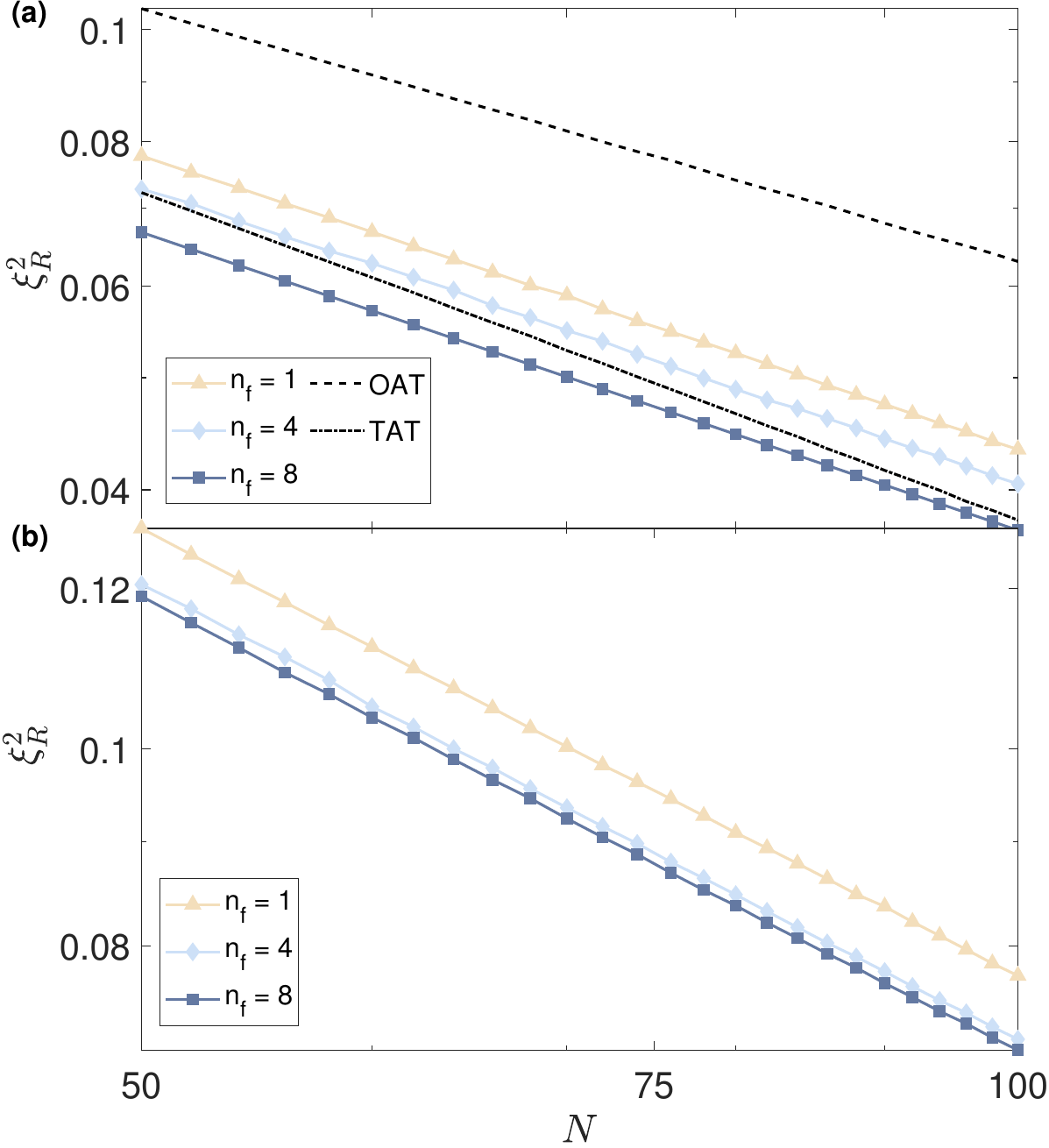}
    \caption{One sequence squeezing with the squeezing parameter $\xi_R^2$ versus the qubit number $N$ in the ideal case (a) and $^{87}\rm Rb$ case with loss (b) with $\eta=200$. Here we fit $\xi^2_R$ with a scaling relation $\xi^2_R=\alpha/N^\beta$. 
    In both panels, $n_f$ is the frequency number in the incident light.  
    In panel (a), $\beta$ equals to 0.843 ($n_f=1$), 0.851 ($n_f=4$), and 0.855 ($n_f=8$). For comparison, we also plot the relation of $\xi^2_R\sim 1/N^\beta$ for the OAT and TAT spin squeezing with dashed and dashed-dotted lines, where $\beta$ equals to 0.724 (OAT) and 0.942 (TAT).
    In panel (b), the decoherence from incident photon scattering is taken into considerations. The scaling factor $\beta$ equals 0.731 ($n_f=1$), 0.745 ($n_f=4$), and 0.751 ($n_f=8$).}
    \label{fig:SSS}
\end{figure}

We first take the creation of SSS as an example. 
We use the Wineland parameter \cite{wineland1992spin}  $\xi_{R}^2=S\left(\Delta S_{\perp}\right)_{\min }^2 /|\langle\mathbf{S}\rangle|^2$ to characterize the entanglement. Compared with the criterion from Kitagawa and Ueda \cite{kitagawa1993squeezed}, $\xi_{R}^2$ is more strict and robust since it includes the considerations of the curvature of the Bloch sphere.
Here, $\left(\Delta S_{\perp}\right)_{\min }^2$ is the minimum of the fluctuation $\left(\Delta S_{\perp}\right)^2=\left\langle S_{\perp}^2\right\rangle-\left\langle S_{\perp}\right\rangle^2$ for the spin component perpendicular to the mean spin direction and $|\langle\mathbf{S}\rangle|$ is the mean spin length.
To show the power and the robustness of this scheme, 
we restrict ourselves by creating SSS within only one sequence in the discussions below.

For a resonator with a cooperativity $\eta=200$,
we control the form of diagonal terms via creating multiple frequencies components for the incident light which can be achieved by an electro-optical modulator (EOM).
We first consider the ideal case without the spontaneous decay $\Gamma$, such as $^{171}$Yb atoms in an optical cavity where the spontaneous decay $\Gamma$ is only 7~mHz. Recent spin squeezing experiments in the optical transition of Ytterbium \cite{braverman2019near} also supports this treatment.
Here we use $n_f$ to label the number of frequencies used in the incident light.
To understand this process more intuitively, 
we plot how the wave function evolves under the $H_{\rm AC}$ term near a certain Dicke states in Fig.~\ref{fig:SSS_demo}.
It is shown that when the population rotates approaching this boundary, it is stretched into a strip shape, corresponding to a SSS.
The whole procedure can be simply written as:
\begin{equation}
    \left|\psi_{\textrm{SSS}}\right\rangle = e^{-i(\Omega S_x + H_{\rm AC})t_2} e^{-i(\Omega S_x)t_1}\left|\downarrow\right\rangle^{\otimes N}.\label{eq:sss}
\end{equation}

We show in the Fig.~\ref{fig:SSS}(a) the scaling of squeezing parameter $\xi_{R}^2$ as a function of total qubit number $N$ under $\eta=200$.
Under the optimized parameters (the optimized parameters can be found in Appendix~\ref{SSSIdeal}), the squeezing parameter shows a similar power-law scaling $\xi^2_R =  \alpha/N^\beta$ with respect to $N$ for different $n_f=1$, $4$, and $8$.

As our numerical calculation reveals, while the offset changes, the scaling factor $\beta$ only varies slightly from 0.843 to 0.855 with the increasing of $n_f$ from 1 to 8.
The relation of squeezing parameter $\xi_R^2$ versus the total qubit number $N$ of the one-axis twisting (OAT) scheme (dashed line) and the two-axis counter-twisting (TAT) scheme (dashed-dotted line with $\beta=0.94$) are also shown in Fig.~\ref{fig:SSS}(a).

\begin{figure}[htb]
    \centering
    \includegraphics[width=0.48\textwidth]{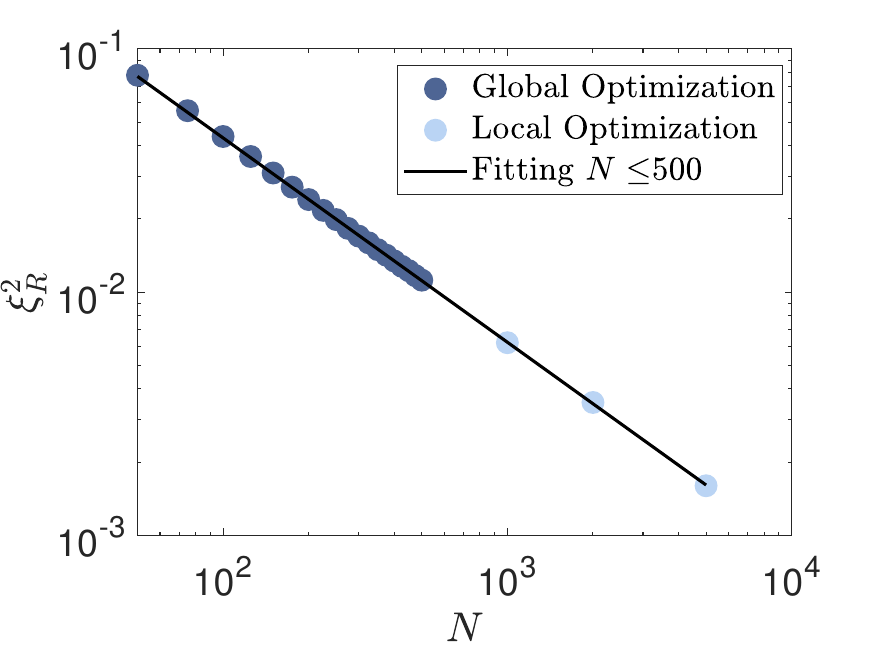}
    \caption{Spin squeezing scalability. We first find global optimized parameters at $N\leq500$ (gray dots). Then we fit $\xi_{R}^2$ based on data found at $N\leq500$ (black solid line) and predict $\xi_{R}^2$ we can achieve at $N>500$. To obtain correct parameters for $N>500$, e.g. $N = 1000$, we fit the optimized parameters found with $N\leq500$, and conduct local optimizations on these parameters extrapolated from small $N\leq500$. We find the squeezing parameters $\xi^2_R$ for large $N=1000$, 2000, and 5000 (light blue) matches the extrapolation from $\xi_{R}^2$ for $N\leq500$.
    }
    \label{fig:SSS_scalability}
\end{figure}

Here we perform the exact calculations and global parameters optimizations for $N$ from 50 to 100 in Fig.~\ref{fig:SSS}. However, our conclusion still works very well for a large $N$. We can extrapolate these parameters by fitting data in small $N$, e.g. in the range of $N=50$ to 500. Once we obtain the extrapolated values for a large $N$, we only need to perform local optimizations and the optimized $\xi^2_R$ exactly matches the relation in Fig.~\ref{fig:SSS}(a) and Fig.~\ref{fig:SSS_scalability}.
We extrapolate the optimized parameters and the scaling relation of $\xi^2_R$. Then, we also calculate the corresponding $\xi^2_{R}$ for $N=1000,$ 2000, and 5000 by local optimization, and the results also matches the extrapolation.
Therefore, our scheme is scalable for generating SSS and the squeezing parameter is better than OAT and close to TAT, i.e. HL, meanwhile this scheme only needs very simple control on/off of the cavity fields.

Furthermore, to show the practical and broad applications of our scheme, we calculate the achievable SSS with a strong spontaneous decay $\Gamma$ and cooperativity $\eta=200$. 
We choose $^{87}\rm Rb$, the most common element in atomic clocks and interferometers, as the candidate.
Here two $5S_{1/2}$ hyperfine ground states $\left|\downarrow\right\rangle=|F=1\rangle$ and $\left|\uparrow\right\rangle=|F=2\rangle$ are separated by $2\pi \times 6.8$~GHz, and an excited state $\left|e\right\rangle$ in $5P_{3/2}$ has a spontaneous-decay rate $\Gamma = 2\pi \times 6$~MHz.
As shown in Fig.~\ref{fig:SSS}(b), the optimized squeezing parameter shows similar power-law scaling for different $n_f$.
Different from the ideal-cavity situation above, when the photon scattering appears, one needs to find a balance between further stretching the spin wave function for better $\xi_R^2$ and the additional variance induced by decoherence from intra-cavity light that deteriorates the metrological gain. When $\Delta$ increases to avoid scattering, cavity light gradually couples both ground states to the excited state. Therefore, we need to include the corrections of the state $\left|\downarrow\right\rangle$ to Eq. \ref{simplifiedEq} and \ref{trans} in the numerical calculations.
When the spontaneous decay is taken into account, the scaling factor $\beta$ is robust and only decreases a little. 
In Fig.~\ref{fig:SSS}(b), we have taken decoherence into considerations, and thus the vertical axis is different.
Moreover, since the OAT and TAT spin squeezing models are ideal without consideration of decoherence, we show these two lines only in Fig.~\ref{fig:SSS}(a).

The detailed parameters and the calculations can be found in Appendix~\ref{SSSIdeal}.
The parameter $\Delta$ is fixed during the optimization  process that do not consider decoherence, and further enhancement on the squeezing parameter can be achieved by adding $\Delta$ into the optimization parameter set.
We also calculate the performance of spin squeezing for superconducting-qubit systems in Appendix~\ref{SSSIdeal}.

For practical applications against noises, we also include the considerations of stochastic errors in either the light intensity or the evolution time. We numerically model and simulate these two types or errors in Appendix~\ref{imperfect}. Under 5\% intensity fluctuations, $\xi^2_R$ changes very slightly which is almost negligible. This supports the robustness of our scheme for spin squeezing. 

\begin{figure}[htbp]
  \centering
  \includegraphics[width=0.5\textwidth]{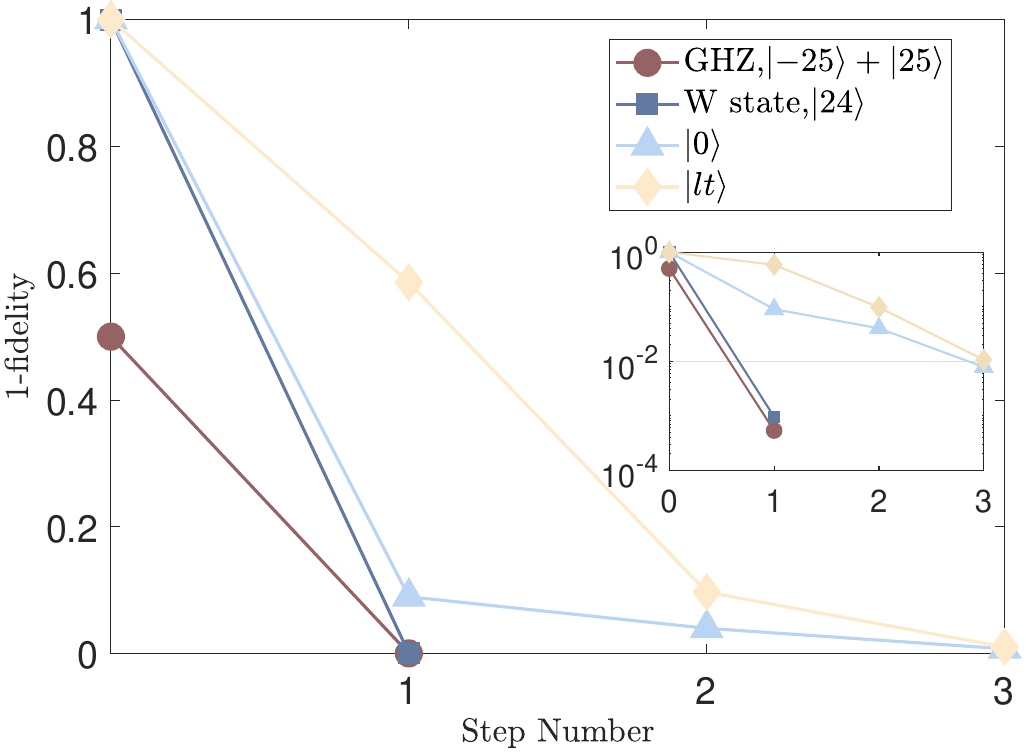}
  \caption{
  The entangled-state fidelity versus the step number. 
  The vertical axis is labeled by the difference between 1 and fidelities.
  The main plot is in a linear scale and the inset is the same data but in a logarithmic scale. 
  Each step contains one non-trivial rotation. Here we use $N=50$ to demonstrate. The target states include the GHZ state (circle), W state (square), Dicke state $|0\rangle$ (triangle), and lantern state $|lt\rangle$ (diamond).
  For the GHZ state and W state, within only one step, the fidelity goes over 0.999. 
  For the other two states $\left|0\right\rangle$ and $\left|lt\right\rangle$, the fidelity goes over 0.99 with three steps.
  }
  \label{fig4}
\end{figure}

\begin{figure*}[htb]
    \centering
    \includegraphics[width=0.9\textwidth]{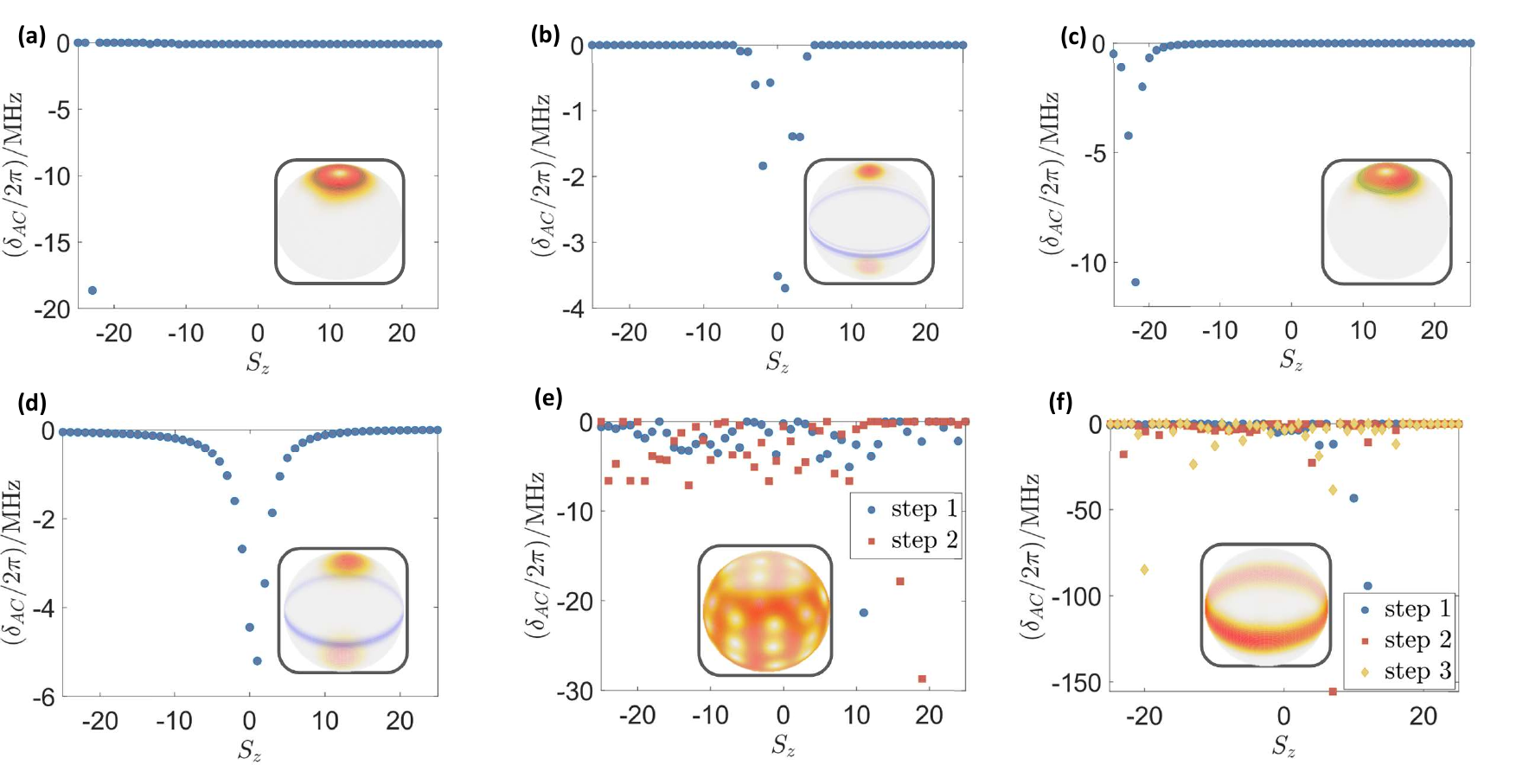}
    \caption{
     The form of $H_{\rm AC}$ with corresponding barriers (blue belts) and wave functions on the Bloch spheres with the rotation Rabi-frequency
    $\Omega =2\pi \times 0.2$~MHz. The scatter plots show the energy shift for different Dicke states, and the subplots show the result of generated entangled states (Husimi-Q functions) with corresponding barriers used in non-trivial rotations.
    (a) The ideal situation generating a W state with one step, $t=0.353\mu s$. (b) The ideal situation generating a GHZ state with one step, $t=2.49\mu s$. (c) The $\eta = 1000$ $^{87}\rm Rb$ cavity system generating a W state with one sequence and monochromatic light, $t_1 = 0.151\mu s, t_2=0.234\mu s$. (d) The $\eta = 1000$ $^{87}\rm Rb$ cavity system generating a GHZ state with one sequence and monochromatic light, $t_1 = 0.414\mu s, t_2=2.17\mu s$. (e) Two steps generating the lantern state $|lt\rangle=(|-15\rangle+|-5\rangle+|5\rangle+|15\rangle)/2$. Blue and red corresponds to each $H_{\rm AC}$ during the first and second steps, $t = (1.55,0.305)\mu s)$. (f) Three steps generating $|0\rangle$. Blue, red, and orange corresponds to each $H_{\rm AC}$ during the first, second, and third steps, $t = (1.03,0.318,0.258)\mu s$.}
    \label{figWallsQ}
\end{figure*}

\subsection{Other entangled states}
We now explore the creation of other entangled states.
In principle, by repeating applying trivial and non-trivial rotations, our scheme can generate any states on the surface of the $N$ spin-1/2 Bloch sphere. 
Equivalently, one can approach a target state at an arbitrarily high fidelity with adequate steps.
In this and next sections, we consider the scenario of only sequentially applying non-trivial rotations for numerical calculation simplicity.
We use $N=50$ qubits for demonstrations and we pick up the GHZ state, W state, Dicke state, and lantern state as examples.
We will use Dicke states $|S_z\rangle$ to express these states.
The GHZ state corresponds to an equal superposition of $\left|-25\right\rangle$ and $\left|25\right\rangle$.
The W state corresponds to $\left|-24\right\rangle$.
For the Dicke state, we choose $|0\rangle$ to exemplify.
The lantern state $\left|lt\right\rangle$, which resembles a lantern in the Husimi-Q representation, is defined as an equal superposition of $\left|-15\right\rangle$, $\left|-5\right\rangle$, $\left|5\right\rangle$ and $\left|15\right\rangle$.
The definition of the fidelity of these four states in our calculation is summarized in Appendix~\ref{fidelity}.
In Fig.~\ref{fig4} we plot the fidelity of these four states versus the step number.
The fidelity of creating a target state increases as we increase total step numbers, where each step only contains an non-trivial step for simplicity.
The fidelities rapidly converge to 1 within only very few steps.
For the entangled states illustrated, we always obtain a fidelity over 0.99 within 3 steps.
Notably, within only one step, a GHZ state with fidelity over $0.999$, and a W state with fidelity over $0.99$ can be obtained.
We will discuss in detail the generation of these two important states in the next section.
We plot the wave function of these states and corresponding $H_{\rm AC}$ in Fig.~\ref{figWallsQ}. 
Since the microwave rotation is easy to implement with negligible loss in real experiments \cite{liu2021infidelity, sheng2018high}, further improvement of the above result includes inserting some trivial rotations between non-trivial ones in order to achieve faster convergence of fidelity approaching unity.

\subsection{GHZ state and W state\label{GHZandW}}
\begin{figure}[htb]
    \includegraphics[width=0.4\textwidth]{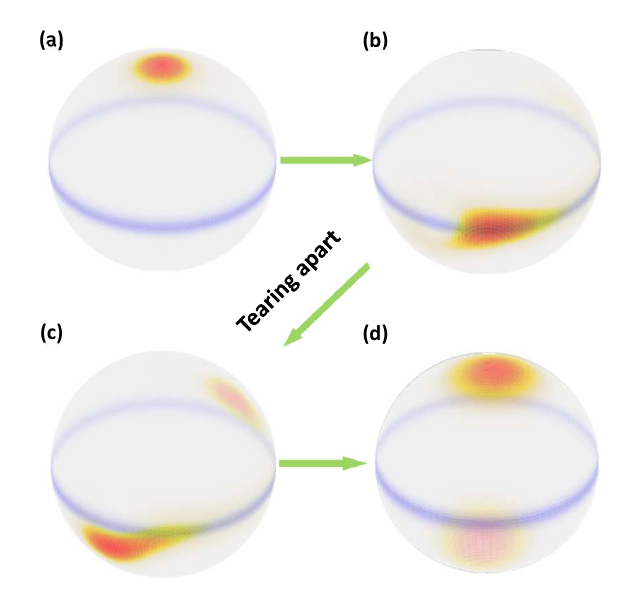}
    \caption{The illustration of creating a GHZ state represented by the Husimi-Q functions on the Bloch sphere with $N=50$.
    (a) A CSS is initialized in $\left|\downarrow\right\rangle ^{\otimes N}$. 
    (b) A non-trivial rotation directly rotates the CSS approaching to 
    the boundary shown as a blue stripe in each panel, and the boundary starts to tear the wave function.
    (c) The CSS is torn into two parts while half is transmitted and half is reflected.
    (d) Continuing the rotation, the wave function is reshaped into the superposition of two opposite CSS, which is a GHZ state.}
    \label{fig:GHZ_demo}
\end{figure}

In this section, we discuss the generation of the GHZ state and W state more in detail.
For GHZ state, we set the boundary near the equator, i.e. non-zero diagonal terms near $\left|S_z=0\right\rangle$.
The intensity of the boundary is set as a weak barrier such that half of the wave function can pass through the barrier and half of the wave function has to be reflected by the barrier (Fig.~\ref{fig:GHZ_demo}).
By continuing the global single-qubit rotations, the transmitted part is rotated into the south pole $|\uparrow\rangle^{\otimes N}$ of the Bloch sphere, and the reflected part takes a U turn around the barrier and comes back to the north pole $|\downarrow\rangle^{\otimes N}$, leading to a GHZ state. 

For the creation of W state, one need to set the boundary at $\left|S_z = -N/2+2\right\rangle$ so that the effective Hilbert subspace consists only of two states, $\left|S_z = -N/2\right\rangle$ and $\left|S_z = -N/2+1\right\rangle$.
Under this condition, any global single-qubit rotations will behave like a two-level Rabi oscillation in between these two states.

\begin{figure}[htb]
    \includegraphics[width=0.5\textwidth]{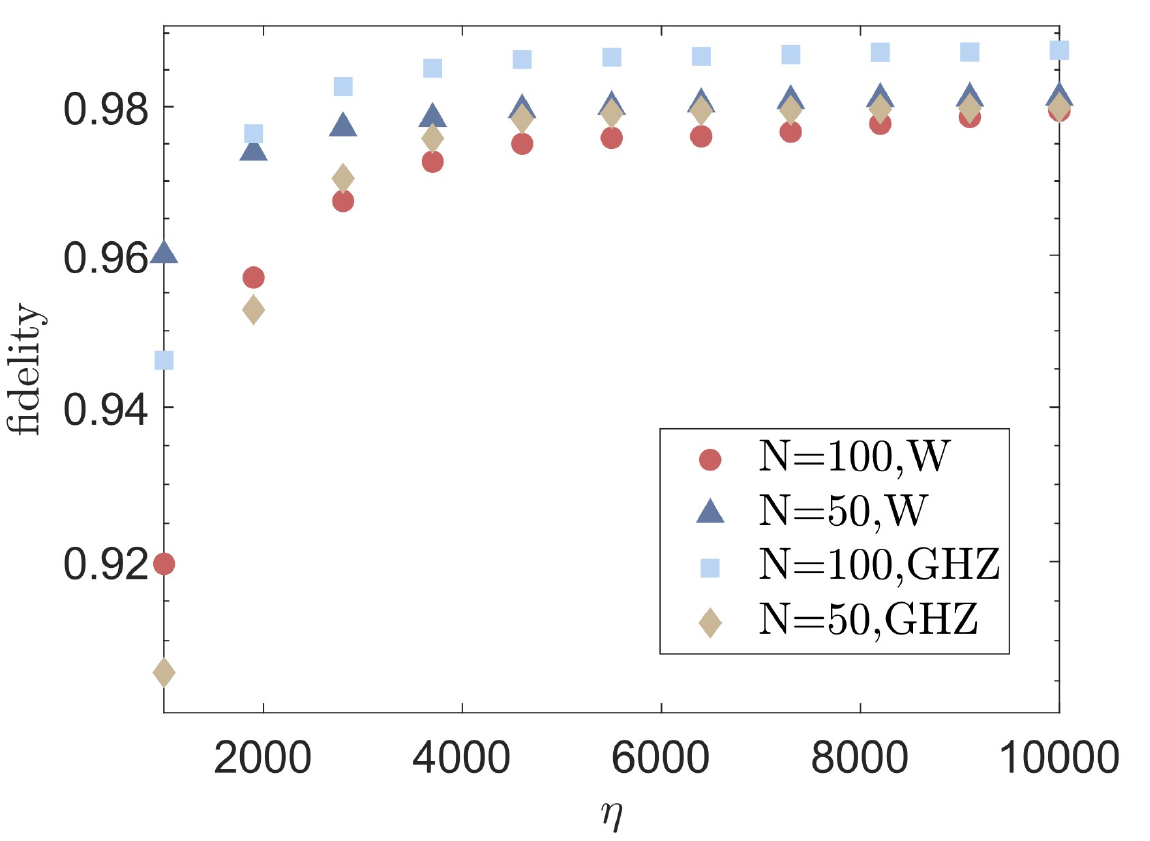}
    \caption{Fidelity for generating the GHZ state and W state with single monochromatic light by one non-trivial rotation under different cooperativity $\eta$ and qubit number $N$. The fidelity of the generated states increases monotonically with $\eta$.}
    \label{figGHZscaling}
\end{figure}

We show the optimized diagonal terms in the Fig.~\ref{figWallsQ}(a) and (b) for creating these two states with only one step.
Here we need to utilize multiple-frequency light to engineer the diagonal terms of $H_{\rm AC}$.
Then, we discuss an experimentally-friendly scheme for creating these two states with only monochromatic light (Fig.~\ref{figWallsQ}(c) and (d)), where the diagonal terms in $H_{\rm AC}$ are in Lorentzian shape according to Eq.~\ref{trans}.
The process of generating the GHZ state is shown in Fig.~\ref{fig:GHZ_demo}. One can notice that comparing to the most optimized case, the GHZ state generated by monochromatic light is a little tilted, and a trivial rotation can further improve the fidelity.
With the decoherence and loss taken into accounts, the fidelity of GHZ states and W states can be higher than 0.9 and 0.96 with a cooperativity $\eta = 1000$, respectively.
We also check the influences of information leakage due to the cavity photons (Appendix~\ref{informationLeakage}) correlated with the Dicke states. We find it does not hurt our scheme where the fidelity only decreases by at most 2.0\%.
Such a cooperativity can be readily achieved in the superconducting-qubit system, where $g \sim 100$~MHz, $\Gamma \sim 1$~kHz and resonator with an effective $\kappa\sim10$~kHz has already been realized in the experiment \cite{rigetti2012superconducting,chakram2021seamless,milul2023superconducting}. 
Since $\eta$ quantifies the ability to resolve adjacent Dicke states, larger $\eta$ helps to obtain these two states with a higher fidelity.
For large enough $\eta$, with single monochromatic light and within one non-trivial rotation, we found the fidelity of the generated GHZ state and W state can reach as high as $0.98$ with loss taken into consideration, as shown in Fig.~\ref{figGHZscaling}. For larger $N=100$, the achievable fidelity of the GHZ state further increases.

\section{Conclusion and Outlook}
In conclusion, we find an efficient yet simple scheme to generate arbitrary symmetric entangled states via single-qubit operations in a torn $N$ spin-1/2 Hilbert space.
With currently available technology, we are able to generate any superposition of Dicke states, including spin squeezed states, W states, and GHZ states.
This scheme utilize the tearing capability of the intra-cavity light on the Bloch sphere, inducing non-linear interaction among these Dicke states, thus generating highly entangled states in combination with single-qubit rotation. The only multiple-qubit operation is just switching on/off light or microwave which is pratically simple in experiments.
We believe our scheme can be of important guidance for experiments in neutral atom-cQED system and superconducting qubits, and variations of this protocol may be used for other system including mechanical oscillators-cavity system and quantum dots ensembles.

\section{Acknowledgement}
This work is supported by National Key Research and Development Program of China (2021YFA0718303, 2021YFA1400904), National Natural Science Foundation of China (92165203, 61975092, 11974202), and Tsinghua University Initiative Scientific Research Program.

\appendix

    \section{Derivation of effective Hamiltonian}
\label{effHamiltonian}
In the main text, we convert Eq.~\ref{originalEq} into Eq.~\ref{simplifiedEq} by the method of adiabatic elimination \cite{brion2007adiabatic}. Here, we give a detailed derivaiton about this transformation.
The original Hamiltonian is
\begin{equation}
    \frac{H'}{\hbar}=\sum_i\left[-\omega_0\left|\downarrow\right\rangle_i\langle \downarrow|_i-\Delta|e\rangle_{i}\langle e|_i+(g\hat{c}^\dagger|\!\uparrow\rangle_i\langle e|_i+h.c.)\right],\label{originalEq}
\end{equation}
All relevant symbols have the same definitions as in the main text. Here we choose a relative reference frame by canceling the photon energy term $\hbar \omega_c\hat{c}^{\dagger}\hat{c}$. Then, we can reduce the excited energy into the detuning $\Delta$. 
Furthermore, because all states $\left|\downarrow\right\rangle_i$ do not have transitions to states $\left|e\right\rangle_i$, we can drop the term $-\omega_0\left|\downarrow\right\rangle_i\left\langle \downarrow\right|_i$ and inspect the remaining Hamiltonian by choosing an interaction picture. Then it becomes
\begin{equation}
    \frac{H''}{\hbar}=\sum_i\left(-\Delta|e\rangle_{i}\langle e|_i+g\hat{c}^\dagger|\!\uparrow\rangle_i\langle e|_i+g\hat{c}|e\rangle_i\langle \uparrow\! |_i\right).
\end{equation}
Here we care about the perturbations on the ground state manifold with $|\uparrow\rangle$ and $|\downarrow\rangle$.
If there are $n$ atoms in the state $|\uparrow\rangle$ and $k$ photons in the cavity, we may write the wave function as $|\uparrow\rangle^{\otimes n}\otimes |k\rangle$. And it will couple to the state $|E\rangle\otimes|k-1\rangle$ where $|E\rangle$ is a superposition state of $n$ atoms and has a form as
\begin{equation}
    |E\rangle=\frac{1}{\sqrt{n}}\left(|e\uparrow\uparrow\ldots\uparrow\rangle+|\uparrow e\uparrow\ldots\uparrow\rangle+\ldots+|\uparrow\uparrow\uparrow\ldots e\rangle\right).
\end{equation}
Then we write out the first-order perturbative wave function as 
\begin{equation}
A_0\left|\uparrow\right\rangle^{\otimes n}\otimes |k\rangle+A_1 |E\rangle\otimes|k-1\rangle
\end{equation}
and the Schr\"odinger equation becomes
\begin{eqnarray}
i\frac{d}{d t} A_0&=&g\sqrt{nk} A_1,
\\
i\frac{d}{d t} A_1&=&-\Delta A_1 + g\sqrt{nk} A_0.
\end{eqnarray}
According to the adiabatic elimination, we set $dA_1/dt=0$ and then obtain $A_1=A_0 g\sqrt{nk}/\Delta$. Therefore, the time evolution equation for $A_0$ becomes
\begin{equation}
    i\frac{d}{dt}A_0=A_0 g^2 nk/\Delta.
\end{equation}
It is equivalent to that the state $|\uparrow\rangle^{\otimes n}\otimes |k\rangle$ experiences an additional energy shift with an amount of $g^2 nk/\Delta$ which is the AC Stark shift. By writing this AC Stark shift into an operators' form where $n$ is the atom number in the state $\left|\uparrow\right\rangle$ and $k$ is the photon number in the cavity, we obtain the form as
\begin{equation}
    \frac{g^2}{\Delta}\hat{c}^\dagger \hat{c}\sum_i \left|\uparrow\right\rangle_i\left\langle\uparrow\right|_i.\label{AppexA1}
\end{equation}
By utilizing the equation 
\begin{eqnarray}
\left|\uparrow\right\rangle_i\left\langle\uparrow\right|_i&=&\frac{1}{2}\left(\left|\uparrow\right\rangle_i\left\langle\uparrow\right|_i+\left|\downarrow\right\rangle_i\left\langle\downarrow\right|_i\right) \nonumber \\
& &
+\frac{1}{2}\left(\left|\uparrow\right\rangle_i\left\langle\uparrow\right|_i-\left|\downarrow\right\rangle_i\left\langle\downarrow\right|_i\right) \nonumber \\
&=&1/2+\hat{s}_{z,i},
\end{eqnarray}
we write $\sum_i \left|\uparrow\right\rangle_i\left\langle\uparrow\right|_i$ as $\sum_i \left|\uparrow\right\rangle_i\left\langle\uparrow\right|_i=\hat{S}_z+S$ in Eq.~\ref{AppexA1} and obtain a new form as
\begin{equation}
\frac{H}{\hbar}= \frac{g^2}{\Delta}\left(\hat{S}_z+S\right) \hat{c}^{\dagger} \hat{c}.
\end{equation}
This proves the transformation from Eq.~\ref{originalEq} to Eq.~\ref{simplifiedEq} in the main text by adiabatic elimination.

\section{Non-Hermitian Hamiltonian and fidelity}
\label{nonHermitian}
In the main text, we use a non-Hermitian Hamiltonian to describe the time evolution of states. Here we show that it is enough to calculate the fidelity. For a complete description of the time evolution, we need to use the quantum jump operator and density matrix in a master equation. Based on this point and the dissipation theory, we can divide the density matrix $\rho$ into two parts
\begin{equation}
    \rho=\rho_{\textrm{coh}}+\rho_{\textrm{jump}},
\end{equation}
where $\rho_{\textrm{coh}}$ corresponds to a coherent evolution without any quantum jump event, and $\rho_{\textrm{jump}}$ describes the state with jumps happened. Here $\rho_{\textrm{coh}}$ can be calculated by a non-Hermitian Hamiltonian by directly adding a loss term into the energy parts, as in the main text.

Based on the classification of whether quantum jumps happen or not, both $\rho_{\textrm{coh}}$ and $\rho_{\textrm{jump}}$ have well-defined physical meaning along with corresponding quantum states. Therefore, $\rho_{\textrm{coh}}$ and $\rho_{\textrm{jump}}$ must be non-negative defined matrices to satisfy the fundamental requirement of a density matrix. It means that for any state $|\psi\rangle$, there must be $\langle\psi|\rho_{\textrm{jump}}|\psi\rangle\ge 0$. And this leads to
\begin{equation}
    \langle\psi|\rho|\psi\rangle\ge\langle\psi|\rho_{\textrm{coh}}|\psi\rangle.
\end{equation}
During the calculation of the fidelity, we use the definition $\mathcal{F}=\langle\psi|\rho_{\textrm{coh}}|\psi\rangle$ in the main text which only counts the contribution of the coherent part. According to the inequality above, we know our method is an underestimation of the actual performance. Therefore, it is safe to apply a non-Hermitian Hamiltonian to calculate the fidelity.

In Appendix \ref{SSSIdeal}, we also explain how we treat the spin squeezing under a non-Hermitian Hamiltonian following the methods from Ref.~\cite{schleier2010squeezing,li2022collective}, which have been tested and verified in many cavity spin-squeezing experiments.

\section{Proof of creating arbitrary states}
\label{proof}

In this section, we will prove mathematically that our scheme is able to generate arbitrary symmetric entangled states. In fact, a similar discussion on the completeness of controlling a quantum system has been investigated in a more general formalism in Ref.~\cite{schirmer_complete_2001}. For a complete readability, we will introduce the proof based on our scheme.
 
During the entanglement generations, we are using both trivial and non-trivial rotations with a form of,
\begin{equation}
    |\psi_j\rangle=e^{-i\left[H_{\textrm{AC,j}}+\tilde{\Omega}_j \hat{S}_{\phi'_j}\right]\tilde{t}_j}e^{-i\Omega \hat{S}_{\phi_j} t_j}|\psi_{j-1}\rangle.
\end{equation}
According to the Baker-Campbell-Hausdorff formula, we can rewrite this operation as an effective Hamiltonian $H_e$ with a form of
\begin{equation}
    |\psi_j\rangle=e^{-i\Omega \hat{S}_{\phi_j} t_j}e^{-iH_e\tilde{t}_j}|\psi_{j-1}\rangle.
\end{equation}
Here $H_e$ is given by
\begin{equation}
    H_e=\sum_k\frac{1}{k!}\left[\left(-i\Omega \hat{S}_{\phi_j} t_j\right)^{(k)},H_{\textrm{AC,j}}+\tilde{\Omega}_j \hat{S}_{\phi'_j}\right],
\end{equation}
where $[A^{(k)},B]=[A,[A^{(k-1)},B]]$ is a recursive form of commutators. Therefore, we want to prove that the commutators $[S^n_x,H_{AC,j}]$ and $[S^n_y,H_{AC,j}]$ as Lie-algebra can generate the full space of the special unitary Lie group SU$(N+1)$, where $n$ is an integer. This idea is similar to the way in Ref.~\cite{lloyd1995almost}.

Generalized Gell-Mann matrices are the Lie-algebra generators of the special unitary group SU$(N+1)$ which have been clearly studied in the maths \cite{kimura2003bloch}. Then our idea converts to that we can use the commutators or their linear combinations to construct this complete set of Gell-Mann matrices. Then based on this, we are able to prove that our operations can lead to any symmetric entangled states under the SU$(N+1)$ form.

First, we would like to give a brief introduction about the formalism of Gell-Mann matrices. Let $E_{(j,k)}$ denote the matrix with a 1 in the $(j,k)$-th entry and 0 elsewhere. This allows one to define three classes of matrices. The first class is symmetric:
\begin{equation}
\lambda^s_{(j,k)}=E_{(k,j)}+E_{(j,k)}.
\end{equation}
The second class is anti-symmetric:
\begin{equation}
\lambda^a_{(j,k)}=-i\left(E_{(k,j)}-E_{(j,k)}\right).
\end{equation}
The third class only contains diagonal matrices:
\begin{equation}
 \lambda^D_m=\sqrt{\frac{2}{m(m-1)}}\left[E_{(m,m)}-(m-1)E_{(m-1,m-1)}\right].
\end{equation}
Here the superscripts $s$, $a$, and $D$ correspond to symmetric, anti-symmetric, and diagonal. All the matrix in these three classes are one complete set of Gell-Mann matrices.

According to our scheme, it is intuitive to find that the third class of matrices is automatically satisfied by tuning different AC Stark shifts in $H_{\rm AC}$. Let's define a matrix set $\{Z_m|m\in \mathbb{Z}^+\}$ with a form of $(Z_m)_{(i,j)}=\delta_{i,j}\delta_{i,m}$ where $\delta_{i,j}$ is the Kronecker delta function and it is equivalent to $H_{\rm AC}$ and $\lambda^D_m$. Then, let's check the first order commutators,
\begin{equation}
    S_xZ_1-Z_1S_x = 
    \left[  
    \begin{array}{c c c c}
        0&-r_1 &... &0\\
        r_1 &0 &... &0\\
        \vdots &\vdots &\ddots &\vdots\\
        0 &0 &\cdots &0
    \end{array}
    \right],
\end{equation}
\begin{equation}
    S_yZ_1-Z_1S_y = i
    \left[
    \begin{array}{c c c c}
        0&r_1 &... &0\\
        r_1 &0 &... &0\\
        \vdots &\vdots &\ddots &\vdots\\
        0 &0 &\cdots &0
    \end{array}
    \right],
\end{equation}

\begin{equation}
    S_xZ_2-Z_2S_x = 
    \left[
    \begin{array}{c c c c c}
        0&r_1 &0 &\cdots &0\\
        -r_1 &0 &-r_2 &\cdots &0\\
        0 &r_2 &0 &\cdots&0\\
        \vdots &\vdots &\vdots &\ddots &\vdots\\
        0 &0 &0 &\cdots &0
    \end{array}
    \right],
\end{equation}

\begin{equation}
    S_yZ_2-Z_2S_y = i
    \left[
    \begin{array}{c c c c c}
        0&r_1 &0 &\cdots &0\\
        r_1 &0 &-r_2 &\cdots &0\\
        0 &-r_2 &0 &\cdots&0\\
        \vdots &\vdots &\vdots &\ddots &\vdots\\
        0 &0 &0 &\cdots &0
    \end{array}
    \right].
\end{equation}
Here $r_i$ are real numbers and matrix elements of $S_x$ and $S_y$. Then by linear combinations of $\left\{S_x Z_m-Z_m S_x|m\in\mathbb{Z}^+\right\}$ and $\left\{S_ y Z_m-Z_m S_y|m\in\mathbb{Z}^+\right\}$, we can construct parts of the first class and the second class of matrices with the restriction of $|i-j|=1$.

Now let's check the second-order commutators such as
\begin{equation}
    S_x^2Z_1-Z_1S_x^2 = 
    \left[
    \begin{array}{c c c c c}
        0&0 &-r^2_1 &\cdots &0\\
        0&0 &0 &\cdots &0\\
        r^2_1 &0 &0 &\cdots&0\\
        \vdots &\vdots &\vdots &\ddots &\vdots\\
        0 &0 &0 &\cdots &0
    \end{array}
    \right].
\end{equation}
Then it is clear that by linear combinations of the second-order commutators $\left\{S^2_x Z_m-Z_m S^2_x\right|m\in\mathbb{Z}^+\}$ and $\left\{S^2_y Z_m-Z_m S^2_y|m\in\mathbb{Z}^+\right\}$, we can construct parts of the first and second classes of matrices with the condition of $|i-j|=2$. Utilizing the same arguments, any $\lambda^s_{(j,k)}$ and $\lambda^a_{(j,k)}$ can be constructed. And this proves the completeness of our operations. Therefore, our scheme can lead to any symmetric entangled states connected by the SU$(N+1)$ group.

\begin{figure*}[htbp]
    \label{fig:SSS_Parameters}
    \centering
    \includegraphics[width=0.85\textwidth]{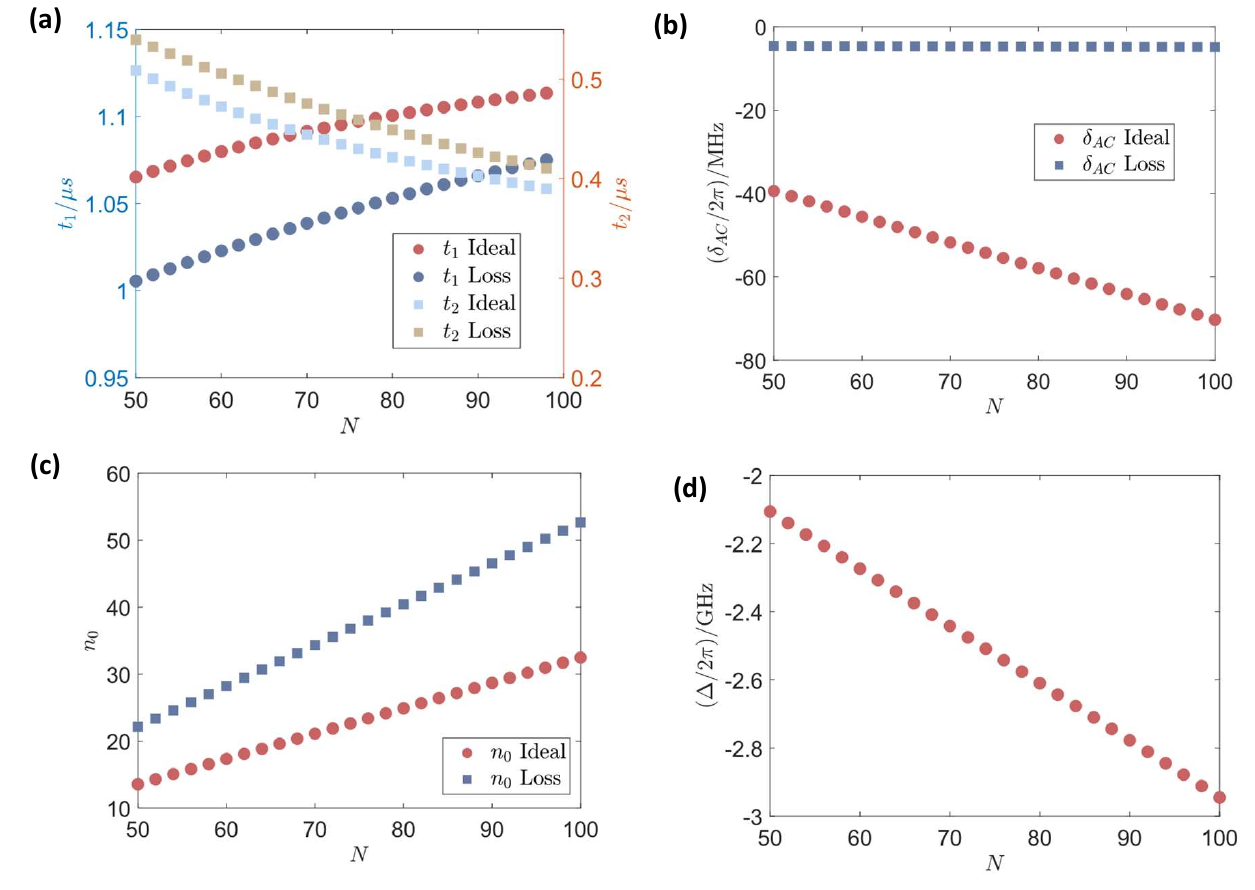}
    \caption{ 
    The globally optimized parameters for $N$ ranging from 50 to 100. In Fig (a), (b), and (c), we show trivial rotation time $t_1$, non-trivial rotation time $t_2$, $\delta_{\rm AC}$ and frequency order parameter $n_0$ for the ideal case and realistic $^{87}\rm Rb$ case with loss from the spontaneous decay taken into consideration. (d) corresponds to the optimization of $\Delta$ for the realistic $^{87}\rm Rb$ case.}
    \label{fig:SSS_Parameters_SC}
\end{figure*}

\section{The applicability in superconducting qubits \label{Superconducting}}
In this section, we discuss the applicability of our scheme in the superconducting-qubit system.
The experimental realization of employing on-chip microwave resonators and coupling them to superconducting qubits has a history of two decades \cite{wallraff2004strong}, and it creates
a solid-state analogue of conventional optical cavity QED with atoms.
The realized Hamiltonian of a driven cavity mode coupled to $N$ qubits is written as \cite{helmer2009measurement}
\begin{equation}
    \begin{aligned}
\hat{H} /\hbar& =\omega\left(\hat{c}^{\dagger} \hat{c}+\frac{1}{2}\right)+\sum_{i=1}^N \epsilon_i\hat{S}_i^z \\
& +\sum_{i=1}^N g_i^0\left(\hat{S}_i^{+} \hat{c}+\hat{S}_i^{-} \hat{c}^{\dagger}\right)+\frac{\kappa}{2}\left(\varepsilon  \hat{c}^{\dagger}+\hat{c} \varepsilon^*\right).
\end{aligned}
\end{equation}
The first term corresponds to a cavity mode with a frequency $\omega$, the second term corresponds to the energy splitting of each qubit, the third term describes the coupling of one qubit to the cavity mode with a coupling strength $g_i^0$, and the last term describes the external driving of the cavity.
Here $\hat{c}^\dagger$ and $\hat{c}$ are the cavity field creation and annihilation operators.

When the energy $\epsilon_i$ of each qubit is mismatched from the cavity resonant frequency $\omega$, the interaction becomes that
each qubit experiences a state-dependent phase shift from the cavity mode where a detail derivation can be found in Ref.~\cite{blais2004cavity}, and the effective Hamiltonian becomes 
\begin{equation}
\begin{aligned}
\hat{H}/\hbar & = \omega\left(\hat{c}^\dagger\hat{c}+\frac{1}{2}\right)+\sum_{i=1}^N \epsilon_i\hat{S}_i^z \\
& +\sum_{i=1}^N \frac{\left(g_i^0\right)^2}{\Delta_i} \hat{S}_i^z \hat{c}^\dagger\hat{c}+\frac{\kappa_{\text {cavity }}}{2}\left(\varepsilon \hat{c}^{\dagger}+\hat{c} \varepsilon^*\right) \\
& =\hbar\left[\omega+\sum_{i=1}^N \frac{\left(g_i^0\right)^2}{\Delta_i} \hat{S}_i^z\right]\left(\hat{c}^\dagger\hat{c}+\frac{1}{2}\right)\\
& +\sum_{i=1}^N \epsilon_i\hat{S}_i^z +\frac{\kappa}{2}\left(\varepsilon \hat{c}^{\dagger}+\hat{c} \varepsilon^*\right),
\end{aligned}
\end{equation}
where $\Delta_i=\omega-\epsilon_i$.
By adjusting the energy splitting $\epsilon_i$ and the coupling strength $g^0_i$ into the same magnitude, the interaction term becomes proportional to $\hat{S}_z \hat{c}^\dagger \hat{c}$ which is only different by a constant shift term comparing to Eq.~\ref{simplifiedEq} in the main text. Therefore, the same entanglement creation scheme can also be applied to the superconducting system.

\begin{figure}[htbp]
    \centering
    \includegraphics[width=0.45\textwidth]{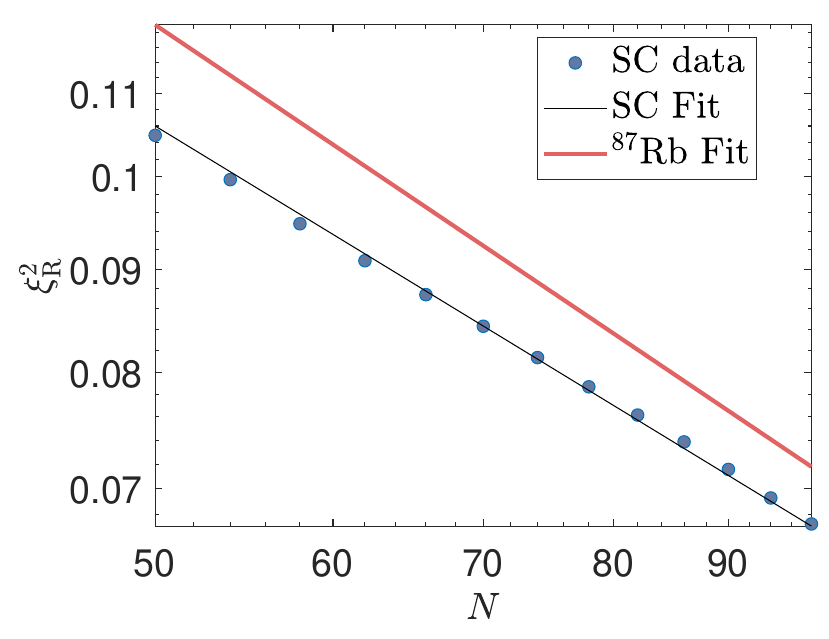}
    \caption{The spin squeezing parameter $\xi^2_{\textrm{R}}$ for the superconducting-qubit system where the qubit number $N$ ranges from 50 to 100 with a scaling $\xi^2_{\textrm{R}}\propto N^{-0.68}$ (black solid line) and $n_f = 1$. 
    Here each point has a better squeezing than the results of $^{87}\rm{Rb}$ system (red solid line) in the range of $N$ from 50 to 100.}
    \label{fig:SSS_SC}
\end{figure}

\section{The fidelity of GHZ, W, Dicke, and lantern  states\label{fidelity}}
In this section, we summarize the fidelity used in the main text for the GHZ state, W state, Dicke state $\left|0\right\rangle$, and lantern state $\left|lt\right\rangle$.
The definition of GHZ state, borrowed from \cite{sackett2000experimental}, is defined as an equal superposition of all spin-up and all spin-down state in our calculations.
For a given state with density matrix $\rho$, the fidelity is calculated as
\[
\mathcal{F} = \max_{\phi} \left( \langle \mathrm{GHZ}, \phi | \rho | \mathrm{GHZ}, \phi \rangle \right)
\]
where $|\mathrm{GHZ}, \phi\rangle=(\left|\uparrow\right\rangle^{\otimes N}+\mathrm{e}^{\mathrm{i} \phi }\left|\downarrow\right\rangle^{\otimes N}) / \sqrt{2}$.
Here we use the phase $\phi$ to match the known phase difference between $\left|\uparrow\right\rangle^{\otimes N}$ and $\left|\downarrow\right\rangle^{\otimes N}$ of the generated $\mathrm{GHZ}$ state. 
The above defined fidelity can also be calculated using the matrix elements $\rho_{m, n}$:
$$
\begin{aligned}
\mathcal{F}&=\frac{1}{2}(\rho_{N / 2, N / 2}+\rho_{-N / 2,-N / 2}\\
&+\left|\rho_{-N / 2, N / 2}\right|+\left|\rho_{N / 2,-N / 2}\right|)
\end{aligned}
$$
where $\rho_{m, n}$ corresponds to the coefficients of $|m\rangle\langle n|$ in $\rho$.
Furthermore, if decoherence and loss is taken into consideration in the state preparation process, as is discussed in Section~\ref{GHZandW}, any qubit undergoing spontaneous decay will destroy the whole state.
We thus describe the generated state with density matrix $\rho$ that is decomposed into two parts as $\rho=\rho_{\text {coh }}+\rho_{\text {jump }}$, where $\rho_{\text {coh }}$ represents the coherent-evolution part, while $\rho_{\text {jump }}$ corresponds to the incoherent-scattered part where the spontaneous decay or quantum jumps introduce qubit loss.
We are only interested in the coherent-evolved part, and we also utilize the underestimated fidelity in Appendix~\ref{nonHermitian}
\[
\mathcal{F}=\max _\phi\left\langle\mathrm{GHZ}, \phi\left|\rho_{\text {coh}}\right| \mathrm{GHZ}, \phi\right\rangle,
\]
where the total density matrix $\rho$ is replaced by its coherent part $\rho_{\mathrm{coh}}$. And it is safe to estimate and characterize the performance of entanglement creation.
The W state is the eigenstate of the angular momentum with an eigenvalue $-N/2+1$ for $N$ qubits. Therefore, we can directly use the matrix element $\rho_{-N/2+1,-N/2+1}$ to characterize the fidelity, where $\mathcal{F}=\rho_{-N/2+1,-N/2+1}$. The same rule also applied to the Dicke state $|0\rangle$ with a fidelity $\mathcal{F}=\rho_{0,0}$.

The lantern state $\left|lt\right\rangle$ is defined as an equal superposition of $\left|-15\right\rangle$, $\left|-5\right\rangle$, $\left|5\right\rangle$ and $\left|15\right\rangle$ in the main text.
The phases between each superposition components are known and determined during the entanglement preparation. 
We thus calculate its fidelity through similar procedure as in the calculation of GHZ states, which is
$$
\begin{aligned}
\mathcal{F}&=\frac{1}{4}(\rho_{-15, -15}+\rho_{-5,-5}+\rho_{5,5}+\rho_{15,15}\\
&+\left|\rho_{-15, -5}\right|+\left|\rho_{-15,5}\right|+\left|\rho_{-15, 15}\right|+\left|\rho_{-5,-15}\right|\\
&+\left|\rho_{-5, 5}\right|+\left|\rho_{-5,15}\right|+\left|\rho_{5, -15}\right|+\left|\rho_{5,-5}\right|\\
&+\left|\rho_{5, 15}\right|+\left|\rho_{15,-15}\right|+\left|\rho_{-15, -5}\right|+\left|\rho_{-15,5}\right|)
\end{aligned}
$$
This can help us to include the contributions of phase coherence.

\section{Spin-squeezed state\label{SSSIdeal}}
In this section, we provide the details about calculating the SSS in both ideal and $^{87}\rm Rb$ cases.

First we show how we calculate squeezing parameter $\xi^2_{R}$ under the Wineland criterion. For a quantum state, we calculate the expectation value of the spin $\langle\hat{\textbf{S}}\rangle = (\langle \hat{S}_x\rangle,\langle \hat{S}_y\rangle,\langle \hat{S}_z\rangle)$ along different directions. This defines the mean spin direction $\vec{n}(\theta,\phi)$ parallel to $\langle\hat{\textbf{S}}\rangle$, where $\theta = \arccos(\langle \hat{S}_z\rangle/|\langle\hat{\textbf{S}}\rangle|)$ and $\phi = \arctan(\langle \hat{S}_y\rangle/\langle \hat{S}_x\rangle)$.
We can choose 2 directions $\vec{n}_1$, $\vec{n}_2$ orthogonal to each other and perpendicular to $\vec{n}(\theta,\phi)$, and calculate $C = \langle \hat{S}_{\vec{n}_1}^2+\hat{S}_{\vec{n}_2}^2\rangle$, $A = \langle \hat{S}_{\vec{n}_1}^2-\hat{S}_{\vec{n}_2}^2\rangle$, and $B = \langle \hat{S}_{\vec{n}_1}\hat{S}_{\vec{n}_2}+\hat{S}_{\vec{n}_2}\hat{S}_{\vec{n}_1}\rangle$. Then the minimum variance of the angular momentum perpendicular to $\vec{n}$ is $v_m = \frac{1}{2}(C-\sqrt{A^2+B^2})$. 

Then we calculate the squeezing parameter $\xi^2_R$ as
\begin{equation}
    \xi^2_R=S v_m/|\langle\hat{\textbf{S}}\rangle|^2.
\end{equation}
Based on Eq.~\ref{eq:sss} in the main text where we also copy it below,
\begin{equation}
    \left|\psi_{\textrm{SSS}}\right\rangle = e^{-i(\Omega S_x + H_{\rm AC})t_2} e^{-i(\Omega S_x)t_1} \left| \downarrow \right\rangle^{\otimes N},
\end{equation}
the parameters we need to optimize are $n_0$, $\delta_{\rm AC}$, $t_2$, and $t_1$ with fixed parameters $\Omega=2\pi\times 0.2$~MHz and $\Delta=2\pi\times 0.65$~GHz. Here $n_0$ corresponds to the light frequency $\omega=\omega_c+n_0 \omega_s$. $\delta_{\rm AC}$ is the AC Stark shift for the resonant Dicke state. In Fig.~\ref{fig:SSS_Parameters} (a) to (d), we plot these globally optimized parameters from $N=50$ to 500. Based on these information, we can extrapolate the optimized SSS parameters for a large $N$, and the result for large $N$ is consistent with the scaling of the $\xi_R^2$ for small $N$, as shown in Fig.~\ref{fig:SSS_scalability}.

\begin{table}
    \caption{Fixed parameters for spin squeezing in the $^{87}\rm Rb$-cavity case}
    \begin{tabular}{llll}
        \hline  
        Parameters & Symbol & Number & Unit\\
        \hline 
        Spin Spontaneous Decay Rate & $\Gamma$&  $2\pi\times6.06$& MHz\\
        Cavity Decay Rate & $\kappa$ &
        $2\pi \times 0.1$ &MHz\\
        Photon-Qubit Coupling & $g$ &
        $2\pi \times 5.4$ & MHz\\
        Hyperfine splitting & $\omega_0$ & $2\pi\times 6.8$ &GHz\\
        Trivial Rotation Angular Velocity & $\Omega$ & $2\pi\times0.2$ & MHz\\
        Cooperativity & $\eta$ & 200&\\
        \hline 
        
    \end{tabular}
    \label{fixparams}
\end{table}

\begin{table}
    \caption{Fixed parameters for spin squeezing in the superconducting-qubit system}
    \begin{tabular}{llll}
        \hline  
        Parameters & Symbol & Number & Unit\\
        \hline 
        Spin Spontaneous Decay Rate & $\Gamma$&  $2\pi\times3.8$& MHz\\
        Cavity Decay Rate & $\kappa$ &
        $2\pi \times 0.3$ &MHz\\
        Photon-Qubit Coupling & $g$ &
        $2\pi \times 78$ & MHz\\
        Ground state splitting & $\omega_0$ & $2\pi\times 47$ &GHz\\
        Trivial Rotation Angular Velocity & $\Omega$ & $2\pi\times0.2$ & MHz\\
        Cooperativity & $\eta$ & 19000&\\
        \hline 
        
    \end{tabular}
    \label{fixparams_SC}
\end{table}

For $^{87}\rm Rb$ atoms, we consider two $5S_{1/2}$ hyperfine ground states $\left|\downarrow\right\rangle=|F=1\rangle$ and $\left|\uparrow\right\rangle=|F=2\rangle$, and an excited state $\left|e\right\rangle$ in $5P_{3/2}$. Fixed parameters are shown in the Table.~\ref{fixparams}. The hyperfine splitting is 6.8~GHz, and the light coupling between $\left|\downarrow\right\rangle$ and $|e\rangle$ must also be included. Therefore, the detuning $\Delta$ between $\left|\uparrow\right\rangle$ and $\left|e\right\rangle$ must be optimized to balance this additional contribution. Then the Hamiltonian becomes a form like following
\begin{equation}
    H_{\rm{exp}} = H^{\uparrow}_{\rm{AC}}+H^{\downarrow}_{\rm{AC}}.
\end{equation}
Here $H^{\uparrow}_{\rm{AC}}$ and $H^{\downarrow}_{\rm{AC}}$ corresponds to the AC Stark shift contributed from the states $\left|\uparrow\right\rangle$ and $\left|\downarrow\right\rangle$.

In order to maintain a better scaling of SSS, the optimal parameters $n_0$, $\delta_{\rm AC}$, $t_2$, and $t_1$ must change under the spontaneous decay. Additionally, we need to optimize $\Delta$.
We still define the AC Stark shift coefficient for the state $\left|\uparrow\right\rangle$ with the form of $C = \hbar \delta_{\rm AC}(1-\frac{i\Gamma}{\Delta})$ to label out the contribution of light. The influence of $\left|\downarrow\right\rangle$ can be easily calculated based on the parameters of $\left|\uparrow\right\rangle$ according to the hyperfine splitting.
For the contribution of the spontaneous decay, we utilize the method in Ref.~\cite{schleier2010squeezing, li2022collective}. If one qubit is scattered randomly by a photon, it will not contribute to the total spin length anymore. Meanwhile, this randomly reshuffled spin will be in a maximal mixed state after the scattering. Any further coherent evolution does not purify this mixed state anymore. Therefore based on the evolution of a non-Hermitian Hamiltonian for a final state $\rho$, we are expecting $\left[1-\rm{Tr}(\rho)\right]N$ qubits randomly scatterred. Therefore the variance contribution becomes $v_m+\left[1-\rm{Tr}(\rho)\right]N/4$. The squeezing parameter becomes
\begin{equation}
    \xi^2_{R,loss} =S \frac{v_m+\left[1-\rm{Tr}(\rho)\right]N/4}{|\langle\hat{\textbf{S}}\rangle|^2}.
\end{equation}
In Fig.~\ref{fig:SSS_Parameters}(a) to (d), we show these parameters used in the calculation in the main text Fig.~\ref{fig:SSS}. 

In Table~\ref{fixparams_SC}, we list typical parameters for superconducting-qubit system which we summarize from Ref.~\cite{rigetti2012superconducting,chakram2021seamless,milul2023superconducting,huo2023gatemon}.
Then we calculated the preparations of three quantum states including W states, GHZ states, and SSS with a monochromatic light in superconducting qubits. The fidelity of a GHZ state with 50 qubits is $0.975$ and the fidelity of a W state with 50 qubits is $0.880$. The SSS is shown in Fig.~\ref{fig:SSS_SC}.

All these calculation of entangled states can be obtained online \cite{gitlink}.

\section{Experimental imperfections}\label{imperfect}

\begin{figure}[htbp]
    \centering
    \includegraphics[width=0.45\textwidth]{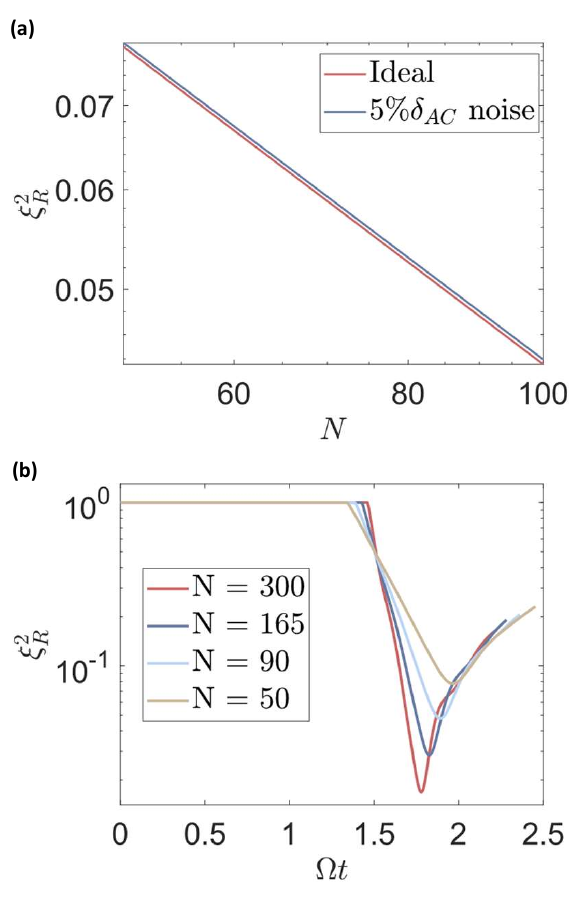}
    \caption{(a) One sequence squeezing with the squeezing parameter $\xi_{R}^2$ versus the qubit number $N$ under 0 or $5\%$ intensity noise. The other parameters are the same as the $n_f=1$ line in Fig.~\ref{fig:SSS}. (b) The evolution of $\xi_{R}^2$ with time for different $N$ under optimized parameters.}
    \label{fig:SSS_Noise}
\end{figure}

Imperfections occur when we consider a real experiment, including inhomogeneous coupling, photon shot noise, and cavity instability. Inhomogeneous coupling caused by cavity standing wave can be avoided by using a double-frequency laser. Such as in $^{87}\rm Rb$ systems, one can use a 780~nm laser to induce the AC Stark shift of Dicke states and a 1560~nm laser to trap atoms. Then,
all atoms are trapped to the anti-nodes of lattice as well as the 780~nm light for entanglement. 
The photon shot noise and the cavity frequency instability randomly change the intra-cavity light intensity, and thus bring fluctuations to AC Stark shifts. Here we simulate this effect with $5\%$ intensity noise in Fig.~\ref{fig:SSS_Noise}(a), and there is almost no influence from this imperfection.
 
Then we also consider the relative time accuracy requirement for $t_1$ and $t_2$.
As shown in Fig.~\ref{fig:SSS_Noise}(b), we find that the temporal region corresponding to an SSS becomes narrower as $N$ increases. When $N$ increases, the relative width of a CSS on the Bloch sphere becomes narrower with a scale of $1/\sqrt{N}$.
Only when the CSS starts to overlap with the barrier, the squeezing effect appears. Therefore, we have to accurately control the time $t_1$ and $t_2$ to exactly force $\xi^2_R$ sitting at the minimal point. For the scenarios in the main text, any relative time uncertainty below 0.1\% can help us to resolve and locate the the minimal point of $\xi^2_R$ which corresponds to 1~ns. To our knowledge, the current microwave atomic clock can safely satisfy these time accuracy requirement.

\begin{figure*}[htbp]
    \centering
    \includegraphics[width=1\textwidth]{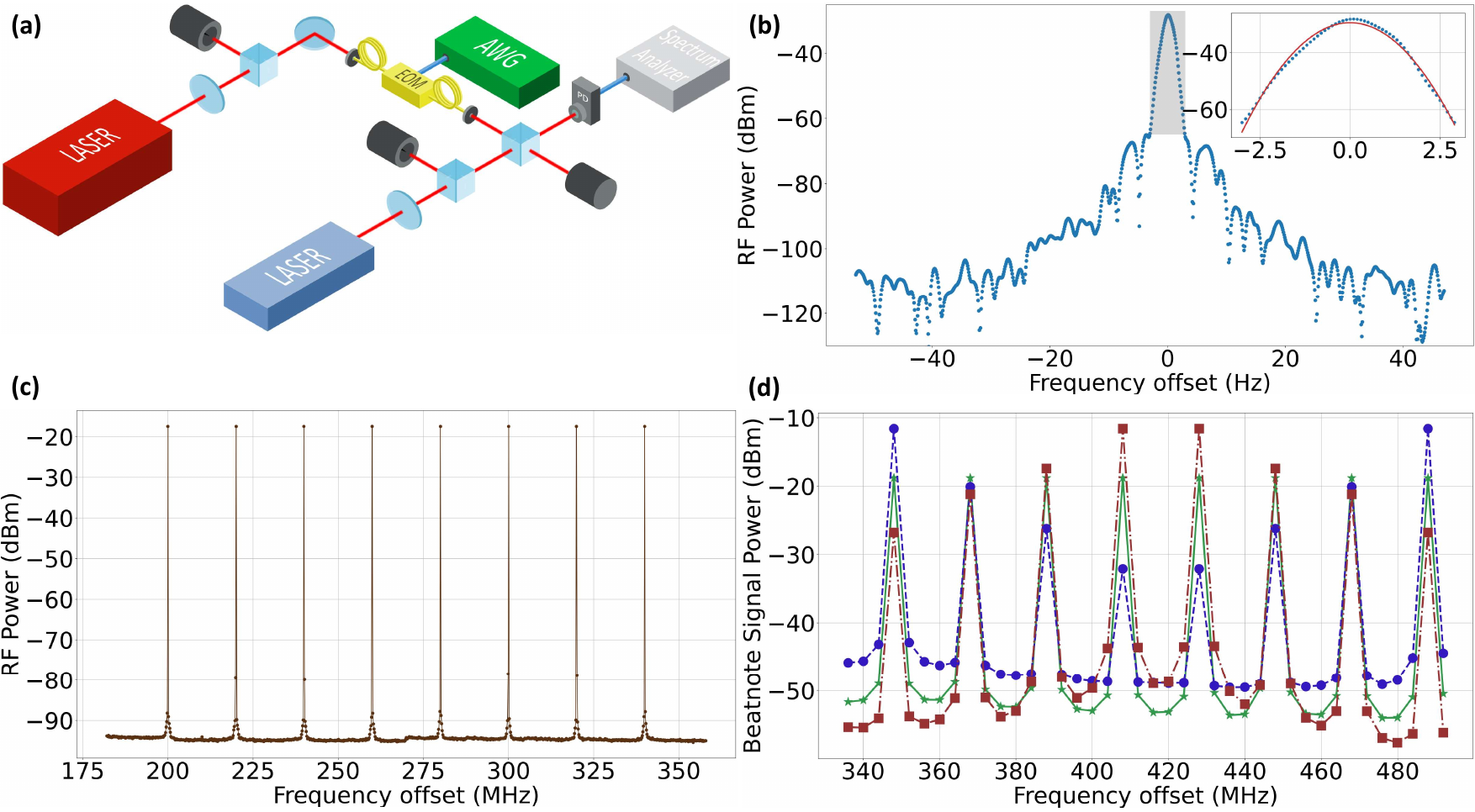}
    \caption{ 
    \textbf{Experimental testing of a multi-frequency laser.} 
    (a) The testing setup. Two individual 780~nm lasers is beated for measurement of relative intensity and linewidth. One of the lasers (red one) is phase modulated by an EOM, and is combined with another reference laser (blue one) and detected by a photodetector. The beatnote signal is then filtered by a bias-tee, and sent to a spectrum analyzer. 
    (b) The linewidth measurement of multi-frequency RF signal generated by the AWG. We pick one of the RF signals for demonstration. Inset shows the zoom in figure of the dashed region. The signal is fitted with a Gaussian function $f(x) = 10 * \log_{10}(e^{-((x - a) / b)^2}) + c$ shown in red line, with 
    fitted width parameter $b=1.424$~Hz.
    (c) The 8-frequency mixed RF signal with equal spacing of 20~MHz.
    (d) The beatnote signals. The red, green, and blue lines are three individual demonstrations with different relative intensity. The relative intensity noise is below 0.74\% over several times of measurement and the linewidth shows no degradation after phase modulation.}
    \label{fig:Imperfections}
\end{figure*}

Other noises may include the relative intensity and frequency noise in the multi-frequency scheme.
We thus experimentally test these noises with optical and radio-frequency (RF) components.
The test setup is shown in Fig.~\ref{fig:Imperfections}(a).
We beat two individual 780~nm lasers (red: DL-pro from Toptica, blue: frequency doubled from 1560~nm a fixed external cavity diode Laser from Precilaser) to test the relative intensity and linewidth of a phase modulated laser.
The 1560-nm laser (blue in Fig.~\ref{fig:Imperfections}(a)) serves as a reference and has a narrow linewidth of 6.1~kHz tested with 50~km fiber delayed self-heterodyne linewidth measurement.
The Totica laser (red in Fig.~\ref{fig:Imperfections}(a)) is modulated by an EOM driven by multi-frequency RF signals to serve the multi-frequency scheme. 
The multi-frequency RF signals are generated by an arbitrary waveform generator (AWG, M4i.6631-x8 by Spectrum) with tested narrow bandwidth of 1.4~Hz shown in Fig.~\ref{fig:Imperfections}(b) and signal-to-noise ratio as high as 76 dB for $8$ frequencies shown in Fig.~\ref{fig:Imperfections}(c), which decays to around 71 dB after RF amplifier (ZFL-500+ by Minicircuits).

After modulating the Toptica laser with an EOM (ixBlue NIR-MPX800-LN-20), we measure the beat note of two lasers experimentally and find it is sufficient to generate multi-frequency laser with adequate modulation depth, as shown in Fig.~\ref{fig:Imperfections}(d), where we show that the relative intensity among different frequency parts can be adjusted by changing the relative intensity of the RF tones from the AWG in spite of the existence of intermodulation by the EOM.
The relative intensity between the beat signals keep steady with standard deviation of $0.74\%$ without intensity locking, and the phase noise, which reflects on the linewidth of the beat signal, shows no sign of degradation (below the measurement accuracy).
It supports that the relative intensity noise and frequency noise is negligibly low and can be ignored during the theoretical calculations. 
Furthermore, our non-trivial rotation stems from the AC-Stark shifts of multi-frequency laser to the qubits which are only sensitive to whether the state is shifted away or not by the laser intensity.
The relative phase noise between those frequencies are thus not so important.

\section{Information leakage from cavity photons}
\label{informationLeakage}
In cavity-QED systems, qubits are entangled with photons, and then the leakage of the photons from a cavity may bring information about qubits and make the qubit state becoming a mixed state. For example, for a state in the form of $|m_1\rangle|k_1\rangle+|m_2\rangle|k_2\rangle$ where $m_i$ is the qubit's angular momentum and $k_i$ is the photon number, then the state of qubits becomes a mixed state $|m_1\rangle\langle m_1|+|m_2\rangle\langle m_2|$ after tracing out photons when $k_1\neq k_2$. This information leakage destroys the entanglement. On the other hand, if $k_1=k_2$, the entanglement can be preserved.

Fortunately, the quantum states discussed here are unaffected by this mechanism for the following reasons.
The major reason is that only the resonant state at the barrier will be populated by many photons. For the other states, the photon occupation is very small due to off resonance and the photon states are almost the same as a vacuum state. Meanwhile, when we create the entanglement, the resonant state is off-coupled from the qubits' energy level. Therefore, the entangled states created from non-trivial rotations are carrying almost the same wave function of photons (very close to a vacuum state). 
Therefore, our scheme is not influenced by the information leakage from photons.

Now we build a more quantitative model to describe this leaking process.  Let's consider the quantum state as $|m_z\rangle\otimes|\psi_i\rangle|\psi_r\rangle|\psi_t\rangle$, in which $|\psi_i\rangle$ correspond to an intra-cavity photon state, $|\psi_r\rangle$ corresponds to a photon state at the cavity reflection side, $|\psi_t\rangle$ corresponds to a photon state at the cavity transmission side, and $|m_z\rangle$ corresponds to qubits' states. 
Our target is to create an entangled state of qubits in a form of $\sum_{m_z} c_{m_z}|m_z\rangle$. During the preparation, we create a state with a form of $\sum_{m_z} c_{m_z}|m_z\rangle\otimes |\alpha_{m_z}\rangle|0\rangle|0\rangle$ and $|\alpha_{m_z}\rangle$ is a coherent state. After the cavity ring down, the intra-cavity photons escape and the state becomes 
\begin{equation}
|\Psi\rangle=\sum_{m_z} c_{m_z}|m_z\rangle\otimes |0\rangle|\mathcal{R}(m_z)\alpha_{m_z}\rangle|\mathcal{T}(m_z)\alpha_{m_z}\rangle, 
\end{equation}
where $\mathcal{R}(m_z)$ (or $\mathcal{T}(m_z)$) is the reflection (or transmission) amplitude. Here we find during the W-state creation shown in Fig.~\ref{figWallsQ}(c) (the scheme with a monochromatic light), $|\alpha_{m_z}\rangle$ contains around 0.099 photons for off-resonant states while these states occupy 99.92\% population in the desired W state. 

For a GHZ state shown in Fig.~\ref{figWallsQ}(d), $|\alpha_{m_z}\rangle$ contains 0.0025 photons for off-resonant states while these states occupy 99.99\% populations, because the desired states are further far away from the barrier. 
The overlap of a weak coherent state and the vacuum state is 0.9975.
Then, the GHZ state fidelity changes from 0.9 to 0.899 and the W state fidelity (0.96) does not change within a 0.1\% accuracy. It supports the previous qualitative argument that the tracing out of photons does not hurt the scheme. For SSS, we test qubits number at 50, 100,  and 5000, squeezing parameters drop as $0.0786\rightarrow0.0801$, $0.0435\rightarrow0.0448$, and $0.00160\rightarrow0.00168$, this barely hurt the scale.
In the next section, we will use the master equation to calculate the dynamics and confirm that the fidelity of the qubit states will not be hurt.

\begin{figure}[htbp]
    \centering
    \includegraphics[width=0.45\textwidth]{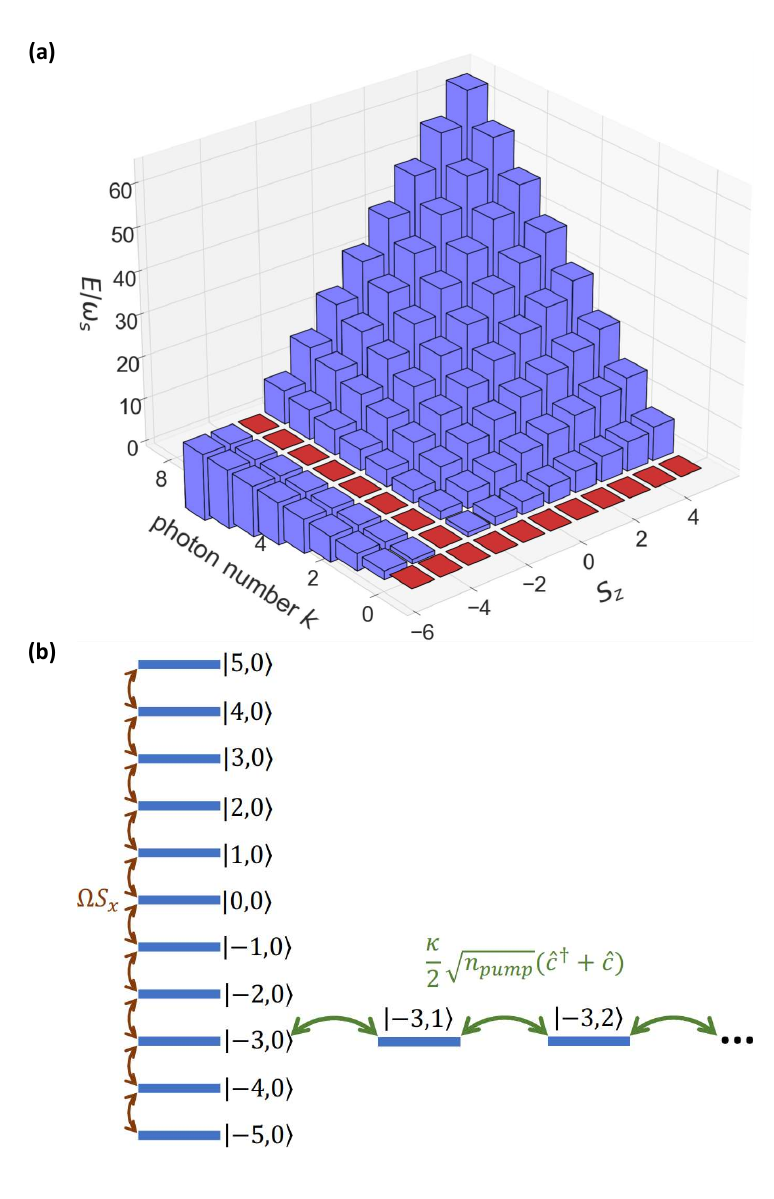}
    \vspace{-5mm}
    \caption{\textbf{The energy level diagram.} 
    Here we choose $n=2$ and $S=5$ (10 qubits) as an example.
    (a) The energy of each state $|S_z\rangle\otimes|k\rangle$ under the Hamiltonian $H_0/\hbar=\omega_s\left(\hat{S}_z+S\right) \hat{c}^{\dagger} \hat{c}-n\omega_s \hat{c}^{\dagger} \hat{c}$. The vertical axis is in the unit of $\omega_s$. Two horizontal axes are labeled by the angular momentum $S_z$ and the photon number $k$. Here the red color highlights the states of $|S_z\rangle\otimes|0\rangle$ and $|-S+n\rangle\otimes|k\rangle$ which are degenerate with an energy 0 under $H_0$.
    (b) Here we plot the degenerate levels under $H_0$.
    The states $|S_z\rangle\otimes |0\rangle$ with zero photons are coupled via the rotation term $\Omega \hat{S}_x$ (shown in brown arrows).
    This rotation can transfer the qubits' populations in this angular momentum ladder.
    Then, a pumping term $\frac{\kappa}{2} \sqrt{n_{\text {pump }}}\left(\hat{c}^{\dagger}+\hat{c}\right)$ (shown in green arrows) couples the states $|-S+n\rangle\otimes|k\rangle$ with one particular angular momentum $-S+n$ and different photon numbers $k$. Because of this pumping term, it will disturb the original state $|-S+n\rangle\otimes|0\rangle$ in the angular momentum ladder and creates a barrier for the population transfer.
    }
    \label{fig:LindbladEnergy}
\end{figure}

\section{Numerical approach based on the master equation in Lindblad form}
\label{informationLeakage}

In the main text, we use the effective Hamiltonian $H_{\textrm{AC}}$ to describe the evolution of the qubit states.
To further support the validity of our scheme, we quantize the intra-cavity light and use the photonic Fock states to describe the dynamics in this appendix.
Here we use the Lindblad master equation to calculate the dynamics of the non-trivial rotation where both quantum states of qubits and photons are counted.
Then we find the barrier state is still working and tearing the Hilbert space into two parts, while any wave functions can hardly penetrate this barrier. We also confirm that the effective Hamiltonian $H_{\textrm{AC}}$ provides consistent results with the master equation.

\begin{figure}[htbp]
    \centering
    \includegraphics[width=0.4\textwidth]{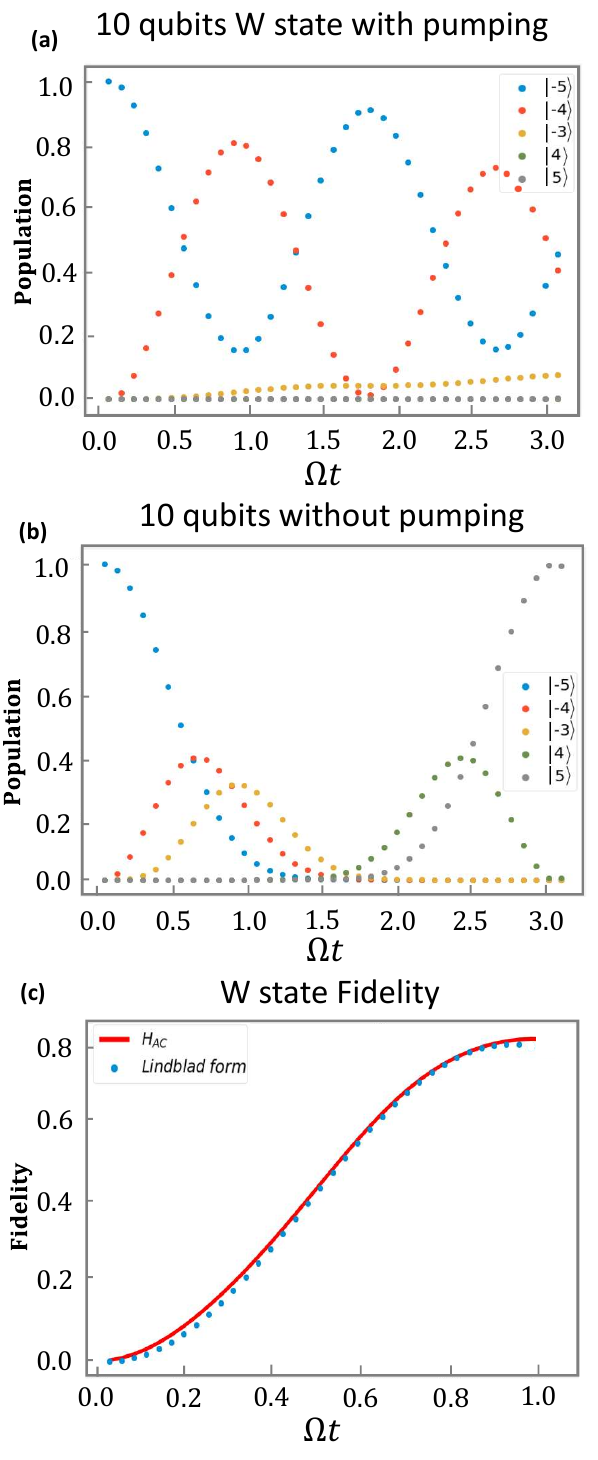}
    \vspace{-5mm}
    \caption{
    \textbf{The Lindblad master equation calculation.} (a) The generation of a W state with one non-trivial rotation for 10 qubits.
    Here we set $n=2$ corresponding to a barrier at the state $\left|-3\right\rangle$. The initial state is $\left|-5\rangle\otimes |0\right\rangle$. Circles with different colors correspond to the populations of different Dicke states $\left|S_z\rangle\otimes|0\right\rangle$ with zero photons. And we find the barrier state at $\left|-3\rangle\otimes|0\right\rangle$ (yellow circles) can hardly be populated.
    (b) The same condition as (a), but without the external incident light. The population can get through the $|-3\rangle$ state without a barrier. The total time evolution is described by a Rabi oscillation. (c) Comparison between the effective Hamiltonian $H_{\textrm{AC}}$ and the master equation. 
    The two results are obtained using the same parameters with $\Delta = -2\pi\times0.957$~GHz, $\Omega = 2\pi\times 0.2$~kHz, $\Gamma = 2\pi\times 6.06$~MHz, and $\kappa = 2\pi\times0.1$~MHz in Table.~\ref{fixparams}. The barrier state is set at $|-3\rangle$. Here the red solid line corresponds to the results of $H_{\textrm{AC}}$ and the blue circles correspond to the master equation. They show a consistency between each other.}
    \label{fig:Wall_demo}
\end{figure}

The Lindblad master equation is written as
\begin{equation}
\label{Lindblad}
\frac{d \rho}{d t}=-i[\hat{H}/\hbar, \rho]+\kappa\left(\hat{c} \rho \hat{c}^{\dagger}-\frac{1}{2}\left\{\hat{c}^{\dagger} \hat{c}, \rho\right\}\right).
\end{equation}
Here the Lindblad term $\kappa\left(\hat{c} \rho \hat{c}^{\dagger}-\frac{1}{2}\left\{\hat{c}^{\dagger} \hat{c}, \rho\right\}\right)$ corresponds to the process that photons continuously leak out from cavity, which also describes the process of information leakage, $\rho$ is the density matrix including both qubits and photons, $\hat{c}$ and $\hat{c}^\dagger$ are the annihilation and creation operator of the photonic Fock state, and $H$ is the Hamiltonian with a form as
\begin{eqnarray}
    H/\hbar
    &=&
    \Omega\hat{S}_x+\omega_s\left(\hat{S}_z+S\right) \hat{c}^{\dagger} \hat{c}-\delta \hat{c}^{\dagger} \hat{c} \nonumber \\
    & &+\frac{\kappa}{2} \sqrt{n_{\text {pump }}}\left(\hat{c}^{\dagger}+\hat{c}\right).
\end{eqnarray}
The first term $\Omega\hat{S}_x$ in the Hamiltonian describes the rotation connecting different Dicke states. 
The last term $\frac{\kappa}{2} \sqrt{n_{\text {pump }}}\left(\hat{c}^{\dagger}+\hat{c}\right)$ is the pumping term describing that there is an incident light sent into the cavity and it can couple the vacuum state to the other Fock states of photons. If the light is resonant with the cavity, the mean intra-cavity photon number will be $n_{\text{pump}}$.
The middle term $H_0/\hbar=\omega_s\left(\hat{S}_z+S\right) \hat{c}^{\dagger} \hat{c}-\delta \hat{c}^{\dagger} \hat{c}$ describes the energy of each state including both qubits and photons. Here $\delta=\omega_l-\omega_c$ is the frequency difference between the light frequency $ \omega_l$ and the cavity resonant frequency $\omega_c$. When $\delta=n\omega_s$, only the particular Dicke state $\left|-S+n\right\rangle$ will make the cavity resonant with the external incident light, and this Dicke state will serve as a barrier which we will show later. Here we use $n=2$ and $S=5$ as an example to plot the energy of each state $\left|S_z\rangle\otimes|k\right\rangle$ under $H_0$ in Fig.~\ref{fig:LindbladEnergy}(a), where $|S_z\rangle$ is the Dicke state of qubits and $\left|k\right\rangle$ is the Fock state with $k$ photons.

First, we would like to give a qualitative discussion about why the state $|-S+n\rangle\otimes|0\rangle$ can serve as a barrier. In Fig.~\ref{fig:LindbladEnergy}(a), we plot the energy of each state and we can find the states of $|S_z\rangle\otimes|0\rangle$ and $|-S+n\rangle\otimes|k\rangle$ are degenerate which are labeled by red color. Then, the rotation term $\Omega \hat{S}_x$ can couple the states $|S_z\rangle\otimes|0\rangle$ together as shown in Fig.~\ref{fig:LindbladEnergy}(b) ($n=2$ and $S=5$ as an example). However, the particular Dicke state $|-S+n\rangle\otimes|0\rangle$ is strongly coupled to the other states of $|-S+n\rangle\otimes|k\neq 0\rangle$ via the pumping term $\frac{\kappa}{2} \sqrt{n_{\text {pump }}}\left(\hat{c}^{\dagger}+\hat{c}\right)$. 
When $n_{\text{pump}}$ becomes larger, this coupling is stronger.
Due to this coupling to other states, the state $\left|-S+n\rangle\otimes|0\right\rangle$ is shifted and cannot serve the original role in the angular momentum ladder anymore. Therefore, this particular Dicke state is a barrier and it makes the rotation term $\Omega\hat{S}_x$ cannot transfer the wave functions through this particular angular momentum $-S+n$ anymore.

Then, we use the numerical tool to further quantitatively support this argument. 
We choose $n=2$ as an example which corresponds to a barrier at $|-3=-5+2\rangle$ in an ensemble of 10 qubits.
We set $\Omega=2\pi\times0.2$~kHz, $\kappa=2\pi\times 0.1$~MHz, $\omega_s=-2\pi\times1.57$~MHz, and $n_{\text{pump}}=0.33$ in the master equation. During the calculations, we cut off the maximal photon number at 10. In Fig.~\ref{fig:Wall_demo}(a), we show our numerical results for an initial state at $|-5\rangle\otimes|0\rangle$. The density matrix becomes a mixed state after the time evolution due to the cavity photon decay. Here we plot the populations of each state $|S_z\rangle\otimes|0\rangle$ versus the evolution time 
because the populations with non-zero photon numbers are almost negligible. Then, we find the population transfer is limited in the angular momentum subspace of $\left|-5\right\rangle$ and $\left|-4\right\rangle$. The barrier state $\left|-3\right\rangle$ and other states separated by the barrier (such as $\left|4\right\rangle$ and $\left|5\right\rangle$) are very slightly populated. 
When $\Omega t=0.895$, the non-trivial rotation finishes a W state preparation corresponding to the state transfer into $\left|-4\right\rangle$.
Therefore, this supports that the state $\left|-3\right\rangle$ is a barrier tearing the Hilbert space into two parts and induces the entanglement under the non-trivial rotation.

For a comparison, we also show the numerical results (Fig.~\ref{fig:Wall_demo}(b)) with $n_{\text{pump}}=0$ which corresponds to a trivial rotation with no incident light. And we find that the time evolution is described by a Rabi oscillation between all Dicke states, and an initial population in $\left|-5\right\rangle$ can be fully transferred into $\left|5\right\rangle$ without any barriers. This supports that the barrier in the panel (a) is indeed introduced by an external incident light. Besides these, we also compare the results of the master equation and the effective Hamiltonian $H_{\textrm{AC}}$ in Fig.~\ref{fig:Wall_demo}(c). Here we use the same parameters for both methods and calculate the W state fidelity. The red solid line corresponds to the fidelity under $H_{\textrm{AC}}$ and blue circles correspond to the fidelity under the master equation. We find the W state fidelity drops from 0.820 (the effective Hamiltonian) to 0.810 (the master equation). Besides the W state, we also check the other states and we find the performance only receives very slight influence.
The squeezing parameter drops from $0.44$ to $0.48$ and the fidelity of a GHZ state drops from $0.891$ to $0.889$. 
These numerical results support the consistency between the effective Hamiltonian and the Lindblad master equation.
All the detailed calculation results and codes can be obtained online \cite{gitlink}.

\bibliography{refs.bib}

\begin{thebibliography}{113}%
\makeatletter
\providecommand \@ifxundefined [1]{%
 \@ifx{#1\undefined}
}%
\providecommand \@ifnum [1]{%
 \ifnum #1\expandafter \@firstoftwo
 \else \expandafter \@secondoftwo
 \fi
}%
\providecommand \@ifx [1]{%
 \ifx #1\expandafter \@firstoftwo
 \else \expandafter \@secondoftwo
 \fi
}%
\providecommand \natexlab [1]{#1}%
\providecommand \enquote  [1]{``#1''}%
\providecommand \bibnamefont  [1]{#1}%
\providecommand \bibfnamefont [1]{#1}%
\providecommand \citenamefont [1]{#1}%
\providecommand \href@noop [0]{\@secondoftwo}%
\providecommand \href [0]{\begingroup \@sanitize@url \@href}%
\providecommand \@href[1]{\@@startlink{#1}\@@href}%
\providecommand \@@href[1]{\endgroup#1\@@endlink}%
\providecommand \@sanitize@url [0]{\catcode `\\12\catcode `\$12\catcode `\&12\catcode `\#12\catcode `\^12\catcode `\_12\catcode `\%12\relax}%
\providecommand \@@startlink[1]{}%
\providecommand \@@endlink[0]{}%
\providecommand \url  [0]{\begingroup\@sanitize@url \@url }%
\providecommand \@url [1]{\endgroup\@href {#1}{\urlprefix }}%
\providecommand \urlprefix  [0]{URL }%
\providecommand \Eprint [0]{\href }%
\providecommand \doibase [0]{https://doi.org/}%
\providecommand \selectlanguage [0]{\@gobble}%
\providecommand \bibinfo  [0]{\@secondoftwo}%
\providecommand \bibfield  [0]{\@secondoftwo}%
\providecommand \translation [1]{[#1]}%
\providecommand \BibitemOpen [0]{}%
\providecommand \bibitemStop [0]{}%
\providecommand \bibitemNoStop [0]{.\EOS\space}%
\providecommand \EOS [0]{\spacefactor3000\relax}%
\providecommand \BibitemShut  [1]{\csname bibitem#1\endcsname}%
\let\auto@bib@innerbib\@empty
\bibitem [{\citenamefont {Giovannetti}\ \emph {et~al.}(2011)\citenamefont {Giovannetti}, \citenamefont {Lloyd},\ and\ \citenamefont {Maccone}}]{giovannetti2011advances}%
  \BibitemOpen
  \bibfield  {author} {\bibinfo {author} {\bibfnamefont {V.}~\bibnamefont {Giovannetti}}, \bibinfo {author} {\bibfnamefont {S.}~\bibnamefont {Lloyd}},\ and\ \bibinfo {author} {\bibfnamefont {L.}~\bibnamefont {Maccone}},\ }\bibfield  {title} {\bibinfo {title} {Advances in quantum metrology},\ }\href@noop {} {\bibfield  {journal} {\bibinfo  {journal} {Nature photonics}\ }\textbf {\bibinfo {volume} {5}},\ \bibinfo {pages} {222} (\bibinfo {year} {2011})}\BibitemShut {NoStop}%
\bibitem [{\citenamefont {Pezze}\ \emph {et~al.}(2018)\citenamefont {Pezze}, \citenamefont {Smerzi}, \citenamefont {Oberthaler}, \citenamefont {Schmied},\ and\ \citenamefont {Treutlein}}]{pezze2018quantum}%
  \BibitemOpen
  \bibfield  {author} {\bibinfo {author} {\bibfnamefont {L.}~\bibnamefont {Pezze}}, \bibinfo {author} {\bibfnamefont {A.}~\bibnamefont {Smerzi}}, \bibinfo {author} {\bibfnamefont {M.~K.}\ \bibnamefont {Oberthaler}}, \bibinfo {author} {\bibfnamefont {R.}~\bibnamefont {Schmied}},\ and\ \bibinfo {author} {\bibfnamefont {P.}~\bibnamefont {Treutlein}},\ }\bibfield  {title} {\bibinfo {title} {Quantum metrology with nonclassical states of atomic ensembles},\ }\href@noop {} {\bibfield  {journal} {\bibinfo  {journal} {Reviews of Modern Physics}\ }\textbf {\bibinfo {volume} {90}},\ \bibinfo {pages} {035005} (\bibinfo {year} {2018})}\BibitemShut {NoStop}%
\bibitem [{\citenamefont {Marciniak}\ \emph {et~al.}(2022)\citenamefont {Marciniak}, \citenamefont {Feldker}, \citenamefont {Pogorelov}, \citenamefont {Kaubruegger}, \citenamefont {Vasilyev}, \citenamefont {van Bijnen}, \citenamefont {Schindler}, \citenamefont {Zoller}, \citenamefont {Blatt},\ and\ \citenamefont {Monz}}]{marciniak2022optimal}%
  \BibitemOpen
  \bibfield  {author} {\bibinfo {author} {\bibfnamefont {C.~D.}\ \bibnamefont {Marciniak}}, \bibinfo {author} {\bibfnamefont {T.}~\bibnamefont {Feldker}}, \bibinfo {author} {\bibfnamefont {I.}~\bibnamefont {Pogorelov}}, \bibinfo {author} {\bibfnamefont {R.}~\bibnamefont {Kaubruegger}}, \bibinfo {author} {\bibfnamefont {D.~V.}\ \bibnamefont {Vasilyev}}, \bibinfo {author} {\bibfnamefont {R.}~\bibnamefont {van Bijnen}}, \bibinfo {author} {\bibfnamefont {P.}~\bibnamefont {Schindler}}, \bibinfo {author} {\bibfnamefont {P.}~\bibnamefont {Zoller}}, \bibinfo {author} {\bibfnamefont {R.}~\bibnamefont {Blatt}},\ and\ \bibinfo {author} {\bibfnamefont {T.}~\bibnamefont {Monz}},\ }\bibfield  {title} {\bibinfo {title} {Optimal metrology with programmable quantum sensors},\ }\href@noop {} {\bibfield  {journal} {\bibinfo  {journal} {Nature}\ }\textbf {\bibinfo {volume} {603}},\ \bibinfo {pages} {604} (\bibinfo {year} {2022})}\BibitemShut {NoStop}%
\bibitem [{\citenamefont {Giovannetti}\ \emph {et~al.}(2006)\citenamefont {Giovannetti}, \citenamefont {Lloyd},\ and\ \citenamefont {Maccone}}]{giovannetti2006quantum}%
  \BibitemOpen
  \bibfield  {author} {\bibinfo {author} {\bibfnamefont {V.}~\bibnamefont {Giovannetti}}, \bibinfo {author} {\bibfnamefont {S.}~\bibnamefont {Lloyd}},\ and\ \bibinfo {author} {\bibfnamefont {L.}~\bibnamefont {Maccone}},\ }\bibfield  {title} {\bibinfo {title} {Quantum metrology},\ }\href@noop {} {\bibfield  {journal} {\bibinfo  {journal} {Physical Reiview Letters}\ }\textbf {\bibinfo {volume} {96}},\ \bibinfo {pages} {010401} (\bibinfo {year} {2006})}\BibitemShut {NoStop}%
\bibitem [{\citenamefont {Anisimov}\ \emph {et~al.}(2010)\citenamefont {Anisimov}, \citenamefont {Raterman}, \citenamefont {Chiruvelli}, \citenamefont {Plick}, \citenamefont {Huver}, \citenamefont {Lee},\ and\ \citenamefont {Dowling}}]{anisimov2010quantum}%
  \BibitemOpen
  \bibfield  {author} {\bibinfo {author} {\bibfnamefont {P.~M.}\ \bibnamefont {Anisimov}}, \bibinfo {author} {\bibfnamefont {G.~M.}\ \bibnamefont {Raterman}}, \bibinfo {author} {\bibfnamefont {A.}~\bibnamefont {Chiruvelli}}, \bibinfo {author} {\bibfnamefont {W.~N.}\ \bibnamefont {Plick}}, \bibinfo {author} {\bibfnamefont {S.~D.}\ \bibnamefont {Huver}}, \bibinfo {author} {\bibfnamefont {H.}~\bibnamefont {Lee}},\ and\ \bibinfo {author} {\bibfnamefont {J.~P.}\ \bibnamefont {Dowling}},\ }\bibfield  {title} {\bibinfo {title} {Quantum metrology with two-mode squeezed vacuum: parity detection beats the heisenberg limit},\ }\href@noop {} {\bibfield  {journal} {\bibinfo  {journal} {Physical Reiview Letters}\ }\textbf {\bibinfo {volume} {104}},\ \bibinfo {pages} {103602} (\bibinfo {year} {2010})}\BibitemShut {NoStop}%
\bibitem [{\citenamefont {Lawrie}\ \emph {et~al.}(2019)\citenamefont {Lawrie}, \citenamefont {Lett}, \citenamefont {Marino},\ and\ \citenamefont {Pooser}}]{lawrie2019quantum}%
  \BibitemOpen
  \bibfield  {author} {\bibinfo {author} {\bibfnamefont {B.~J.}\ \bibnamefont {Lawrie}}, \bibinfo {author} {\bibfnamefont {P.~D.}\ \bibnamefont {Lett}}, \bibinfo {author} {\bibfnamefont {A.~M.}\ \bibnamefont {Marino}},\ and\ \bibinfo {author} {\bibfnamefont {R.~C.}\ \bibnamefont {Pooser}},\ }\bibfield  {title} {\bibinfo {title} {Quantum sensing with squeezed light},\ }\href@noop {} {\bibfield  {journal} {\bibinfo  {journal} {Acs Photonics}\ }\textbf {\bibinfo {volume} {6}},\ \bibinfo {pages} {1307} (\bibinfo {year} {2019})}\BibitemShut {NoStop}%
\bibitem [{\citenamefont {Kaubruegger}\ \emph {et~al.}(2019)\citenamefont {Kaubruegger}, \citenamefont {Silvi}, \citenamefont {Kokail}, \citenamefont {van Bijnen}, \citenamefont {Rey}, \citenamefont {Ye}, \citenamefont {Kaufman},\ and\ \citenamefont {Zoller}}]{kaubruegger2019variational}%
  \BibitemOpen
  \bibfield  {author} {\bibinfo {author} {\bibfnamefont {R.}~\bibnamefont {Kaubruegger}}, \bibinfo {author} {\bibfnamefont {P.}~\bibnamefont {Silvi}}, \bibinfo {author} {\bibfnamefont {C.}~\bibnamefont {Kokail}}, \bibinfo {author} {\bibfnamefont {R.}~\bibnamefont {van Bijnen}}, \bibinfo {author} {\bibfnamefont {A.~M.}\ \bibnamefont {Rey}}, \bibinfo {author} {\bibfnamefont {J.}~\bibnamefont {Ye}}, \bibinfo {author} {\bibfnamefont {A.~M.}\ \bibnamefont {Kaufman}},\ and\ \bibinfo {author} {\bibfnamefont {P.}~\bibnamefont {Zoller}},\ }\bibfield  {title} {\bibinfo {title} {Variational spin-squeezing algorithms on programmable quantum sensors},\ }\href@noop {} {\bibfield  {journal} {\bibinfo  {journal} {Physical Reiview Letters}\ }\textbf {\bibinfo {volume} {123}},\ \bibinfo {pages} {260505} (\bibinfo {year} {2019})}\BibitemShut {NoStop}%
\bibitem [{\citenamefont {Kaubruegger}\ \emph {et~al.}(2021)\citenamefont {Kaubruegger}, \citenamefont {Vasilyev}, \citenamefont {Schulte}, \citenamefont {Hammerer},\ and\ \citenamefont {Zoller}}]{kaubruegger2021quantum}%
  \BibitemOpen
  \bibfield  {author} {\bibinfo {author} {\bibfnamefont {R.}~\bibnamefont {Kaubruegger}}, \bibinfo {author} {\bibfnamefont {D.~V.}\ \bibnamefont {Vasilyev}}, \bibinfo {author} {\bibfnamefont {M.}~\bibnamefont {Schulte}}, \bibinfo {author} {\bibfnamefont {K.}~\bibnamefont {Hammerer}},\ and\ \bibinfo {author} {\bibfnamefont {P.}~\bibnamefont {Zoller}},\ }\bibfield  {title} {\bibinfo {title} {Quantum variational optimization of ramsey interferometry and atomic clocks},\ }\href@noop {} {\bibfield  {journal} {\bibinfo  {journal} {Physical review X}\ }\textbf {\bibinfo {volume} {11}},\ \bibinfo {pages} {041045} (\bibinfo {year} {2021})}\BibitemShut {NoStop}%
\bibitem [{\citenamefont {Kessler}\ \emph {et~al.}(2014)\citenamefont {Kessler}, \citenamefont {Komar}, \citenamefont {Bishof}, \citenamefont {Jiang}, \citenamefont {S{\o}rensen}, \citenamefont {Ye},\ and\ \citenamefont {Lukin}}]{kessler2014heisenberg}%
  \BibitemOpen
  \bibfield  {author} {\bibinfo {author} {\bibfnamefont {E.~M.}\ \bibnamefont {Kessler}}, \bibinfo {author} {\bibfnamefont {P.}~\bibnamefont {Komar}}, \bibinfo {author} {\bibfnamefont {M.}~\bibnamefont {Bishof}}, \bibinfo {author} {\bibfnamefont {L.}~\bibnamefont {Jiang}}, \bibinfo {author} {\bibfnamefont {A.~S.}\ \bibnamefont {S{\o}rensen}}, \bibinfo {author} {\bibfnamefont {J.}~\bibnamefont {Ye}},\ and\ \bibinfo {author} {\bibfnamefont {M.~D.}\ \bibnamefont {Lukin}},\ }\bibfield  {title} {\bibinfo {title} {Heisenberg-limited atom clocks based on entangled qubits},\ }\href@noop {} {\bibfield  {journal} {\bibinfo  {journal} {Physical Reiview Letters}\ }\textbf {\bibinfo {volume} {112}},\ \bibinfo {pages} {190403} (\bibinfo {year} {2014})}\BibitemShut {NoStop}%
\bibitem [{\citenamefont {Pezz{\`e}}\ and\ \citenamefont {Smerzi}(2020)}]{pezze2020heisenberg}%
  \BibitemOpen
  \bibfield  {author} {\bibinfo {author} {\bibfnamefont {L.}~\bibnamefont {Pezz{\`e}}}\ and\ \bibinfo {author} {\bibfnamefont {A.}~\bibnamefont {Smerzi}},\ }\bibfield  {title} {\bibinfo {title} {Heisenberg-limited noisy atomic clock using a hybrid coherent and squeezed state protocol},\ }\href@noop {} {\bibfield  {journal} {\bibinfo  {journal} {Physical Review Letters}\ }\textbf {\bibinfo {volume} {125}},\ \bibinfo {pages} {210503} (\bibinfo {year} {2020})}\BibitemShut {NoStop}%
\bibitem [{\citenamefont {Hu}\ \emph {et~al.}(2017)\citenamefont {Hu}, \citenamefont {Chen}, \citenamefont {Vendeiro}, \citenamefont {Urvoy}, \citenamefont {Braverman},\ and\ \citenamefont {Vuleti{\'c}}}]{hu2017vacuum}%
  \BibitemOpen
  \bibfield  {author} {\bibinfo {author} {\bibfnamefont {J.}~\bibnamefont {Hu}}, \bibinfo {author} {\bibfnamefont {W.}~\bibnamefont {Chen}}, \bibinfo {author} {\bibfnamefont {Z.}~\bibnamefont {Vendeiro}}, \bibinfo {author} {\bibfnamefont {A.}~\bibnamefont {Urvoy}}, \bibinfo {author} {\bibfnamefont {B.}~\bibnamefont {Braverman}},\ and\ \bibinfo {author} {\bibfnamefont {V.}~\bibnamefont {Vuleti{\'c}}},\ }\bibfield  {title} {\bibinfo {title} {Vacuum spin squeezing},\ }\href@noop {} {\bibfield  {journal} {\bibinfo  {journal} {Physical Review A}\ }\textbf {\bibinfo {volume} {96}},\ \bibinfo {pages} {050301} (\bibinfo {year} {2017})}\BibitemShut {NoStop}%
\bibitem [{\citenamefont {Chen}\ \emph {et~al.}(2015)\citenamefont {Chen}, \citenamefont {Hu}, \citenamefont {Duan}, \citenamefont {Braverman}, \citenamefont {Zhang},\ and\ \citenamefont {Vuleti{\'c}}}]{chen2015carving}%
  \BibitemOpen
  \bibfield  {author} {\bibinfo {author} {\bibfnamefont {W.}~\bibnamefont {Chen}}, \bibinfo {author} {\bibfnamefont {J.}~\bibnamefont {Hu}}, \bibinfo {author} {\bibfnamefont {Y.}~\bibnamefont {Duan}}, \bibinfo {author} {\bibfnamefont {B.}~\bibnamefont {Braverman}}, \bibinfo {author} {\bibfnamefont {H.}~\bibnamefont {Zhang}},\ and\ \bibinfo {author} {\bibfnamefont {V.}~\bibnamefont {Vuleti{\'c}}},\ }\bibfield  {title} {\bibinfo {title} {Carving complex many-atom entangled states by single-photon detection},\ }\href@noop {} {\bibfield  {journal} {\bibinfo  {journal} {Physical Reiview Letters}\ }\textbf {\bibinfo {volume} {115}},\ \bibinfo {pages} {250502} (\bibinfo {year} {2015})}\BibitemShut {NoStop}%
\bibitem [{\citenamefont {Hosten}\ \emph {et~al.}(2016)\citenamefont {Hosten}, \citenamefont {Krishnakumar}, \citenamefont {Engelsen},\ and\ \citenamefont {Kasevich}}]{hosten2016quantum}%
  \BibitemOpen
  \bibfield  {author} {\bibinfo {author} {\bibfnamefont {O.}~\bibnamefont {Hosten}}, \bibinfo {author} {\bibfnamefont {R.}~\bibnamefont {Krishnakumar}}, \bibinfo {author} {\bibfnamefont {N.~J.}\ \bibnamefont {Engelsen}},\ and\ \bibinfo {author} {\bibfnamefont {M.~A.}\ \bibnamefont {Kasevich}},\ }\bibfield  {title} {\bibinfo {title} {Quantum phase magnification},\ }\href@noop {} {\bibfield  {journal} {\bibinfo  {journal} {Science}\ }\textbf {\bibinfo {volume} {352}},\ \bibinfo {pages} {1552} (\bibinfo {year} {2016})}\BibitemShut {NoStop}%
\bibitem [{\citenamefont {Wang}\ \emph {et~al.}(2019{\natexlab{a}})\citenamefont {Wang}, \citenamefont {Wu}, \citenamefont {Yuwei}, \citenamefont {Cai}, \citenamefont {Hu}, \citenamefont {Mu}, \citenamefont {Xu}, \citenamefont {Chen}, \citenamefont {Wang}, \citenamefont {Song}, \citenamefont {Yuan}, \citenamefont {Zou}, \citenamefont {Duan},\ and\ \citenamefont {Sun}}]{wang2019heisenberg}%
  \BibitemOpen
  \bibfield  {author} {\bibinfo {author} {\bibfnamefont {W.}~\bibnamefont {Wang}}, \bibinfo {author} {\bibfnamefont {Y.}~\bibnamefont {Wu}}, \bibinfo {author} {\bibfnamefont {M.}~\bibnamefont {Yuwei}}, \bibinfo {author} {\bibfnamefont {W.}~\bibnamefont {Cai}}, \bibinfo {author} {\bibfnamefont {L.}~\bibnamefont {Hu}}, \bibinfo {author} {\bibfnamefont {X.}~\bibnamefont {Mu}}, \bibinfo {author} {\bibfnamefont {Y.}~\bibnamefont {Xu}}, \bibinfo {author} {\bibfnamefont {Z.-J.}\ \bibnamefont {Chen}}, \bibinfo {author} {\bibfnamefont {H.}~\bibnamefont {Wang}}, \bibinfo {author} {\bibfnamefont {Y.}~\bibnamefont {Song}}, \bibinfo {author} {\bibfnamefont {H.}~\bibnamefont {Yuan}}, \bibinfo {author} {\bibfnamefont {C.-L.}\ \bibnamefont {Zou}}, \bibinfo {author} {\bibfnamefont {L.-M.}\ \bibnamefont {Duan}},\ and\ \bibinfo {author} {\bibfnamefont {L.}~\bibnamefont {Sun}},\ }\bibfield  {title} {\bibinfo {title} {Heisenberg-limited single-mode quantum metrology in a superconducting circuit},\ }\href@noop {} {\bibfield  {journal} {\bibinfo  {journal} {Nature communications}\ }\textbf {\bibinfo {volume} {10}},\ \bibinfo {pages} {4382} (\bibinfo {year} {2019}{\natexlab{a}})}\BibitemShut {NoStop}%
\bibitem [{\citenamefont {Bennett}\ and\ \citenamefont {DiVincenzo}(2000)}]{bennett2000quantum}%
  \BibitemOpen
  \bibfield  {author} {\bibinfo {author} {\bibfnamefont {C.~H.}\ \bibnamefont {Bennett}}\ and\ \bibinfo {author} {\bibfnamefont {D.~P.}\ \bibnamefont {DiVincenzo}},\ }\bibfield  {title} {\bibinfo {title} {Quantum information and computation},\ }\href@noop {} {\bibfield  {journal} {\bibinfo  {journal} {Nature}\ }\textbf {\bibinfo {volume} {404}},\ \bibinfo {pages} {247} (\bibinfo {year} {2000})}\BibitemShut {NoStop}%
\bibitem [{\citenamefont {Horodecki}\ \emph {et~al.}(2022)\citenamefont {Horodecki}, \citenamefont {Rudnicki},\ and\ \citenamefont {{\.Z}yczkowski}}]{horodecki2022five}%
  \BibitemOpen
  \bibfield  {author} {\bibinfo {author} {\bibfnamefont {P.}~\bibnamefont {Horodecki}}, \bibinfo {author} {\bibfnamefont {{\L}.}~\bibnamefont {Rudnicki}},\ and\ \bibinfo {author} {\bibfnamefont {K.}~\bibnamefont {{\.Z}yczkowski}},\ }\bibfield  {title} {\bibinfo {title} {Five open problems in quantum information theory},\ }\href@noop {} {\bibfield  {journal} {\bibinfo  {journal} {PRX Quantum}\ }\textbf {\bibinfo {volume} {3}},\ \bibinfo {pages} {010101} (\bibinfo {year} {2022})}\BibitemShut {NoStop}%
\bibitem [{\citenamefont {Slussarenko}\ and\ \citenamefont {Pryde}(2019)}]{slussarenko2019photonic}%
  \BibitemOpen
  \bibfield  {author} {\bibinfo {author} {\bibfnamefont {S.}~\bibnamefont {Slussarenko}}\ and\ \bibinfo {author} {\bibfnamefont {G.~J.}\ \bibnamefont {Pryde}},\ }\bibfield  {title} {\bibinfo {title} {Photonic quantum information processing: A concise review},\ }\href@noop {} {\bibfield  {journal} {\bibinfo  {journal} {Applied Physics Reviews}\ }\textbf {\bibinfo {volume} {6}} (\bibinfo {year} {2019})}\BibitemShut {NoStop}%
\bibitem [{\citenamefont {Xavier}\ and\ \citenamefont {Lima}(2020)}]{xavier2020quantum}%
  \BibitemOpen
  \bibfield  {author} {\bibinfo {author} {\bibfnamefont {G.~B.}\ \bibnamefont {Xavier}}\ and\ \bibinfo {author} {\bibfnamefont {G.}~\bibnamefont {Lima}},\ }\bibfield  {title} {\bibinfo {title} {Quantum information processing with space-division multiplexing optical fibres},\ }\href@noop {} {\bibfield  {journal} {\bibinfo  {journal} {Communications Physics}\ }\textbf {\bibinfo {volume} {3}},\ \bibinfo {pages} {9} (\bibinfo {year} {2020})}\BibitemShut {NoStop}%
\bibitem [{\citenamefont {Ladd}\ \emph {et~al.}(2010)\citenamefont {Ladd}, \citenamefont {Jelezko}, \citenamefont {Laflamme}, \citenamefont {Nakamura}, \citenamefont {Monroe},\ and\ \citenamefont {O’Brien}}]{ladd2010quantum}%
  \BibitemOpen
  \bibfield  {author} {\bibinfo {author} {\bibfnamefont {T.~D.}\ \bibnamefont {Ladd}}, \bibinfo {author} {\bibfnamefont {F.}~\bibnamefont {Jelezko}}, \bibinfo {author} {\bibfnamefont {R.}~\bibnamefont {Laflamme}}, \bibinfo {author} {\bibfnamefont {Y.}~\bibnamefont {Nakamura}}, \bibinfo {author} {\bibfnamefont {C.}~\bibnamefont {Monroe}},\ and\ \bibinfo {author} {\bibfnamefont {J.~L.}\ \bibnamefont {O’Brien}},\ }\bibfield  {title} {\bibinfo {title} {Quantum computers},\ }\href@noop {} {\bibfield  {journal} {\bibinfo  {journal} {Nature}\ }\textbf {\bibinfo {volume} {464}},\ \bibinfo {pages} {45} (\bibinfo {year} {2010})}\BibitemShut {NoStop}%
\bibitem [{\citenamefont {DiVincenzo}(1995)}]{divincenzo1995quantum}%
  \BibitemOpen
  \bibfield  {author} {\bibinfo {author} {\bibfnamefont {D.~P.}\ \bibnamefont {DiVincenzo}},\ }\bibfield  {title} {\bibinfo {title} {Quantum computation},\ }\href@noop {} {\bibfield  {journal} {\bibinfo  {journal} {Science}\ }\textbf {\bibinfo {volume} {270}},\ \bibinfo {pages} {255} (\bibinfo {year} {1995})}\BibitemShut {NoStop}%
\bibitem [{\citenamefont {Albash}\ and\ \citenamefont {Lidar}(2018)}]{albash2018adiabatic}%
  \BibitemOpen
  \bibfield  {author} {\bibinfo {author} {\bibfnamefont {T.}~\bibnamefont {Albash}}\ and\ \bibinfo {author} {\bibfnamefont {D.~A.}\ \bibnamefont {Lidar}},\ }\bibfield  {title} {\bibinfo {title} {Adiabatic quantum computation},\ }\href@noop {} {\bibfield  {journal} {\bibinfo  {journal} {Reviews of Modern Physics}\ }\textbf {\bibinfo {volume} {90}},\ \bibinfo {pages} {015002} (\bibinfo {year} {2018})}\BibitemShut {NoStop}%
\bibitem [{\citenamefont {Guo}\ \emph {et~al.}(2019)\citenamefont {Guo}, \citenamefont {Liu}, \citenamefont {Li},\ and\ \citenamefont {Guo}}]{guo2019advances}%
  \BibitemOpen
  \bibfield  {author} {\bibinfo {author} {\bibfnamefont {Y.}~\bibnamefont {Guo}}, \bibinfo {author} {\bibfnamefont {B.-H.}\ \bibnamefont {Liu}}, \bibinfo {author} {\bibfnamefont {C.-F.}\ \bibnamefont {Li}},\ and\ \bibinfo {author} {\bibfnamefont {G.-C.}\ \bibnamefont {Guo}},\ }\bibfield  {title} {\bibinfo {title} {Advances in quantum dense coding},\ }\href@noop {} {\bibfield  {journal} {\bibinfo  {journal} {Advanced Quantum Technologies}\ }\textbf {\bibinfo {volume} {2}},\ \bibinfo {pages} {1900011} (\bibinfo {year} {2019})}\BibitemShut {NoStop}%
\bibitem [{\citenamefont {Mattle}\ \emph {et~al.}(1996)\citenamefont {Mattle}, \citenamefont {Weinfurter}, \citenamefont {Kwiat},\ and\ \citenamefont {Zeilinger}}]{mattle1996dense}%
  \BibitemOpen
  \bibfield  {author} {\bibinfo {author} {\bibfnamefont {K.}~\bibnamefont {Mattle}}, \bibinfo {author} {\bibfnamefont {H.}~\bibnamefont {Weinfurter}}, \bibinfo {author} {\bibfnamefont {P.~G.}\ \bibnamefont {Kwiat}},\ and\ \bibinfo {author} {\bibfnamefont {A.}~\bibnamefont {Zeilinger}},\ }\bibfield  {title} {\bibinfo {title} {Dense coding in experimental quantum communication},\ }\href@noop {} {\bibfield  {journal} {\bibinfo  {journal} {Physical Reiview Letters}\ }\textbf {\bibinfo {volume} {76}},\ \bibinfo {pages} {4656} (\bibinfo {year} {1996})}\BibitemShut {NoStop}%
\bibitem [{\citenamefont {Braunstein}\ and\ \citenamefont {Kimble}(2000)}]{braunstein2000dense}%
  \BibitemOpen
  \bibfield  {author} {\bibinfo {author} {\bibfnamefont {S.~L.}\ \bibnamefont {Braunstein}}\ and\ \bibinfo {author} {\bibfnamefont {H.~J.}\ \bibnamefont {Kimble}},\ }\bibfield  {title} {\bibinfo {title} {Dense coding for continuous variables},\ }\href@noop {} {\bibfield  {journal} {\bibinfo  {journal} {Physical Review A}\ }\textbf {\bibinfo {volume} {61}},\ \bibinfo {pages} {042302} (\bibinfo {year} {2000})}\BibitemShut {NoStop}%
\bibitem [{\citenamefont {Wang}\ \emph {et~al.}(2018)\citenamefont {Wang}, \citenamefont {Luo}, \citenamefont {Huang}, \citenamefont {Chen}, \citenamefont {Su}, \citenamefont {Liu}, \citenamefont {Chen}, \citenamefont {Li}, \citenamefont {Fang}, \citenamefont {Jiang}, \citenamefont {Zhang}, \citenamefont {Li}, \citenamefont {Liu}, \citenamefont {Lu},\ and\ \citenamefont {Pan}}]{wang201818}%
  \BibitemOpen
  \bibfield  {author} {\bibinfo {author} {\bibfnamefont {X.-L.}\ \bibnamefont {Wang}}, \bibinfo {author} {\bibfnamefont {Y.-H.}\ \bibnamefont {Luo}}, \bibinfo {author} {\bibfnamefont {H.-L.}\ \bibnamefont {Huang}}, \bibinfo {author} {\bibfnamefont {M.-C.}\ \bibnamefont {Chen}}, \bibinfo {author} {\bibfnamefont {Z.-E.}\ \bibnamefont {Su}}, \bibinfo {author} {\bibfnamefont {C.}~\bibnamefont {Liu}}, \bibinfo {author} {\bibfnamefont {C.}~\bibnamefont {Chen}}, \bibinfo {author} {\bibfnamefont {W.}~\bibnamefont {Li}}, \bibinfo {author} {\bibfnamefont {Y.-Q.}\ \bibnamefont {Fang}}, \bibinfo {author} {\bibfnamefont {X.}~\bibnamefont {Jiang}}, \bibinfo {author} {\bibfnamefont {J.}~\bibnamefont {Zhang}}, \bibinfo {author} {\bibfnamefont {L.}~\bibnamefont {Li}}, \bibinfo {author} {\bibfnamefont {N.-L.}\ \bibnamefont {Liu}}, \bibinfo {author} {\bibfnamefont {C.-Y.}\ \bibnamefont {Lu}},\ and\ \bibinfo {author} {\bibfnamefont {J.-W.}\ \bibnamefont {Pan}},\ }\bibfield  {title} {\bibinfo {title} {18-qubit entanglement with six photons’ three degrees of freedom},\ }\href@noop {} {\bibfield  {journal} {\bibinfo  {journal} {Physical Reiview Letters}\ }\textbf {\bibinfo {volume} {120}},\ \bibinfo {pages} {260502} (\bibinfo {year} {2018})}\BibitemShut {NoStop}%
\bibitem [{\citenamefont {Ouyang}(2014)}]{ouyang_permutation-invariant_2014}%
  \BibitemOpen
  \bibfield  {author} {\bibinfo {author} {\bibfnamefont {Y.}~\bibnamefont {Ouyang}},\ }\bibfield  {title} {\bibinfo {title} {Permutation-invariant quantum codes},\ }\href@noop {} {\bibfield  {journal} {\bibinfo  {journal} {Physical Review A}\ }\textbf {\bibinfo {volume} {90}},\ \bibinfo {pages} {062317} (\bibinfo {year} {2014})}\BibitemShut {NoStop}%
\bibitem [{\citenamefont {Ouyang}\ and\ \citenamefont {Fitzsimons}(2016)}]{ouyang_permutation-invariant_2016}%
  \BibitemOpen
  \bibfield  {author} {\bibinfo {author} {\bibfnamefont {Y.}~\bibnamefont {Ouyang}}\ and\ \bibinfo {author} {\bibfnamefont {J.}~\bibnamefont {Fitzsimons}},\ }\bibfield  {title} {\bibinfo {title} {Permutation-invariant codes encoding more than one qubit},\ }\href@noop {} {\bibfield  {journal} {\bibinfo  {journal} {Physical Review A}\ }\textbf {\bibinfo {volume} {93}},\ \bibinfo {pages} {042340} (\bibinfo {year} {2016})}\BibitemShut {NoStop}%
\bibitem [{\citenamefont {Pirandola}\ \emph {et~al.}(2015)\citenamefont {Pirandola}, \citenamefont {Eisert}, \citenamefont {Weedbrook}, \citenamefont {Furusawa},\ and\ \citenamefont {Braunstein}}]{pirandola2015advances}%
  \BibitemOpen
  \bibfield  {author} {\bibinfo {author} {\bibfnamefont {S.}~\bibnamefont {Pirandola}}, \bibinfo {author} {\bibfnamefont {J.}~\bibnamefont {Eisert}}, \bibinfo {author} {\bibfnamefont {C.}~\bibnamefont {Weedbrook}}, \bibinfo {author} {\bibfnamefont {A.}~\bibnamefont {Furusawa}},\ and\ \bibinfo {author} {\bibfnamefont {S.~L.}\ \bibnamefont {Braunstein}},\ }\bibfield  {title} {\bibinfo {title} {Advances in quantum teleportation},\ }\href@noop {} {\bibfield  {journal} {\bibinfo  {journal} {Nature photonics}\ }\textbf {\bibinfo {volume} {9}},\ \bibinfo {pages} {641} (\bibinfo {year} {2015})}\BibitemShut {NoStop}%
\bibitem [{\citenamefont {Bouwmeester}\ \emph {et~al.}(1997)\citenamefont {Bouwmeester}, \citenamefont {Pan}, \citenamefont {Mattle}, \citenamefont {Eibl}, \citenamefont {Weinfurter},\ and\ \citenamefont {Zeilinger}}]{bouwmeester1997experimental}%
  \BibitemOpen
  \bibfield  {author} {\bibinfo {author} {\bibfnamefont {D.}~\bibnamefont {Bouwmeester}}, \bibinfo {author} {\bibfnamefont {J.-W.}\ \bibnamefont {Pan}}, \bibinfo {author} {\bibfnamefont {K.}~\bibnamefont {Mattle}}, \bibinfo {author} {\bibfnamefont {M.}~\bibnamefont {Eibl}}, \bibinfo {author} {\bibfnamefont {H.}~\bibnamefont {Weinfurter}},\ and\ \bibinfo {author} {\bibfnamefont {A.}~\bibnamefont {Zeilinger}},\ }\bibfield  {title} {\bibinfo {title} {Experimental quantum teleportation},\ }\href@noop {} {\bibfield  {journal} {\bibinfo  {journal} {Nature}\ }\textbf {\bibinfo {volume} {390}},\ \bibinfo {pages} {575} (\bibinfo {year} {1997})}\BibitemShut {NoStop}%
\bibitem [{\citenamefont {Xia}\ \emph {et~al.}(2017)\citenamefont {Xia}, \citenamefont {Sun}, \citenamefont {Zhang},\ and\ \citenamefont {Pan}}]{xia2017long}%
  \BibitemOpen
  \bibfield  {author} {\bibinfo {author} {\bibfnamefont {X.-X.}\ \bibnamefont {Xia}}, \bibinfo {author} {\bibfnamefont {Q.-C.}\ \bibnamefont {Sun}}, \bibinfo {author} {\bibfnamefont {Q.}~\bibnamefont {Zhang}},\ and\ \bibinfo {author} {\bibfnamefont {J.-W.}\ \bibnamefont {Pan}},\ }\bibfield  {title} {\bibinfo {title} {Long distance quantum teleportation},\ }\href@noop {} {\bibfield  {journal} {\bibinfo  {journal} {Quantum Science and Technology}\ }\textbf {\bibinfo {volume} {3}},\ \bibinfo {pages} {014012} (\bibinfo {year} {2017})}\BibitemShut {NoStop}%
\bibitem [{\citenamefont {Hu}\ \emph {et~al.}(2023)\citenamefont {Hu}, \citenamefont {Guo}, \citenamefont {Liu}, \citenamefont {Li},\ and\ \citenamefont {Guo}}]{hu2023progress}%
  \BibitemOpen
  \bibfield  {author} {\bibinfo {author} {\bibfnamefont {X.-M.}\ \bibnamefont {Hu}}, \bibinfo {author} {\bibfnamefont {Y.}~\bibnamefont {Guo}}, \bibinfo {author} {\bibfnamefont {B.-H.}\ \bibnamefont {Liu}}, \bibinfo {author} {\bibfnamefont {C.-F.}\ \bibnamefont {Li}},\ and\ \bibinfo {author} {\bibfnamefont {G.-C.}\ \bibnamefont {Guo}},\ }\bibfield  {title} {\bibinfo {title} {Progress in quantum teleportation},\ }\href@noop {} {\bibfield  {journal} {\bibinfo  {journal} {Nature Reviews Physics}\ ,\ \bibinfo {pages} {1}} (\bibinfo {year} {2023})}\BibitemShut {NoStop}%
\bibitem [{\citenamefont {Gisin}\ \emph {et~al.}(2002)\citenamefont {Gisin}, \citenamefont {Ribordy}, \citenamefont {Tittel},\ and\ \citenamefont {Zbinden}}]{gisin2002quantum}%
  \BibitemOpen
  \bibfield  {author} {\bibinfo {author} {\bibfnamefont {N.}~\bibnamefont {Gisin}}, \bibinfo {author} {\bibfnamefont {G.}~\bibnamefont {Ribordy}}, \bibinfo {author} {\bibfnamefont {W.}~\bibnamefont {Tittel}},\ and\ \bibinfo {author} {\bibfnamefont {H.}~\bibnamefont {Zbinden}},\ }\bibfield  {title} {\bibinfo {title} {Quantum cryptography},\ }\href@noop {} {\bibfield  {journal} {\bibinfo  {journal} {Reviews of Modern Physics}\ }\textbf {\bibinfo {volume} {74}},\ \bibinfo {pages} {145} (\bibinfo {year} {2002})}\BibitemShut {NoStop}%
\bibitem [{\citenamefont {Bennett}\ \emph {et~al.}(1992)\citenamefont {Bennett}, \citenamefont {Bessette}, \citenamefont {Brassard}, \citenamefont {Salvail},\ and\ \citenamefont {Smolin}}]{bennett1992experimental}%
  \BibitemOpen
  \bibfield  {author} {\bibinfo {author} {\bibfnamefont {C.~H.}\ \bibnamefont {Bennett}}, \bibinfo {author} {\bibfnamefont {F.}~\bibnamefont {Bessette}}, \bibinfo {author} {\bibfnamefont {G.}~\bibnamefont {Brassard}}, \bibinfo {author} {\bibfnamefont {L.}~\bibnamefont {Salvail}},\ and\ \bibinfo {author} {\bibfnamefont {J.}~\bibnamefont {Smolin}},\ }\bibfield  {title} {\bibinfo {title} {Experimental quantum cryptography},\ }\href@noop {} {\bibfield  {journal} {\bibinfo  {journal} {Journal of Cryptology}\ }\textbf {\bibinfo {volume} {5}},\ \bibinfo {pages} {3} (\bibinfo {year} {1992})}\BibitemShut {NoStop}%
\bibitem [{\citenamefont {Yin}\ \emph {et~al.}(2020)\citenamefont {Yin}, \citenamefont {Li}, \citenamefont {Shengkai}, \citenamefont {Yang}, \citenamefont {Cao}, \citenamefont {Zhang}, \citenamefont {Wang}, \citenamefont {Cai}, \citenamefont {Liu}, \citenamefont {Li}, \citenamefont {Huang}, \citenamefont {Deng}, \citenamefont {Li}, \citenamefont {Zhang}, \citenamefont {Liu}, \citenamefont {Chen}, \citenamefont {Lu}, \citenamefont {Wang},\ and\ \citenamefont {Pan}}]{yin2020entanglement}%
  \BibitemOpen
  \bibfield  {author} {\bibinfo {author} {\bibfnamefont {J.}~\bibnamefont {Yin}}, \bibinfo {author} {\bibfnamefont {Y.-H.}\ \bibnamefont {Li}}, \bibinfo {author} {\bibfnamefont {L.}~\bibnamefont {Shengkai}}, \bibinfo {author} {\bibfnamefont {M.}~\bibnamefont {Yang}}, \bibinfo {author} {\bibfnamefont {Y.}~\bibnamefont {Cao}}, \bibinfo {author} {\bibfnamefont {L.}~\bibnamefont {Zhang}}, \bibinfo {author} {\bibfnamefont {J.}~\bibnamefont {Wang}}, \bibinfo {author} {\bibfnamefont {W.-Q.}\ \bibnamefont {Cai}}, \bibinfo {author} {\bibfnamefont {W.-Y.}\ \bibnamefont {Liu}}, \bibinfo {author} {\bibfnamefont {S.-L.}\ \bibnamefont {Li}}, \bibinfo {author} {\bibfnamefont {Y.-M.}\ \bibnamefont {Huang}}, \bibinfo {author} {\bibfnamefont {L.}~\bibnamefont {Deng}}, \bibinfo {author} {\bibfnamefont {L.}~\bibnamefont {Li}}, \bibinfo {author} {\bibfnamefont {Q.}~\bibnamefont {Zhang}}, \bibinfo {author} {\bibfnamefont {N.-L.}\ \bibnamefont {Liu}}, \bibinfo {author} {\bibfnamefont {Y.-A.}\ \bibnamefont {Chen}}, \bibinfo {author} {\bibfnamefont {C.-Y.}\ \bibnamefont {Lu}}, \bibinfo {author} {\bibfnamefont {X.-B.}\ \bibnamefont {Wang}},\ and\ \bibinfo {author} {\bibfnamefont {J.-W.}\ \bibnamefont {Pan}},\ }\bibfield  {title} {\bibinfo {title} {Entanglement-based secure quantum cryptography over 1,120 kilometres},\ }\href@noop {} {\bibfield  {journal} {\bibinfo  {journal} {Nature}\ }\textbf {\bibinfo {volume} {582}},\ \bibinfo {pages} {501} (\bibinfo {year} {2020})}\BibitemShut {NoStop}%
\bibitem [{\citenamefont {Daley}\ \emph {et~al.}(2022)\citenamefont {Daley}, \citenamefont {Bloch}, \citenamefont {Kokail}, \citenamefont {Flannigan}, \citenamefont {Pearson}, \citenamefont {Troyer},\ and\ \citenamefont {Zoller}}]{daley2022practical}%
  \BibitemOpen
  \bibfield  {author} {\bibinfo {author} {\bibfnamefont {A.~J.}\ \bibnamefont {Daley}}, \bibinfo {author} {\bibfnamefont {I.}~\bibnamefont {Bloch}}, \bibinfo {author} {\bibfnamefont {C.}~\bibnamefont {Kokail}}, \bibinfo {author} {\bibfnamefont {S.}~\bibnamefont {Flannigan}}, \bibinfo {author} {\bibfnamefont {N.}~\bibnamefont {Pearson}}, \bibinfo {author} {\bibfnamefont {M.}~\bibnamefont {Troyer}},\ and\ \bibinfo {author} {\bibfnamefont {P.}~\bibnamefont {Zoller}},\ }\bibfield  {title} {\bibinfo {title} {Practical quantum advantage in quantum simulation},\ }\href@noop {} {\bibfield  {journal} {\bibinfo  {journal} {Nature}\ }\textbf {\bibinfo {volume} {607}},\ \bibinfo {pages} {667} (\bibinfo {year} {2022})}\BibitemShut {NoStop}%
\bibitem [{\citenamefont {Rist{\`e}}\ \emph {et~al.}(2017)\citenamefont {Rist{\`e}}, \citenamefont {Da~Silva}, \citenamefont {Ryan}, \citenamefont {Cross}, \citenamefont {C{\'o}rcoles}, \citenamefont {Smolin}, \citenamefont {Gambetta}, \citenamefont {Chow},\ and\ \citenamefont {Johnson}}]{riste2017demonstration}%
  \BibitemOpen
  \bibfield  {author} {\bibinfo {author} {\bibfnamefont {D.}~\bibnamefont {Rist{\`e}}}, \bibinfo {author} {\bibfnamefont {M.~P.}\ \bibnamefont {Da~Silva}}, \bibinfo {author} {\bibfnamefont {C.~A.}\ \bibnamefont {Ryan}}, \bibinfo {author} {\bibfnamefont {A.~W.}\ \bibnamefont {Cross}}, \bibinfo {author} {\bibfnamefont {A.~D.}\ \bibnamefont {C{\'o}rcoles}}, \bibinfo {author} {\bibfnamefont {J.~A.}\ \bibnamefont {Smolin}}, \bibinfo {author} {\bibfnamefont {J.~M.}\ \bibnamefont {Gambetta}}, \bibinfo {author} {\bibfnamefont {J.~M.}\ \bibnamefont {Chow}},\ and\ \bibinfo {author} {\bibfnamefont {B.~R.}\ \bibnamefont {Johnson}},\ }\bibfield  {title} {\bibinfo {title} {Demonstration of quantum advantage in machine learning},\ }\href@noop {} {\bibfield  {journal} {\bibinfo  {journal} {npj Quantum Information}\ }\textbf {\bibinfo {volume} {3}},\ \bibinfo {pages} {16} (\bibinfo {year} {2017})}\BibitemShut {NoStop}%
\bibitem [{\citenamefont {Horodecki}\ \emph {et~al.}(2009)\citenamefont {Horodecki}, \citenamefont {Horodecki}, \citenamefont {Horodecki},\ and\ \citenamefont {Horodecki}}]{horodecki2009quantum}%
  \BibitemOpen
  \bibfield  {author} {\bibinfo {author} {\bibfnamefont {R.}~\bibnamefont {Horodecki}}, \bibinfo {author} {\bibfnamefont {P.}~\bibnamefont {Horodecki}}, \bibinfo {author} {\bibfnamefont {M.}~\bibnamefont {Horodecki}},\ and\ \bibinfo {author} {\bibfnamefont {K.}~\bibnamefont {Horodecki}},\ }\bibfield  {title} {\bibinfo {title} {Quantum entanglement},\ }\href@noop {} {\bibfield  {journal} {\bibinfo  {journal} {Reviews of Modern Physics}\ }\textbf {\bibinfo {volume} {81}},\ \bibinfo {pages} {865} (\bibinfo {year} {2009})}\BibitemShut {NoStop}%
\bibitem [{\citenamefont {Erhard}\ \emph {et~al.}(2020)\citenamefont {Erhard}, \citenamefont {Krenn},\ and\ \citenamefont {Zeilinger}}]{erhard2020advances}%
  \BibitemOpen
  \bibfield  {author} {\bibinfo {author} {\bibfnamefont {M.}~\bibnamefont {Erhard}}, \bibinfo {author} {\bibfnamefont {M.}~\bibnamefont {Krenn}},\ and\ \bibinfo {author} {\bibfnamefont {A.}~\bibnamefont {Zeilinger}},\ }\bibfield  {title} {\bibinfo {title} {Advances in high-dimensional quantum entanglement},\ }\href@noop {} {\bibfield  {journal} {\bibinfo  {journal} {Nature Reviews Physics}\ }\textbf {\bibinfo {volume} {2}},\ \bibinfo {pages} {365} (\bibinfo {year} {2020})}\BibitemShut {NoStop}%
\bibitem [{\citenamefont {McConnell}\ \emph {et~al.}(2015)\citenamefont {McConnell}, \citenamefont {Zhang}, \citenamefont {Hu}, \citenamefont {{\'C}uk},\ and\ \citenamefont {Vuleti{\'c}}}]{mcconnell2015entanglement}%
  \BibitemOpen
  \bibfield  {author} {\bibinfo {author} {\bibfnamefont {R.}~\bibnamefont {McConnell}}, \bibinfo {author} {\bibfnamefont {H.}~\bibnamefont {Zhang}}, \bibinfo {author} {\bibfnamefont {J.}~\bibnamefont {Hu}}, \bibinfo {author} {\bibfnamefont {S.}~\bibnamefont {{\'C}uk}},\ and\ \bibinfo {author} {\bibfnamefont {V.}~\bibnamefont {Vuleti{\'c}}},\ }\bibfield  {title} {\bibinfo {title} {Entanglement with negative wigner function of almost 3,000 atoms heralded by one photon},\ }\href@noop {} {\bibfield  {journal} {\bibinfo  {journal} {Nature}\ }\textbf {\bibinfo {volume} {519}},\ \bibinfo {pages} {439} (\bibinfo {year} {2015})}\BibitemShut {NoStop}%
\bibitem [{\citenamefont {Haas}\ \emph {et~al.}(2014)\citenamefont {Haas}, \citenamefont {Volz}, \citenamefont {Gehr}, \citenamefont {Reichel},\ and\ \citenamefont {Est{\`e}ve}}]{haas2014entangled}%
  \BibitemOpen
  \bibfield  {author} {\bibinfo {author} {\bibfnamefont {F.}~\bibnamefont {Haas}}, \bibinfo {author} {\bibfnamefont {J.}~\bibnamefont {Volz}}, \bibinfo {author} {\bibfnamefont {R.}~\bibnamefont {Gehr}}, \bibinfo {author} {\bibfnamefont {J.}~\bibnamefont {Reichel}},\ and\ \bibinfo {author} {\bibfnamefont {J.}~\bibnamefont {Est{\`e}ve}},\ }\bibfield  {title} {\bibinfo {title} {Entangled states of more than 40 atoms in an optical fiber cavity},\ }\href@noop {} {\bibfield  {journal} {\bibinfo  {journal} {Science}\ }\textbf {\bibinfo {volume} {344}},\ \bibinfo {pages} {180} (\bibinfo {year} {2014})}\BibitemShut {NoStop}%
\bibitem [{\citenamefont {Barontini}\ \emph {et~al.}(2015)\citenamefont {Barontini}, \citenamefont {Hohmann}, \citenamefont {Haas}, \citenamefont {Est{\`e}ve},\ and\ \citenamefont {Reichel}}]{barontini2015deterministic}%
  \BibitemOpen
  \bibfield  {author} {\bibinfo {author} {\bibfnamefont {G.}~\bibnamefont {Barontini}}, \bibinfo {author} {\bibfnamefont {L.}~\bibnamefont {Hohmann}}, \bibinfo {author} {\bibfnamefont {F.}~\bibnamefont {Haas}}, \bibinfo {author} {\bibfnamefont {J.}~\bibnamefont {Est{\`e}ve}},\ and\ \bibinfo {author} {\bibfnamefont {J.}~\bibnamefont {Reichel}},\ }\bibfield  {title} {\bibinfo {title} {Deterministic generation of multiparticle entanglement by quantum zeno dynamics},\ }\href@noop {} {\bibfield  {journal} {\bibinfo  {journal} {Science}\ }\textbf {\bibinfo {volume} {349}},\ \bibinfo {pages} {1317} (\bibinfo {year} {2015})}\BibitemShut {NoStop}%
\bibitem [{\citenamefont {Kuzmich}\ \emph {et~al.}(1997)\citenamefont {Kuzmich}, \citenamefont {M{\o}lmer},\ and\ \citenamefont {Polzik}}]{kuzmich1997spin}%
  \BibitemOpen
  \bibfield  {author} {\bibinfo {author} {\bibfnamefont {A.}~\bibnamefont {Kuzmich}}, \bibinfo {author} {\bibfnamefont {K.}~\bibnamefont {M{\o}lmer}},\ and\ \bibinfo {author} {\bibfnamefont {E.}~\bibnamefont {Polzik}},\ }\bibfield  {title} {\bibinfo {title} {Spin squeezing in an ensemble of atoms illuminated with squeezed light},\ }\href@noop {} {\bibfield  {journal} {\bibinfo  {journal} {Physical Reiview Letters}\ }\textbf {\bibinfo {volume} {79}},\ \bibinfo {pages} {4782} (\bibinfo {year} {1997})}\BibitemShut {NoStop}%
\bibitem [{\citenamefont {Fernholz}\ \emph {et~al.}(2008)\citenamefont {Fernholz}, \citenamefont {Krauter}, \citenamefont {Jensen}, \citenamefont {Sherson}, \citenamefont {S{\o}rensen},\ and\ \citenamefont {Polzik}}]{fernholz2008spin}%
  \BibitemOpen
  \bibfield  {author} {\bibinfo {author} {\bibfnamefont {T.}~\bibnamefont {Fernholz}}, \bibinfo {author} {\bibfnamefont {H.}~\bibnamefont {Krauter}}, \bibinfo {author} {\bibfnamefont {K.}~\bibnamefont {Jensen}}, \bibinfo {author} {\bibfnamefont {J.~F.}\ \bibnamefont {Sherson}}, \bibinfo {author} {\bibfnamefont {A.~S.}\ \bibnamefont {S{\o}rensen}},\ and\ \bibinfo {author} {\bibfnamefont {E.~S.}\ \bibnamefont {Polzik}},\ }\bibfield  {title} {\bibinfo {title} {Spin squeezing of atomic ensembles via nuclear-electronic spin entanglement},\ }\href@noop {} {\bibfield  {journal} {\bibinfo  {journal} {Physical Reiview Letters}\ }\textbf {\bibinfo {volume} {101}},\ \bibinfo {pages} {073601} (\bibinfo {year} {2008})}\BibitemShut {NoStop}%
\bibitem [{\citenamefont {Oblak}\ \emph {et~al.}(2005)\citenamefont {Oblak}, \citenamefont {Petrov}, \citenamefont {Alzar}, \citenamefont {Tittel}, \citenamefont {Vershovski}, \citenamefont {Mikkelsen}, \citenamefont {S{\o}rensen},\ and\ \citenamefont {Polzik}}]{oblak2005quantum}%
  \BibitemOpen
  \bibfield  {author} {\bibinfo {author} {\bibfnamefont {D.}~\bibnamefont {Oblak}}, \bibinfo {author} {\bibfnamefont {P.~G.}\ \bibnamefont {Petrov}}, \bibinfo {author} {\bibfnamefont {C.~L.~G.}\ \bibnamefont {Alzar}}, \bibinfo {author} {\bibfnamefont {W.}~\bibnamefont {Tittel}}, \bibinfo {author} {\bibfnamefont {A.~K.}\ \bibnamefont {Vershovski}}, \bibinfo {author} {\bibfnamefont {J.~K.}\ \bibnamefont {Mikkelsen}}, \bibinfo {author} {\bibfnamefont {J.~L.}\ \bibnamefont {S{\o}rensen}},\ and\ \bibinfo {author} {\bibfnamefont {E.~S.}\ \bibnamefont {Polzik}},\ }\bibfield  {title} {\bibinfo {title} {Quantum-noise-limited interferometric measurement of atomic noise: Towards spin squeezing on the cs clock transition},\ }\href@noop {} {\bibfield  {journal} {\bibinfo  {journal} {Physical Review A}\ }\textbf {\bibinfo {volume} {71}},\ \bibinfo {pages} {043807} (\bibinfo {year} {2005})}\BibitemShut {NoStop}%
\bibitem [{\citenamefont {Saffman}\ \emph {et~al.}(2009)\citenamefont {Saffman}, \citenamefont {Oblak}, \citenamefont {Appel},\ and\ \citenamefont {Polzik}}]{saffman2009spin}%
  \BibitemOpen
  \bibfield  {author} {\bibinfo {author} {\bibfnamefont {M.}~\bibnamefont {Saffman}}, \bibinfo {author} {\bibfnamefont {D.}~\bibnamefont {Oblak}}, \bibinfo {author} {\bibfnamefont {J.}~\bibnamefont {Appel}},\ and\ \bibinfo {author} {\bibfnamefont {E.}~\bibnamefont {Polzik}},\ }\bibfield  {title} {\bibinfo {title} {Spin squeezing of atomic ensembles by multicolor quantum nondemolition measurements},\ }\href@noop {} {\bibfield  {journal} {\bibinfo  {journal} {Physical Review A}\ }\textbf {\bibinfo {volume} {79}},\ \bibinfo {pages} {023831} (\bibinfo {year} {2009})}\BibitemShut {NoStop}%
\bibitem [{\citenamefont {Polzik}\ and\ \citenamefont {Ye}(2016)}]{polzik2016entanglement}%
  \BibitemOpen
  \bibfield  {author} {\bibinfo {author} {\bibfnamefont {E.~S.}\ \bibnamefont {Polzik}}\ and\ \bibinfo {author} {\bibfnamefont {J.}~\bibnamefont {Ye}},\ }\bibfield  {title} {\bibinfo {title} {Entanglement and spin squeezing in a network of distant optical lattice clocks},\ }\href@noop {} {\bibfield  {journal} {\bibinfo  {journal} {Physical Review A}\ }\textbf {\bibinfo {volume} {93}},\ \bibinfo {pages} {021404} (\bibinfo {year} {2016})}\BibitemShut {NoStop}%
\bibitem [{\citenamefont {Chen}\ \emph {et~al.}(2014)\citenamefont {Chen}, \citenamefont {Bohnet}, \citenamefont {Weiner}, \citenamefont {Cox},\ and\ \citenamefont {Thompson}}]{chen2014cavity}%
  \BibitemOpen
  \bibfield  {author} {\bibinfo {author} {\bibfnamefont {Z.}~\bibnamefont {Chen}}, \bibinfo {author} {\bibfnamefont {J.~G.}\ \bibnamefont {Bohnet}}, \bibinfo {author} {\bibfnamefont {J.~M.}\ \bibnamefont {Weiner}}, \bibinfo {author} {\bibfnamefont {K.~C.}\ \bibnamefont {Cox}},\ and\ \bibinfo {author} {\bibfnamefont {J.~K.}\ \bibnamefont {Thompson}},\ }\bibfield  {title} {\bibinfo {title} {Cavity-aided nondemolition measurements for atom counting and spin squeezing},\ }\href@noop {} {\bibfield  {journal} {\bibinfo  {journal} {Physical Review A}\ }\textbf {\bibinfo {volume} {89}},\ \bibinfo {pages} {043837} (\bibinfo {year} {2014})}\BibitemShut {NoStop}%
\bibitem [{\citenamefont {De~Echaniz}\ \emph {et~al.}(2005)\citenamefont {De~Echaniz}, \citenamefont {Mitchell}, \citenamefont {Kubasik}, \citenamefont {Koschorreck}, \citenamefont {Crepaz}, \citenamefont {Eschner},\ and\ \citenamefont {Polzik}}]{de2005conditions}%
  \BibitemOpen
  \bibfield  {author} {\bibinfo {author} {\bibfnamefont {S.}~\bibnamefont {De~Echaniz}}, \bibinfo {author} {\bibfnamefont {M.~W.}\ \bibnamefont {Mitchell}}, \bibinfo {author} {\bibfnamefont {M.}~\bibnamefont {Kubasik}}, \bibinfo {author} {\bibfnamefont {M.}~\bibnamefont {Koschorreck}}, \bibinfo {author} {\bibfnamefont {H.}~\bibnamefont {Crepaz}}, \bibinfo {author} {\bibfnamefont {J.}~\bibnamefont {Eschner}},\ and\ \bibinfo {author} {\bibfnamefont {E.~S.}\ \bibnamefont {Polzik}},\ }\bibfield  {title} {\bibinfo {title} {Conditions for spin squeezing in a cold87rb ensemble},\ }\href@noop {} {\bibfield  {journal} {\bibinfo  {journal} {Journal of Optics B: Quantum and Semiclassical Optics}\ }\textbf {\bibinfo {volume} {7}},\ \bibinfo {pages} {S548} (\bibinfo {year} {2005})}\BibitemShut {NoStop}%
\bibitem [{\citenamefont {Vitagliano}\ \emph {et~al.}(2018)\citenamefont {Vitagliano}, \citenamefont {Colangelo}, \citenamefont {Ciurana}, \citenamefont {Mitchell}, \citenamefont {Sewell},\ and\ \citenamefont {T{\'o}th}}]{vitagliano2018entanglement}%
  \BibitemOpen
  \bibfield  {author} {\bibinfo {author} {\bibfnamefont {G.}~\bibnamefont {Vitagliano}}, \bibinfo {author} {\bibfnamefont {G.}~\bibnamefont {Colangelo}}, \bibinfo {author} {\bibfnamefont {F.~M.}\ \bibnamefont {Ciurana}}, \bibinfo {author} {\bibfnamefont {M.~W.}\ \bibnamefont {Mitchell}}, \bibinfo {author} {\bibfnamefont {R.~J.}\ \bibnamefont {Sewell}},\ and\ \bibinfo {author} {\bibfnamefont {G.}~\bibnamefont {T{\'o}th}},\ }\bibfield  {title} {\bibinfo {title} {Entanglement and extreme planar spin squeezing},\ }\href@noop {} {\bibfield  {journal} {\bibinfo  {journal} {Physical Review A}\ }\textbf {\bibinfo {volume} {97}},\ \bibinfo {pages} {020301} (\bibinfo {year} {2018})}\BibitemShut {NoStop}%
\bibitem [{\citenamefont {Wang}\ \emph {et~al.}(2017)\citenamefont {Wang}, \citenamefont {Qu}, \citenamefont {Li}, \citenamefont {Bao}, \citenamefont {Vuleti{\'c}},\ and\ \citenamefont {Xiao}}]{wang2017two}%
  \BibitemOpen
  \bibfield  {author} {\bibinfo {author} {\bibfnamefont {M.}~\bibnamefont {Wang}}, \bibinfo {author} {\bibfnamefont {W.}~\bibnamefont {Qu}}, \bibinfo {author} {\bibfnamefont {P.}~\bibnamefont {Li}}, \bibinfo {author} {\bibfnamefont {H.}~\bibnamefont {Bao}}, \bibinfo {author} {\bibfnamefont {V.}~\bibnamefont {Vuleti{\'c}}},\ and\ \bibinfo {author} {\bibfnamefont {Y.}~\bibnamefont {Xiao}},\ }\bibfield  {title} {\bibinfo {title} {Two-axis-twisting spin squeezing by multipass quantum erasure},\ }\href@noop {} {\bibfield  {journal} {\bibinfo  {journal} {Physical Review A}\ }\textbf {\bibinfo {volume} {96}},\ \bibinfo {pages} {013823} (\bibinfo {year} {2017})}\BibitemShut {NoStop}%
\bibitem [{\citenamefont {Pedrozo}\ \emph {et~al.}(2020)\citenamefont {Pedrozo}, \citenamefont {Colombo}, \citenamefont {Shu}, \citenamefont {Adiyatullin}, \citenamefont {Li}, \citenamefont {Mendez}, \citenamefont {Braverman}, \citenamefont {Kawasaki}, \citenamefont {Akamatsu}, \citenamefont {Xiao},\ and\ \citenamefont {Vuletic}}]{pedrozo2020entanglement}%
  \BibitemOpen
  \bibfield  {author} {\bibinfo {author} {\bibfnamefont {E.}~\bibnamefont {Pedrozo}}, \bibinfo {author} {\bibfnamefont {S.}~\bibnamefont {Colombo}}, \bibinfo {author} {\bibfnamefont {C.}~\bibnamefont {Shu}}, \bibinfo {author} {\bibfnamefont {A.}~\bibnamefont {Adiyatullin}}, \bibinfo {author} {\bibfnamefont {Z.}~\bibnamefont {Li}}, \bibinfo {author} {\bibfnamefont {E.}~\bibnamefont {Mendez}}, \bibinfo {author} {\bibfnamefont {B.}~\bibnamefont {Braverman}}, \bibinfo {author} {\bibfnamefont {A.}~\bibnamefont {Kawasaki}}, \bibinfo {author} {\bibfnamefont {D.}~\bibnamefont {Akamatsu}}, \bibinfo {author} {\bibfnamefont {Y.}~\bibnamefont {Xiao}},\ and\ \bibinfo {author} {\bibfnamefont {V.}~\bibnamefont {Vuletic}},\ }\bibfield  {title} {\bibinfo {title} {Entanglement on an optical atomic-clock transition},\ }\href@noop {} {\bibfield  {journal} {\bibinfo  {journal} {Nature}\ }\textbf {\bibinfo {volume} {588}},\ \bibinfo {pages} {414} (\bibinfo {year} {2020})}\BibitemShut {NoStop}%
\bibitem [{\citenamefont {Colombo}\ \emph {et~al.}(2022)\citenamefont {Colombo}, \citenamefont {Pedrozo-Pe{\~n}afiel}, \citenamefont {Adiyatullin}, \citenamefont {Li}, \citenamefont {Mendez}, \citenamefont {Shu},\ and\ \citenamefont {Vuleti{\'c}}}]{colombo2022time}%
  \BibitemOpen
  \bibfield  {author} {\bibinfo {author} {\bibfnamefont {S.}~\bibnamefont {Colombo}}, \bibinfo {author} {\bibfnamefont {E.}~\bibnamefont {Pedrozo-Pe{\~n}afiel}}, \bibinfo {author} {\bibfnamefont {A.~F.}\ \bibnamefont {Adiyatullin}}, \bibinfo {author} {\bibfnamefont {Z.}~\bibnamefont {Li}}, \bibinfo {author} {\bibfnamefont {E.}~\bibnamefont {Mendez}}, \bibinfo {author} {\bibfnamefont {C.}~\bibnamefont {Shu}},\ and\ \bibinfo {author} {\bibfnamefont {V.}~\bibnamefont {Vuleti{\'c}}},\ }\bibfield  {title} {\bibinfo {title} {Time-reversal-based quantum metrology with many-body entangled states},\ }\href@noop {} {\bibfield  {journal} {\bibinfo  {journal} {Nature Physics}\ }\textbf {\bibinfo {volume} {18}},\ \bibinfo {pages} {925} (\bibinfo {year} {2022})}\BibitemShut {NoStop}%
\bibitem [{\citenamefont {Yao}\ \emph {et~al.}(2012)\citenamefont {Yao}, \citenamefont {Wang}, \citenamefont {Xu}, \citenamefont {Lu}, \citenamefont {Pan}, \citenamefont {Bao}, \citenamefont {Peng}, \citenamefont {Lu}, \citenamefont {Chen},\ and\ \citenamefont {Pan}}]{yao2012observation}%
  \BibitemOpen
  \bibfield  {author} {\bibinfo {author} {\bibfnamefont {X.-C.}\ \bibnamefont {Yao}}, \bibinfo {author} {\bibfnamefont {T.-X.}\ \bibnamefont {Wang}}, \bibinfo {author} {\bibfnamefont {P.}~\bibnamefont {Xu}}, \bibinfo {author} {\bibfnamefont {H.}~\bibnamefont {Lu}}, \bibinfo {author} {\bibfnamefont {G.-S.}\ \bibnamefont {Pan}}, \bibinfo {author} {\bibfnamefont {X.-H.}\ \bibnamefont {Bao}}, \bibinfo {author} {\bibfnamefont {C.-Z.}\ \bibnamefont {Peng}}, \bibinfo {author} {\bibfnamefont {C.-Y.}\ \bibnamefont {Lu}}, \bibinfo {author} {\bibfnamefont {Y.-A.}\ \bibnamefont {Chen}},\ and\ \bibinfo {author} {\bibfnamefont {J.-W.}\ \bibnamefont {Pan}},\ }\bibfield  {title} {\bibinfo {title} {Observation of eight-photon entanglement},\ }\href@noop {} {\bibfield  {journal} {\bibinfo  {journal} {Nature photonics}\ }\textbf {\bibinfo {volume} {6}},\ \bibinfo {pages} {225} (\bibinfo {year} {2012})}\BibitemShut {NoStop}%
\bibitem [{\citenamefont {Bao}\ \emph {et~al.}(2022)\citenamefont {Bao}, \citenamefont {Wang}, \citenamefont {Wu}, \citenamefont {Li}, \citenamefont {Cai}, \citenamefont {Wang}, \citenamefont {Ma}, \citenamefont {Cai}, \citenamefont {Han}, \citenamefont {Wang}, \citenamefont {Song}, \citenamefont {Sun}, \citenamefont {Zhang},\ and\ \citenamefont {Duan}}]{bao2022experimental}%
  \BibitemOpen
  \bibfield  {author} {\bibinfo {author} {\bibfnamefont {Z.}~\bibnamefont {Bao}}, \bibinfo {author} {\bibfnamefont {Z.}~\bibnamefont {Wang}}, \bibinfo {author} {\bibfnamefont {Y.}~\bibnamefont {Wu}}, \bibinfo {author} {\bibfnamefont {Y.}~\bibnamefont {Li}}, \bibinfo {author} {\bibfnamefont {W.}~\bibnamefont {Cai}}, \bibinfo {author} {\bibfnamefont {W.}~\bibnamefont {Wang}}, \bibinfo {author} {\bibfnamefont {Y.}~\bibnamefont {Ma}}, \bibinfo {author} {\bibfnamefont {T.}~\bibnamefont {Cai}}, \bibinfo {author} {\bibfnamefont {X.}~\bibnamefont {Han}}, \bibinfo {author} {\bibfnamefont {J.}~\bibnamefont {Wang}}, \bibinfo {author} {\bibfnamefont {Y.}~\bibnamefont {Song}}, \bibinfo {author} {\bibfnamefont {L.}~\bibnamefont {Sun}}, \bibinfo {author} {\bibfnamefont {H.}~\bibnamefont {Zhang}},\ and\ \bibinfo {author} {\bibfnamefont {L.}~\bibnamefont {Duan}},\ }\bibfield  {title} {\bibinfo {title} {Experimental preparation of generalized cat states for itinerant microwave photons},\ }\href@noop {} {\bibfield  {journal} {\bibinfo  {journal} {Physical Review A}\ }\textbf {\bibinfo {volume} {105}},\ \bibinfo {pages} {063717} (\bibinfo {year} {2022})}\BibitemShut {NoStop}%
\bibitem [{\citenamefont {Wang}\ \emph {et~al.}(2022)\citenamefont {Wang}, \citenamefont {Bao}, \citenamefont {Wu}, \citenamefont {Li}, \citenamefont {Cai}, \citenamefont {Wang}, \citenamefont {Yuwei}, \citenamefont {Cai}, \citenamefont {Han}, \citenamefont {Wang}, \citenamefont {Song}, \citenamefont {Sun}, \citenamefont {Zhang},\ and\ \citenamefont {Duan}}]{wang2022flying}%
  \BibitemOpen
  \bibfield  {author} {\bibinfo {author} {\bibfnamefont {Z.}~\bibnamefont {Wang}}, \bibinfo {author} {\bibfnamefont {Z.}~\bibnamefont {Bao}}, \bibinfo {author} {\bibfnamefont {Y.}~\bibnamefont {Wu}}, \bibinfo {author} {\bibfnamefont {Y.}~\bibnamefont {Li}}, \bibinfo {author} {\bibfnamefont {W.}~\bibnamefont {Cai}}, \bibinfo {author} {\bibfnamefont {W.}~\bibnamefont {Wang}}, \bibinfo {author} {\bibfnamefont {M.}~\bibnamefont {Yuwei}}, \bibinfo {author} {\bibfnamefont {T.}~\bibnamefont {Cai}}, \bibinfo {author} {\bibfnamefont {X.}~\bibnamefont {Han}}, \bibinfo {author} {\bibfnamefont {J.}~\bibnamefont {Wang}}, \bibinfo {author} {\bibfnamefont {Y.}~\bibnamefont {Song}}, \bibinfo {author} {\bibfnamefont {L.}~\bibnamefont {Sun}}, \bibinfo {author} {\bibfnamefont {H.}~\bibnamefont {Zhang}},\ and\ \bibinfo {author} {\bibfnamefont {L.}~\bibnamefont {Duan}},\ }\bibfield  {title} {\bibinfo {title} {A flying schr{\"o}dinger’s cat in multipartite entangled states},\ }\href@noop {} {\bibfield  {journal} {\bibinfo  {journal} {Science advances}\ }\textbf {\bibinfo {volume} {8}},\ \bibinfo {pages} {eabn1778} (\bibinfo {year} {2022})}\BibitemShut {NoStop}%
\bibitem [{\citenamefont {Cao}\ \emph {et~al.}(2023)\citenamefont {Cao}, \citenamefont {Wu}, \citenamefont {Chen}, \citenamefont {Gong}, \citenamefont {Wu}, \citenamefont {Ye}, \citenamefont {Zha}, \citenamefont {Qian}, \citenamefont {Ying}, \citenamefont {Guo}, \citenamefont {Zhu}, \citenamefont {Huang}, \citenamefont {Zhao}, \citenamefont {Li}, \citenamefont {Wang}, \citenamefont {Yu}, \citenamefont {Fan}, \citenamefont {Wu}, \citenamefont {Su},\ and\ \citenamefont {Pan}}]{cao2023generation}%
  \BibitemOpen
  \bibfield  {author} {\bibinfo {author} {\bibfnamefont {S.}~\bibnamefont {Cao}}, \bibinfo {author} {\bibfnamefont {B.}~\bibnamefont {Wu}}, \bibinfo {author} {\bibfnamefont {F.}~\bibnamefont {Chen}}, \bibinfo {author} {\bibfnamefont {M.}~\bibnamefont {Gong}}, \bibinfo {author} {\bibfnamefont {Y.}~\bibnamefont {Wu}}, \bibinfo {author} {\bibfnamefont {Y.}~\bibnamefont {Ye}}, \bibinfo {author} {\bibfnamefont {C.}~\bibnamefont {Zha}}, \bibinfo {author} {\bibfnamefont {H.}~\bibnamefont {Qian}}, \bibinfo {author} {\bibfnamefont {C.}~\bibnamefont {Ying}}, \bibinfo {author} {\bibfnamefont {S.}~\bibnamefont {Guo}}, \bibinfo {author} {\bibfnamefont {Q.}~\bibnamefont {Zhu}}, \bibinfo {author} {\bibfnamefont {H.-L.}\ \bibnamefont {Huang}}, \bibinfo {author} {\bibfnamefont {Y.}~\bibnamefont {Zhao}}, \bibinfo {author} {\bibfnamefont {S.}~\bibnamefont {Li}}, \bibinfo {author} {\bibfnamefont {S.}~\bibnamefont {Wang}}, \bibinfo {author} {\bibfnamefont {J.}~\bibnamefont {Yu}}, \bibinfo {author} {\bibfnamefont {D.}~\bibnamefont {Fan}}, \bibinfo {author} {\bibfnamefont {D.}~\bibnamefont {Wu}}, \bibinfo {author} {\bibfnamefont {H.}~\bibnamefont {Su}},\ and\ \bibinfo {author} {\bibfnamefont {J.-W.}\ \bibnamefont {Pan}},\ }\bibfield  {title} {\bibinfo {title} {Generation of genuine entanglement up to 51 superconducting qubits},\ }\href@noop {} {\bibfield  {journal} {\bibinfo  {journal} {Nature}\ }\textbf {\bibinfo {volume} {619}},\ \bibinfo {pages} {738} (\bibinfo {year} {2023})}\BibitemShut {NoStop}%
\bibitem [{\citenamefont {Gong}\ \emph {et~al.}(2019)\citenamefont {Gong}, \citenamefont {Chen}, \citenamefont {Zheng}, \citenamefont {Wang}, \citenamefont {Zha}, \citenamefont {Deng}, \citenamefont {Yan}, \citenamefont {Rong}, \citenamefont {Wu}, \citenamefont {Li}, \citenamefont {Chen}, \citenamefont {Zhao}, \citenamefont {Liang}, \citenamefont {Lin}, \citenamefont {Xu}, \citenamefont {Guo}, \citenamefont {Sun}, \citenamefont {Castellano}, \citenamefont {Wang}, \citenamefont {Peng}, \citenamefont {Lu}, \citenamefont {Zhu},\ and\ \citenamefont {Pan}}]{gong2019genuine}%
  \BibitemOpen
  \bibfield  {author} {\bibinfo {author} {\bibfnamefont {M.}~\bibnamefont {Gong}}, \bibinfo {author} {\bibfnamefont {M.-C.}\ \bibnamefont {Chen}}, \bibinfo {author} {\bibfnamefont {Y.}~\bibnamefont {Zheng}}, \bibinfo {author} {\bibfnamefont {S.}~\bibnamefont {Wang}}, \bibinfo {author} {\bibfnamefont {C.}~\bibnamefont {Zha}}, \bibinfo {author} {\bibfnamefont {H.}~\bibnamefont {Deng}}, \bibinfo {author} {\bibfnamefont {Z.}~\bibnamefont {Yan}}, \bibinfo {author} {\bibfnamefont {H.}~\bibnamefont {Rong}}, \bibinfo {author} {\bibfnamefont {Y.}~\bibnamefont {Wu}}, \bibinfo {author} {\bibfnamefont {S.}~\bibnamefont {Li}}, \bibinfo {author} {\bibfnamefont {F.}~\bibnamefont {Chen}}, \bibinfo {author} {\bibfnamefont {Y.}~\bibnamefont {Zhao}}, \bibinfo {author} {\bibfnamefont {F.}~\bibnamefont {Liang}}, \bibinfo {author} {\bibfnamefont {J.}~\bibnamefont {Lin}}, \bibinfo {author} {\bibfnamefont {Y.}~\bibnamefont {Xu}}, \bibinfo {author} {\bibfnamefont {C.}~\bibnamefont {Guo}}, \bibinfo {author} {\bibfnamefont {L.}~\bibnamefont {Sun}}, \bibinfo {author} {\bibfnamefont {A.~D.}\ \bibnamefont {Castellano}}, \bibinfo {author} {\bibfnamefont {H.}~\bibnamefont {Wang}}, \bibinfo {author} {\bibfnamefont {C.}~\bibnamefont {Peng}}, \bibinfo {author} {\bibfnamefont {C.-Y.}\ \bibnamefont {Lu}}, \bibinfo {author} {\bibfnamefont {X.}~\bibnamefont {Zhu}},\ and\ \bibinfo {author} {\bibfnamefont {J.-W.}\ \bibnamefont {Pan}},\ }\bibfield  {title} {\bibinfo {title} {Genuine 12-qubit entanglement on a superconducting quantum processor},\ }\href@noop {} {\bibfield  {journal} {\bibinfo  {journal} {Physical Reiview Letters}\ }\textbf {\bibinfo {volume} {122}},\ \bibinfo {pages} {110501} (\bibinfo {year} {2019})}\BibitemShut {NoStop}%
\bibitem [{\citenamefont {Lu}\ and\ \citenamefont {Pan}(2014)}]{lu2014push}%
  \BibitemOpen
  \bibfield  {author} {\bibinfo {author} {\bibfnamefont {C.-Y.}\ \bibnamefont {Lu}}\ and\ \bibinfo {author} {\bibfnamefont {J.-W.}\ \bibnamefont {Pan}},\ }\bibfield  {title} {\bibinfo {title} {Push-button photon entanglement},\ }\href@noop {} {\bibfield  {journal} {\bibinfo  {journal} {Nature Photonics}\ }\textbf {\bibinfo {volume} {8}},\ \bibinfo {pages} {174} (\bibinfo {year} {2014})}\BibitemShut {NoStop}%
\bibitem [{\citenamefont {Zhao}\ \emph {et~al.}(2021)\citenamefont {Zhao}, \citenamefont {Zhang}, \citenamefont {Chen}, \citenamefont {Wang},\ and\ \citenamefont {Hu}}]{zhao2021creation}%
  \BibitemOpen
  \bibfield  {author} {\bibinfo {author} {\bibfnamefont {Y.}~\bibnamefont {Zhao}}, \bibinfo {author} {\bibfnamefont {R.}~\bibnamefont {Zhang}}, \bibinfo {author} {\bibfnamefont {W.}~\bibnamefont {Chen}}, \bibinfo {author} {\bibfnamefont {X.-B.}\ \bibnamefont {Wang}},\ and\ \bibinfo {author} {\bibfnamefont {J.}~\bibnamefont {Hu}},\ }\bibfield  {title} {\bibinfo {title} {Creation of greenberger-horne-zeilinger states with thousands of atoms by entanglement amplification},\ }\href@noop {} {\bibfield  {journal} {\bibinfo  {journal} {NPJ Quantum Information}\ }\textbf {\bibinfo {volume} {7}},\ \bibinfo {pages} {24} (\bibinfo {year} {2021})}\BibitemShut {NoStop}%
\bibitem [{\citenamefont {Shankar}\ \emph {et~al.}(2021)\citenamefont {Shankar}, \citenamefont {Reilly}, \citenamefont {J{\"a}ger},\ and\ \citenamefont {Holland}}]{shankar2021subradiant}%
  \BibitemOpen
  \bibfield  {author} {\bibinfo {author} {\bibfnamefont {A.}~\bibnamefont {Shankar}}, \bibinfo {author} {\bibfnamefont {J.~T.}\ \bibnamefont {Reilly}}, \bibinfo {author} {\bibfnamefont {S.~B.}\ \bibnamefont {J{\"a}ger}},\ and\ \bibinfo {author} {\bibfnamefont {M.~J.}\ \bibnamefont {Holland}},\ }\bibfield  {title} {\bibinfo {title} {Subradiant-to-subradiant phase transition in the bad cavity laser},\ }\href@noop {} {\bibfield  {journal} {\bibinfo  {journal} {Physical Review Letters}\ }\textbf {\bibinfo {volume} {127}},\ \bibinfo {pages} {073603} (\bibinfo {year} {2021})}\BibitemShut {NoStop}%
\bibitem [{\citenamefont {Dalla~Torre}\ \emph {et~al.}(2013)\citenamefont {Dalla~Torre}, \citenamefont {Otterbach}, \citenamefont {Demler}, \citenamefont {Vuleti{\'c}},\ and\ \citenamefont {Lukin}}]{dalla2013dissipative}%
  \BibitemOpen
  \bibfield  {author} {\bibinfo {author} {\bibfnamefont {E.~G.}\ \bibnamefont {Dalla~Torre}}, \bibinfo {author} {\bibfnamefont {J.}~\bibnamefont {Otterbach}}, \bibinfo {author} {\bibfnamefont {E.}~\bibnamefont {Demler}}, \bibinfo {author} {\bibfnamefont {V.}~\bibnamefont {Vuleti{\'c}}},\ and\ \bibinfo {author} {\bibfnamefont {M.~D.}\ \bibnamefont {Lukin}},\ }\bibfield  {title} {\bibinfo {title} {Dissipative preparation of spin squeezed atomic ensembles in a steady state},\ }\href@noop {} {\bibfield  {journal} {\bibinfo  {journal} {Physical Reiview Letters}\ }\textbf {\bibinfo {volume} {110}},\ \bibinfo {pages} {120402} (\bibinfo {year} {2013})}\BibitemShut {NoStop}%
\bibitem [{\citenamefont {Krauter}\ \emph {et~al.}(2011)\citenamefont {Krauter}, \citenamefont {Muschik}, \citenamefont {Jensen}, \citenamefont {Wasilewski}, \citenamefont {Petersen}, \citenamefont {Cirac},\ and\ \citenamefont {Polzik}}]{krauter2011entanglement}%
  \BibitemOpen
  \bibfield  {author} {\bibinfo {author} {\bibfnamefont {H.}~\bibnamefont {Krauter}}, \bibinfo {author} {\bibfnamefont {C.~A.}\ \bibnamefont {Muschik}}, \bibinfo {author} {\bibfnamefont {K.}~\bibnamefont {Jensen}}, \bibinfo {author} {\bibfnamefont {W.}~\bibnamefont {Wasilewski}}, \bibinfo {author} {\bibfnamefont {J.~M.}\ \bibnamefont {Petersen}}, \bibinfo {author} {\bibfnamefont {J.~I.}\ \bibnamefont {Cirac}},\ and\ \bibinfo {author} {\bibfnamefont {E.~S.}\ \bibnamefont {Polzik}},\ }\bibfield  {title} {\bibinfo {title} {Entanglement generated by dissipation and steady state entanglement of two macroscopic objects},\ }\href@noop {} {\bibfield  {journal} {\bibinfo  {journal} {Physical review letters}\ }\textbf {\bibinfo {volume} {107}},\ \bibinfo {pages} {080503} (\bibinfo {year} {2011})}\BibitemShut {NoStop}%
\bibitem [{\citenamefont {Kastoryano}\ \emph {et~al.}(2011)\citenamefont {Kastoryano}, \citenamefont {Reiter},\ and\ \citenamefont {S{\o}rensen}}]{kastoryano2011dissipative}%
  \BibitemOpen
  \bibfield  {author} {\bibinfo {author} {\bibfnamefont {M.~J.}\ \bibnamefont {Kastoryano}}, \bibinfo {author} {\bibfnamefont {F.}~\bibnamefont {Reiter}},\ and\ \bibinfo {author} {\bibfnamefont {A.~S.}\ \bibnamefont {S{\o}rensen}},\ }\bibfield  {title} {\bibinfo {title} {Dissipative preparation of entanglement in optical cavities},\ }\href@noop {} {\bibfield  {journal} {\bibinfo  {journal} {Physical review letters}\ }\textbf {\bibinfo {volume} {106}},\ \bibinfo {pages} {090502} (\bibinfo {year} {2011})}\BibitemShut {NoStop}%
\bibitem [{\citenamefont {Reiter}\ \emph {et~al.}(2016)\citenamefont {Reiter}, \citenamefont {Reeb},\ and\ \citenamefont {S{\o}rensen}}]{reiter2016scalable}%
  \BibitemOpen
  \bibfield  {author} {\bibinfo {author} {\bibfnamefont {F.}~\bibnamefont {Reiter}}, \bibinfo {author} {\bibfnamefont {D.}~\bibnamefont {Reeb}},\ and\ \bibinfo {author} {\bibfnamefont {A.~S.}\ \bibnamefont {S{\o}rensen}},\ }\bibfield  {title} {\bibinfo {title} {Scalable dissipative preparation of many-body entanglement},\ }\href@noop {} {\bibfield  {journal} {\bibinfo  {journal} {Physical review letters}\ }\textbf {\bibinfo {volume} {117}},\ \bibinfo {pages} {040501} (\bibinfo {year} {2016})}\BibitemShut {NoStop}%
\bibitem [{\citenamefont {Shen}\ \emph {et~al.}(2011)\citenamefont {Shen}, \citenamefont {Chen}, \citenamefont {Yang}, \citenamefont {Wu},\ and\ \citenamefont {Zheng}}]{shen2011steady}%
  \BibitemOpen
  \bibfield  {author} {\bibinfo {author} {\bibfnamefont {L.-T.}\ \bibnamefont {Shen}}, \bibinfo {author} {\bibfnamefont {X.-Y.}\ \bibnamefont {Chen}}, \bibinfo {author} {\bibfnamefont {Z.-B.}\ \bibnamefont {Yang}}, \bibinfo {author} {\bibfnamefont {H.-Z.}\ \bibnamefont {Wu}},\ and\ \bibinfo {author} {\bibfnamefont {S.-B.}\ \bibnamefont {Zheng}},\ }\bibfield  {title} {\bibinfo {title} {Steady-state entanglement for distant atoms by dissipation in coupled cavities},\ }\href@noop {} {\bibfield  {journal} {\bibinfo  {journal} {Physical Review A}\ }\textbf {\bibinfo {volume} {84}},\ \bibinfo {pages} {064302} (\bibinfo {year} {2011})}\BibitemShut {NoStop}%
\bibitem [{\citenamefont {Malia}\ \emph {et~al.}(2022)\citenamefont {Malia}, \citenamefont {Wu}, \citenamefont {Mart{\'\i}nez-Rinc{\'o}n},\ and\ \citenamefont {Kasevich}}]{malia2022distributed}%
  \BibitemOpen
  \bibfield  {author} {\bibinfo {author} {\bibfnamefont {B.~K.}\ \bibnamefont {Malia}}, \bibinfo {author} {\bibfnamefont {Y.}~\bibnamefont {Wu}}, \bibinfo {author} {\bibfnamefont {J.}~\bibnamefont {Mart{\'\i}nez-Rinc{\'o}n}},\ and\ \bibinfo {author} {\bibfnamefont {M.~A.}\ \bibnamefont {Kasevich}},\ }\bibfield  {title} {\bibinfo {title} {Distributed quantum sensing with mode-entangled spin-squeezed atomic states},\ }\href@noop {} {\bibfield  {journal} {\bibinfo  {journal} {Nature}\ }\textbf {\bibinfo {volume} {612}},\ \bibinfo {pages} {661} (\bibinfo {year} {2022})}\BibitemShut {NoStop}%
\bibitem [{\citenamefont {Zhao}\ \emph {et~al.}(2020)\citenamefont {Zhao}, \citenamefont {Okawachi}, \citenamefont {Jang}, \citenamefont {Ji}, \citenamefont {Lipson},\ and\ \citenamefont {Gaeta}}]{zhao2020near}%
  \BibitemOpen
  \bibfield  {author} {\bibinfo {author} {\bibfnamefont {Y.}~\bibnamefont {Zhao}}, \bibinfo {author} {\bibfnamefont {Y.}~\bibnamefont {Okawachi}}, \bibinfo {author} {\bibfnamefont {J.~K.}\ \bibnamefont {Jang}}, \bibinfo {author} {\bibfnamefont {X.}~\bibnamefont {Ji}}, \bibinfo {author} {\bibfnamefont {M.}~\bibnamefont {Lipson}},\ and\ \bibinfo {author} {\bibfnamefont {A.~L.}\ \bibnamefont {Gaeta}},\ }\bibfield  {title} {\bibinfo {title} {Near-degenerate quadrature-squeezed vacuum generation on a silicon-nitride chip},\ }\href@noop {} {\bibfield  {journal} {\bibinfo  {journal} {Physical Review Letters}\ }\textbf {\bibinfo {volume} {124}},\ \bibinfo {pages} {193601} (\bibinfo {year} {2020})}\BibitemShut {NoStop}%
\bibitem [{\citenamefont {Eckner}\ \emph {et~al.}(2023)\citenamefont {Eckner}, \citenamefont {Darkwah~Oppong}, \citenamefont {Cao}, \citenamefont {Young}, \citenamefont {Milner}, \citenamefont {Robinson}, \citenamefont {Ye},\ and\ \citenamefont {Kaufman}}]{eckner2023realizing}%
  \BibitemOpen
  \bibfield  {author} {\bibinfo {author} {\bibfnamefont {W.~J.}\ \bibnamefont {Eckner}}, \bibinfo {author} {\bibfnamefont {N.}~\bibnamefont {Darkwah~Oppong}}, \bibinfo {author} {\bibfnamefont {A.}~\bibnamefont {Cao}}, \bibinfo {author} {\bibfnamefont {A.~W.}\ \bibnamefont {Young}}, \bibinfo {author} {\bibfnamefont {W.~R.}\ \bibnamefont {Milner}}, \bibinfo {author} {\bibfnamefont {J.~M.}\ \bibnamefont {Robinson}}, \bibinfo {author} {\bibfnamefont {J.}~\bibnamefont {Ye}},\ and\ \bibinfo {author} {\bibfnamefont {A.~M.}\ \bibnamefont {Kaufman}},\ }\bibfield  {title} {\bibinfo {title} {Realizing spin squeezing with rydberg interactions in an optical clock},\ }\href@noop {} {\bibfield  {journal} {\bibinfo  {journal} {Nature}\ }\textbf {\bibinfo {volume} {621}},\ \bibinfo {pages} {734} (\bibinfo {year} {2023})}\BibitemShut {NoStop}%
\bibitem [{\citenamefont {Chen}\ \emph {et~al.}(2011)\citenamefont {Chen}, \citenamefont {Bohnet}, \citenamefont {Sankar}, \citenamefont {Dai},\ and\ \citenamefont {Thompson}}]{chen2011conditional}%
  \BibitemOpen
  \bibfield  {author} {\bibinfo {author} {\bibfnamefont {Z.}~\bibnamefont {Chen}}, \bibinfo {author} {\bibfnamefont {J.~G.}\ \bibnamefont {Bohnet}}, \bibinfo {author} {\bibfnamefont {S.~R.}\ \bibnamefont {Sankar}}, \bibinfo {author} {\bibfnamefont {J.}~\bibnamefont {Dai}},\ and\ \bibinfo {author} {\bibfnamefont {J.~K.}\ \bibnamefont {Thompson}},\ }\bibfield  {title} {\bibinfo {title} {Conditional spin squeezing of a large ensemble via the vacuum rabi splitting},\ }\href@noop {} {\bibfield  {journal} {\bibinfo  {journal} {Physical Reiview Letters}\ }\textbf {\bibinfo {volume} {106}},\ \bibinfo {pages} {133601} (\bibinfo {year} {2011})}\BibitemShut {NoStop}%
\bibitem [{\citenamefont {Braverman}\ \emph {et~al.}(2019)\citenamefont {Braverman}, \citenamefont {Kawasaki}, \citenamefont {Pedrozo-Pe\~nafiel}, \citenamefont {Colombo}, \citenamefont {Shu}, \citenamefont {Li}, \citenamefont {Mendez}, \citenamefont {Yamoah}, \citenamefont {Salvi}, \citenamefont {Akamatsu}, \citenamefont {Xiao},\ and\ \citenamefont {Vuleti\ifmmode~\acute{c}\else \'{c}\fi{}}}]{braverman2019near}%
  \BibitemOpen
  \bibfield  {author} {\bibinfo {author} {\bibfnamefont {B.}~\bibnamefont {Braverman}}, \bibinfo {author} {\bibfnamefont {A.}~\bibnamefont {Kawasaki}}, \bibinfo {author} {\bibfnamefont {E.}~\bibnamefont {Pedrozo-Pe\~nafiel}}, \bibinfo {author} {\bibfnamefont {S.}~\bibnamefont {Colombo}}, \bibinfo {author} {\bibfnamefont {C.}~\bibnamefont {Shu}}, \bibinfo {author} {\bibfnamefont {Z.}~\bibnamefont {Li}}, \bibinfo {author} {\bibfnamefont {E.}~\bibnamefont {Mendez}}, \bibinfo {author} {\bibfnamefont {M.}~\bibnamefont {Yamoah}}, \bibinfo {author} {\bibfnamefont {L.}~\bibnamefont {Salvi}}, \bibinfo {author} {\bibfnamefont {D.}~\bibnamefont {Akamatsu}}, \bibinfo {author} {\bibfnamefont {Y.}~\bibnamefont {Xiao}},\ and\ \bibinfo {author} {\bibfnamefont {V.}~\bibnamefont {Vuleti\ifmmode~\acute{c}\else \'{c}\fi{}}},\ }\bibfield  {title} {\bibinfo {title} {Near-unitary spin squeezing in yb 171},\ }\href@noop {} {\bibfield  {journal} {\bibinfo  {journal} {Physical Reiview Letters}\ }\textbf {\bibinfo {volume} {122}},\ \bibinfo {pages} {223203} (\bibinfo {year} {2019})}\BibitemShut {NoStop}%
\bibitem [{\citenamefont {Tashima}\ \emph {et~al.}(2009)\citenamefont {Tashima}, \citenamefont {Wakatsuki}, \citenamefont {{\"O}zdemir}, \citenamefont {Yamamoto}, \citenamefont {Koashi},\ and\ \citenamefont {Imoto}}]{tashima2009local}%
  \BibitemOpen
  \bibfield  {author} {\bibinfo {author} {\bibfnamefont {T.}~\bibnamefont {Tashima}}, \bibinfo {author} {\bibfnamefont {T.}~\bibnamefont {Wakatsuki}}, \bibinfo {author} {\bibfnamefont {{\c{S}}.~K.}\ \bibnamefont {{\"O}zdemir}}, \bibinfo {author} {\bibfnamefont {T.}~\bibnamefont {Yamamoto}}, \bibinfo {author} {\bibfnamefont {M.}~\bibnamefont {Koashi}},\ and\ \bibinfo {author} {\bibfnamefont {N.}~\bibnamefont {Imoto}},\ }\bibfield  {title} {\bibinfo {title} {Local transformation of two einstein-podolsky-rosen photon pairs into a three-photon w state},\ }\href@noop {} {\bibfield  {journal} {\bibinfo  {journal} {Physical Reiview Letters}\ }\textbf {\bibinfo {volume} {102}},\ \bibinfo {pages} {130502} (\bibinfo {year} {2009})}\BibitemShut {NoStop}%
\bibitem [{\citenamefont {Mikami}\ \emph {et~al.}(2005)\citenamefont {Mikami}, \citenamefont {Li}, \citenamefont {Fukuoka},\ and\ \citenamefont {Kobayashi}}]{mikami2005new}%
  \BibitemOpen
  \bibfield  {author} {\bibinfo {author} {\bibfnamefont {H.}~\bibnamefont {Mikami}}, \bibinfo {author} {\bibfnamefont {Y.}~\bibnamefont {Li}}, \bibinfo {author} {\bibfnamefont {K.}~\bibnamefont {Fukuoka}},\ and\ \bibinfo {author} {\bibfnamefont {T.}~\bibnamefont {Kobayashi}},\ }\bibfield  {title} {\bibinfo {title} {New high-efficiency source of a three-photon w state and its full characterization using quantum state tomography},\ }\href@noop {} {\bibfield  {journal} {\bibinfo  {journal} {Physical Reiview Letters}\ }\textbf {\bibinfo {volume} {95}},\ \bibinfo {pages} {150404} (\bibinfo {year} {2005})}\BibitemShut {NoStop}%
\bibitem [{\citenamefont {Fang}\ \emph {et~al.}(2019)\citenamefont {Fang}, \citenamefont {Menotti}, \citenamefont {Liscidini}, \citenamefont {Sipe},\ and\ \citenamefont {Lorenz}}]{fang2019three}%
  \BibitemOpen
  \bibfield  {author} {\bibinfo {author} {\bibfnamefont {B.}~\bibnamefont {Fang}}, \bibinfo {author} {\bibfnamefont {M.}~\bibnamefont {Menotti}}, \bibinfo {author} {\bibfnamefont {M.}~\bibnamefont {Liscidini}}, \bibinfo {author} {\bibfnamefont {J.}~\bibnamefont {Sipe}},\ and\ \bibinfo {author} {\bibfnamefont {V.}~\bibnamefont {Lorenz}},\ }\bibfield  {title} {\bibinfo {title} {Three-photon discrete-energy-entangled w state in an optical fiber},\ }\href@noop {} {\bibfield  {journal} {\bibinfo  {journal} {Physical Reiview Letters}\ }\textbf {\bibinfo {volume} {123}},\ \bibinfo {pages} {070508} (\bibinfo {year} {2019})}\BibitemShut {NoStop}%
\bibitem [{\citenamefont {Eibl}\ \emph {et~al.}(2004)\citenamefont {Eibl}, \citenamefont {Kiesel}, \citenamefont {Bourennane}, \citenamefont {Kurtsiefer},\ and\ \citenamefont {Weinfurter}}]{eibl2004experimental}%
  \BibitemOpen
  \bibfield  {author} {\bibinfo {author} {\bibfnamefont {M.}~\bibnamefont {Eibl}}, \bibinfo {author} {\bibfnamefont {N.}~\bibnamefont {Kiesel}}, \bibinfo {author} {\bibfnamefont {M.}~\bibnamefont {Bourennane}}, \bibinfo {author} {\bibfnamefont {C.}~\bibnamefont {Kurtsiefer}},\ and\ \bibinfo {author} {\bibfnamefont {H.}~\bibnamefont {Weinfurter}},\ }\bibfield  {title} {\bibinfo {title} {Experimental realization of a three-qubit entangled w state},\ }\href@noop {} {\bibfield  {journal} {\bibinfo  {journal} {Physical Reiview Letters}\ }\textbf {\bibinfo {volume} {92}},\ \bibinfo {pages} {077901} (\bibinfo {year} {2004})}\BibitemShut {NoStop}%
\bibitem [{\citenamefont {Sewell}\ \emph {et~al.}(2012)\citenamefont {Sewell}, \citenamefont {Koschorreck}, \citenamefont {Napolitano}, \citenamefont {Dubost}, \citenamefont {Behbood},\ and\ \citenamefont {Mitchell}}]{sewell2012magnetic}%
  \BibitemOpen
  \bibfield  {author} {\bibinfo {author} {\bibfnamefont {R.~J.}\ \bibnamefont {Sewell}}, \bibinfo {author} {\bibfnamefont {M.}~\bibnamefont {Koschorreck}}, \bibinfo {author} {\bibfnamefont {M.}~\bibnamefont {Napolitano}}, \bibinfo {author} {\bibfnamefont {B.}~\bibnamefont {Dubost}}, \bibinfo {author} {\bibfnamefont {N.}~\bibnamefont {Behbood}},\ and\ \bibinfo {author} {\bibfnamefont {M.~W.}\ \bibnamefont {Mitchell}},\ }\bibfield  {title} {\bibinfo {title} {Magnetic sensitivity beyond the projection noise limit by spin squeezing},\ }\href@noop {} {\bibfield  {journal} {\bibinfo  {journal} {Physical Reiview Letters}\ }\textbf {\bibinfo {volume} {109}},\ \bibinfo {pages} {253605} (\bibinfo {year} {2012})}\BibitemShut {NoStop}%
\bibitem [{\citenamefont {Omran}\ \emph {et~al.}(2019)\citenamefont {Omran}, \citenamefont {Levine}, \citenamefont {Keesling}, \citenamefont {Semeghini}, \citenamefont {Wang}, \citenamefont {Ebadi}, \citenamefont {Bernien}, \citenamefont {Zibrov}, \citenamefont {Pichler}, \citenamefont {Choi}, \citenamefont {Cui}, \citenamefont {Rossignolo}, \citenamefont {Rembold}, \citenamefont {Montangero}, \citenamefont {Calarco}, \citenamefont {Endres}, \citenamefont {Greiner}, \citenamefont {Vuletic},\ and\ \citenamefont {Lukin}}]{omran2019generation}%
  \BibitemOpen
  \bibfield  {author} {\bibinfo {author} {\bibfnamefont {A.}~\bibnamefont {Omran}}, \bibinfo {author} {\bibfnamefont {H.}~\bibnamefont {Levine}}, \bibinfo {author} {\bibfnamefont {A.}~\bibnamefont {Keesling}}, \bibinfo {author} {\bibfnamefont {G.}~\bibnamefont {Semeghini}}, \bibinfo {author} {\bibfnamefont {T.}~\bibnamefont {Wang}}, \bibinfo {author} {\bibfnamefont {S.}~\bibnamefont {Ebadi}}, \bibinfo {author} {\bibfnamefont {H.}~\bibnamefont {Bernien}}, \bibinfo {author} {\bibfnamefont {A.}~\bibnamefont {Zibrov}}, \bibinfo {author} {\bibfnamefont {H.}~\bibnamefont {Pichler}}, \bibinfo {author} {\bibfnamefont {S.}~\bibnamefont {Choi}}, \bibinfo {author} {\bibfnamefont {J.}~\bibnamefont {Cui}}, \bibinfo {author} {\bibfnamefont {M.}~\bibnamefont {Rossignolo}}, \bibinfo {author} {\bibfnamefont {P.}~\bibnamefont {Rembold}}, \bibinfo {author} {\bibfnamefont {S.}~\bibnamefont {Montangero}}, \bibinfo {author} {\bibfnamefont {T.}~\bibnamefont {Calarco}}, \bibinfo {author} {\bibfnamefont {M.}~\bibnamefont {Endres}}, \bibinfo {author} {\bibfnamefont {M.}~\bibnamefont {Greiner}}, \bibinfo {author} {\bibfnamefont {V.}~\bibnamefont {Vuletic}},\ and\ \bibinfo {author} {\bibfnamefont {M.}~\bibnamefont {Lukin}},\ }\bibfield  {title} {\bibinfo {title} {Generation and manipulation of schr{\"o}dinger cat states in rydberg atom arrays},\ }\href@noop {} {\bibfield  {journal} {\bibinfo  {journal} {Science}\ }\textbf {\bibinfo {volume} {365}},\ \bibinfo {pages} {570} (\bibinfo {year} {2019})}\BibitemShut {NoStop}%
\bibitem [{\citenamefont {Huang}\ \emph {et~al.}(2011)\citenamefont {Huang}, \citenamefont {Liu}, \citenamefont {Peng}, \citenamefont {Li}, \citenamefont {Li}, \citenamefont {Li},\ and\ \citenamefont {Guo}}]{huang2011experimental}%
  \BibitemOpen
  \bibfield  {author} {\bibinfo {author} {\bibfnamefont {Y.-F.}\ \bibnamefont {Huang}}, \bibinfo {author} {\bibfnamefont {B.-H.}\ \bibnamefont {Liu}}, \bibinfo {author} {\bibfnamefont {L.}~\bibnamefont {Peng}}, \bibinfo {author} {\bibfnamefont {Y.-H.}\ \bibnamefont {Li}}, \bibinfo {author} {\bibfnamefont {L.}~\bibnamefont {Li}}, \bibinfo {author} {\bibfnamefont {C.-F.}\ \bibnamefont {Li}},\ and\ \bibinfo {author} {\bibfnamefont {G.-C.}\ \bibnamefont {Guo}},\ }\bibfield  {title} {\bibinfo {title} {Experimental generation of an eight-photon greenberger--horne--zeilinger state},\ }\href@noop {} {\bibfield  {journal} {\bibinfo  {journal} {Nature communications}\ }\textbf {\bibinfo {volume} {2}},\ \bibinfo {pages} {546} (\bibinfo {year} {2011})}\BibitemShut {NoStop}%
\bibitem [{\citenamefont {Su}\ \emph {et~al.}(2007)\citenamefont {Su}, \citenamefont {Tan}, \citenamefont {Jia}, \citenamefont {Zhang}, \citenamefont {Xie},\ and\ \citenamefont {Peng}}]{su2007experimental}%
  \BibitemOpen
  \bibfield  {author} {\bibinfo {author} {\bibfnamefont {X.}~\bibnamefont {Su}}, \bibinfo {author} {\bibfnamefont {A.}~\bibnamefont {Tan}}, \bibinfo {author} {\bibfnamefont {X.}~\bibnamefont {Jia}}, \bibinfo {author} {\bibfnamefont {J.}~\bibnamefont {Zhang}}, \bibinfo {author} {\bibfnamefont {C.}~\bibnamefont {Xie}},\ and\ \bibinfo {author} {\bibfnamefont {K.}~\bibnamefont {Peng}},\ }\bibfield  {title} {\bibinfo {title} {Experimental preparation of quadripartite cluster and greenberger-horne-zeilinger entangled states for continuous variables},\ }\href@noop {} {\bibfield  {journal} {\bibinfo  {journal} {Physical Reiview Letters}\ }\textbf {\bibinfo {volume} {98}},\ \bibinfo {pages} {070502} (\bibinfo {year} {2007})}\BibitemShut {NoStop}%
\bibitem [{\citenamefont {Takeda}\ \emph {et~al.}(2021)\citenamefont {Takeda}, \citenamefont {Noiri}, \citenamefont {Nakajima}, \citenamefont {Yoneda}, \citenamefont {Kobayashi},\ and\ \citenamefont {Tarucha}}]{takeda2021quantum}%
  \BibitemOpen
  \bibfield  {author} {\bibinfo {author} {\bibfnamefont {K.}~\bibnamefont {Takeda}}, \bibinfo {author} {\bibfnamefont {A.}~\bibnamefont {Noiri}}, \bibinfo {author} {\bibfnamefont {T.}~\bibnamefont {Nakajima}}, \bibinfo {author} {\bibfnamefont {J.}~\bibnamefont {Yoneda}}, \bibinfo {author} {\bibfnamefont {T.}~\bibnamefont {Kobayashi}},\ and\ \bibinfo {author} {\bibfnamefont {S.}~\bibnamefont {Tarucha}},\ }\bibfield  {title} {\bibinfo {title} {Quantum tomography of an entangled three-qubit state in silicon},\ }\href@noop {} {\bibfield  {journal} {\bibinfo  {journal} {Nature Nanotechnology}\ }\textbf {\bibinfo {volume} {16}},\ \bibinfo {pages} {965} (\bibinfo {year} {2021})}\BibitemShut {NoStop}%
\bibitem [{\citenamefont {Zhang}\ \emph {et~al.}(2022)\citenamefont {Zhang}, \citenamefont {Li}, \citenamefont {Zhang}, \citenamefont {Yuan}, \citenamefont {Chen}, \citenamefont {Ren}, \citenamefont {Wang}, \citenamefont {Song}, \citenamefont {Wang}, \citenamefont {Wang}, \citenamefont {Zhu}, \citenamefont {Agarwal},\ and\ \citenamefont {Scully}}]{zhang2022synthesizing}%
  \BibitemOpen
  \bibfield  {author} {\bibinfo {author} {\bibfnamefont {K.}~\bibnamefont {Zhang}}, \bibinfo {author} {\bibfnamefont {H.}~\bibnamefont {Li}}, \bibinfo {author} {\bibfnamefont {P.}~\bibnamefont {Zhang}}, \bibinfo {author} {\bibfnamefont {J.}~\bibnamefont {Yuan}}, \bibinfo {author} {\bibfnamefont {J.}~\bibnamefont {Chen}}, \bibinfo {author} {\bibfnamefont {W.}~\bibnamefont {Ren}}, \bibinfo {author} {\bibfnamefont {Z.}~\bibnamefont {Wang}}, \bibinfo {author} {\bibfnamefont {C.}~\bibnamefont {Song}}, \bibinfo {author} {\bibfnamefont {D.-W.}\ \bibnamefont {Wang}}, \bibinfo {author} {\bibfnamefont {H.}~\bibnamefont {Wang}}, \bibinfo {author} {\bibfnamefont {S.}~\bibnamefont {Zhu}}, \bibinfo {author} {\bibfnamefont {G.~S.}\ \bibnamefont {Agarwal}},\ and\ \bibinfo {author} {\bibfnamefont {M.~O.}\ \bibnamefont {Scully}},\ }\bibfield  {title} {\bibinfo {title} {Synthesizing five-body interaction in a superconducting quantum circuit},\ }\href@noop {} {\bibfield  {journal} {\bibinfo  {journal} {Physical Review Letters}\ }\textbf {\bibinfo {volume} {128}},\ \bibinfo {pages} {190502} (\bibinfo {year} {2022})}\BibitemShut {NoStop}%
\bibitem [{\citenamefont {Wang}\ \emph {et~al.}(2019{\natexlab{b}})\citenamefont {Wang}, \citenamefont {Song}, \citenamefont {Feng}, \citenamefont {Cai}, \citenamefont {Xu}, \citenamefont {Deng}, \citenamefont {Li}, \citenamefont {Zheng}, \citenamefont {Zhu}, \citenamefont {Wang}, \citenamefont {Zhu},\ and\ \citenamefont {Scully}}]{wang2019synthesis}%
  \BibitemOpen
  \bibfield  {author} {\bibinfo {author} {\bibfnamefont {D.-W.}\ \bibnamefont {Wang}}, \bibinfo {author} {\bibfnamefont {C.}~\bibnamefont {Song}}, \bibinfo {author} {\bibfnamefont {W.}~\bibnamefont {Feng}}, \bibinfo {author} {\bibfnamefont {H.}~\bibnamefont {Cai}}, \bibinfo {author} {\bibfnamefont {D.}~\bibnamefont {Xu}}, \bibinfo {author} {\bibfnamefont {H.}~\bibnamefont {Deng}}, \bibinfo {author} {\bibfnamefont {H.}~\bibnamefont {Li}}, \bibinfo {author} {\bibfnamefont {D.}~\bibnamefont {Zheng}}, \bibinfo {author} {\bibfnamefont {X.}~\bibnamefont {Zhu}}, \bibinfo {author} {\bibfnamefont {H.}~\bibnamefont {Wang}}, \bibinfo {author} {\bibfnamefont {S.-Y.}\ \bibnamefont {Zhu}},\ and\ \bibinfo {author} {\bibfnamefont {M.~O.}\ \bibnamefont {Scully}},\ }\bibfield  {title} {\bibinfo {title} {Synthesis of antisymmetric spin exchange interaction and chiral spin clusters in superconducting circuits},\ }\href@noop {} {\bibfield  {journal} {\bibinfo  {journal} {Nature Physics}\ }\textbf {\bibinfo {volume} {15}},\ \bibinfo {pages} {382} (\bibinfo {year} {2019}{\natexlab{b}})}\BibitemShut {NoStop}%
\bibitem [{\citenamefont {Song}\ \emph {et~al.}(2019)\citenamefont {Song}, \citenamefont {Xu}, \citenamefont {Li}, \citenamefont {Zhang}, \citenamefont {Zhang}, \citenamefont {Liu}, \citenamefont {Guo}, \citenamefont {Wang}, \citenamefont {Ren}, \citenamefont {Hao}, \citenamefont {Feng}, \citenamefont {Fan}, \citenamefont {Zheng}, \citenamefont {Wang}, \citenamefont {Wang},\ and\ \citenamefont {Zhu}}]{song2019generation}%
  \BibitemOpen
  \bibfield  {author} {\bibinfo {author} {\bibfnamefont {C.}~\bibnamefont {Song}}, \bibinfo {author} {\bibfnamefont {K.}~\bibnamefont {Xu}}, \bibinfo {author} {\bibfnamefont {H.}~\bibnamefont {Li}}, \bibinfo {author} {\bibfnamefont {Y.-R.}\ \bibnamefont {Zhang}}, \bibinfo {author} {\bibfnamefont {X.}~\bibnamefont {Zhang}}, \bibinfo {author} {\bibfnamefont {W.}~\bibnamefont {Liu}}, \bibinfo {author} {\bibfnamefont {Q.}~\bibnamefont {Guo}}, \bibinfo {author} {\bibfnamefont {Z.}~\bibnamefont {Wang}}, \bibinfo {author} {\bibfnamefont {W.}~\bibnamefont {Ren}}, \bibinfo {author} {\bibfnamefont {J.}~\bibnamefont {Hao}}, \bibinfo {author} {\bibfnamefont {H.}~\bibnamefont {Feng}}, \bibinfo {author} {\bibfnamefont {H.}~\bibnamefont {Fan}}, \bibinfo {author} {\bibfnamefont {D.}~\bibnamefont {Zheng}}, \bibinfo {author} {\bibfnamefont {D.}~\bibnamefont {Wang}}, \bibinfo {author} {\bibfnamefont {H.}~\bibnamefont {Wang}},\ and\ \bibinfo {author} {\bibfnamefont {S.-Y.}\ \bibnamefont {Zhu}},\ }\bibfield  {title} {\bibinfo {title} {Generation of multicomponent atomic schr{\"o}dinger cat states of up to 20 qubits},\ }\href@noop {} {\bibfield  {journal} {\bibinfo  {journal} {Science}\ }\textbf {\bibinfo {volume} {365}},\ \bibinfo {pages} {574} (\bibinfo {year} {2019})}\BibitemShut {NoStop}%
\bibitem [{\citenamefont {Ren}\ \emph {et~al.}(2020)\citenamefont {Ren}, \citenamefont {Liu}, \citenamefont {Song}, \citenamefont {Li}, \citenamefont {Guo}, \citenamefont {Wang}, \citenamefont {Zheng}, \citenamefont {Agarwal}, \citenamefont {Scully}, \citenamefont {Zhu}, \citenamefont {Wang},\ and\ \citenamefont {Wang}}]{ren2020simultaneous}%
  \BibitemOpen
  \bibfield  {author} {\bibinfo {author} {\bibfnamefont {W.}~\bibnamefont {Ren}}, \bibinfo {author} {\bibfnamefont {W.}~\bibnamefont {Liu}}, \bibinfo {author} {\bibfnamefont {C.}~\bibnamefont {Song}}, \bibinfo {author} {\bibfnamefont {H.}~\bibnamefont {Li}}, \bibinfo {author} {\bibfnamefont {Q.}~\bibnamefont {Guo}}, \bibinfo {author} {\bibfnamefont {Z.}~\bibnamefont {Wang}}, \bibinfo {author} {\bibfnamefont {D.}~\bibnamefont {Zheng}}, \bibinfo {author} {\bibfnamefont {G.~S.}\ \bibnamefont {Agarwal}}, \bibinfo {author} {\bibfnamefont {M.~O.}\ \bibnamefont {Scully}}, \bibinfo {author} {\bibfnamefont {S.-Y.}\ \bibnamefont {Zhu}}, \bibinfo {author} {\bibfnamefont {H.}~\bibnamefont {Wang}},\ and\ \bibinfo {author} {\bibfnamefont {D.-W.}\ \bibnamefont {Wang}},\ }\bibfield  {title} {\bibinfo {title} {Simultaneous excitation of two noninteracting atoms with time-frequency correlated photon pairs in a superconducting circuit},\ }\href@noop {} {\bibfield  {journal} {\bibinfo  {journal} {Physical Review Letters}\ }\textbf {\bibinfo {volume} {125}},\ \bibinfo {pages} {133601} (\bibinfo {year} {2020})}\BibitemShut {NoStop}%
\bibitem [{\citenamefont {Bretheau}\ \emph {et~al.}(2015)\citenamefont {Bretheau}, \citenamefont {Campagne-Ibarcq}, \citenamefont {Flurin}, \citenamefont {Mallet},\ and\ \citenamefont {Huard}}]{bretheau_quantum_2015}%
  \BibitemOpen
  \bibfield  {author} {\bibinfo {author} {\bibfnamefont {L.}~\bibnamefont {Bretheau}}, \bibinfo {author} {\bibfnamefont {P.}~\bibnamefont {Campagne-Ibarcq}}, \bibinfo {author} {\bibfnamefont {E.}~\bibnamefont {Flurin}}, \bibinfo {author} {\bibfnamefont {F.}~\bibnamefont {Mallet}},\ and\ \bibinfo {author} {\bibfnamefont {B.}~\bibnamefont {Huard}},\ }\bibfield  {title} {\bibinfo {title} {Quantum dynamics of an electromagnetic mode that cannot contain \textit{{N}} photons},\ }\href@noop {} {\bibfield  {journal} {\bibinfo  {journal} {Science}\ }\textbf {\bibinfo {volume} {348}},\ \bibinfo {pages} {776} (\bibinfo {year} {2015})}\BibitemShut {NoStop}%
\bibitem [{\citenamefont {Chakram}\ \emph {et~al.}(2022)\citenamefont {Chakram}, \citenamefont {He}, \citenamefont {Dixit}, \citenamefont {Oriani}, \citenamefont {Naik}, \citenamefont {Leung}, \citenamefont {Kwon}, \citenamefont {Ma}, \citenamefont {Jiang},\ and\ \citenamefont {Schuster}}]{chakram_multimode_2022}%
  \BibitemOpen
  \bibfield  {author} {\bibinfo {author} {\bibfnamefont {S.}~\bibnamefont {Chakram}}, \bibinfo {author} {\bibfnamefont {K.}~\bibnamefont {He}}, \bibinfo {author} {\bibfnamefont {A.~V.}\ \bibnamefont {Dixit}}, \bibinfo {author} {\bibfnamefont {A.~E.}\ \bibnamefont {Oriani}}, \bibinfo {author} {\bibfnamefont {R.~K.}\ \bibnamefont {Naik}}, \bibinfo {author} {\bibfnamefont {N.}~\bibnamefont {Leung}}, \bibinfo {author} {\bibfnamefont {H.}~\bibnamefont {Kwon}}, \bibinfo {author} {\bibfnamefont {W.-L.}\ \bibnamefont {Ma}}, \bibinfo {author} {\bibfnamefont {L.}~\bibnamefont {Jiang}},\ and\ \bibinfo {author} {\bibfnamefont {D.~I.}\ \bibnamefont {Schuster}},\ }\bibfield  {title} {\bibinfo {title} {Multimode photon blockade},\ }\href@noop {} {\bibfield  {journal} {\bibinfo  {journal} {{Nature Physics}}\ }\textbf {\bibinfo {volume} {18}},\ \bibinfo {pages} {879} (\bibinfo {year} {2022})}\BibitemShut {NoStop}%
\bibitem [{\citenamefont {Dicke}(1954)}]{dicke1954coherence}%
  \BibitemOpen
  \bibfield  {author} {\bibinfo {author} {\bibfnamefont {R.~H.}\ \bibnamefont {Dicke}},\ }\bibfield  {title} {\bibinfo {title} {Coherence in spontaneous radiation processes},\ }\href@noop {} {\bibfield  {journal} {\bibinfo  {journal} {{Physical Review}}\ }\textbf {\bibinfo {volume} {93}},\ \bibinfo {pages} {99} (\bibinfo {year} {1954})}\BibitemShut {NoStop}%
\bibitem [{\citenamefont {Kroeze}\ \emph {et~al.}(2023)\citenamefont {Kroeze}, \citenamefont {Marsh}, \citenamefont {Lin}, \citenamefont {Keeling},\ and\ \citenamefont {Lev}}]{kroeze2023high}%
  \BibitemOpen
  \bibfield  {author} {\bibinfo {author} {\bibfnamefont {R.~M.}\ \bibnamefont {Kroeze}}, \bibinfo {author} {\bibfnamefont {B.~P.}\ \bibnamefont {Marsh}}, \bibinfo {author} {\bibfnamefont {K.-Y.}\ \bibnamefont {Lin}}, \bibinfo {author} {\bibfnamefont {J.}~\bibnamefont {Keeling}},\ and\ \bibinfo {author} {\bibfnamefont {B.~L.}\ \bibnamefont {Lev}},\ }\bibfield  {title} {\bibinfo {title} {High cooperativity using a confocal-cavity--qed microscope},\ }\href@noop {} {\bibfield  {journal} {\bibinfo  {journal} {PRX Quantum}\ }\textbf {\bibinfo {volume} {4}},\ \bibinfo {pages} {020326} (\bibinfo {year} {2023})}\BibitemShut {NoStop}%
\bibitem [{\citenamefont {Huo}\ \emph {et~al.}(2023)\citenamefont {Huo}, \citenamefont {Xia}, \citenamefont {Li}, \citenamefont {Zhang}, \citenamefont {Wang}, \citenamefont {Pan}, \citenamefont {Liu}, \citenamefont {Liu}, \citenamefont {Wang}, \citenamefont {Gao}, \citenamefont {Zhao}, \citenamefont {Li}, \citenamefont {Ying}, \citenamefont {Shang},\ and\ \citenamefont {Zhang}}]{huo2023gatemon}%
  \BibitemOpen
  \bibfield  {author} {\bibinfo {author} {\bibfnamefont {J.}~\bibnamefont {Huo}}, \bibinfo {author} {\bibfnamefont {Z.}~\bibnamefont {Xia}}, \bibinfo {author} {\bibfnamefont {Z.}~\bibnamefont {Li}}, \bibinfo {author} {\bibfnamefont {S.}~\bibnamefont {Zhang}}, \bibinfo {author} {\bibfnamefont {Y.}~\bibnamefont {Wang}}, \bibinfo {author} {\bibfnamefont {D.}~\bibnamefont {Pan}}, \bibinfo {author} {\bibfnamefont {Q.}~\bibnamefont {Liu}}, \bibinfo {author} {\bibfnamefont {Y.}~\bibnamefont {Liu}}, \bibinfo {author} {\bibfnamefont {Z.}~\bibnamefont {Wang}}, \bibinfo {author} {\bibfnamefont {Y.}~\bibnamefont {Gao}}, \bibinfo {author} {\bibfnamefont {J.}~\bibnamefont {Zhao}}, \bibinfo {author} {\bibfnamefont {T.}~\bibnamefont {Li}}, \bibinfo {author} {\bibfnamefont {J.}~\bibnamefont {Ying}}, \bibinfo {author} {\bibfnamefont {R.}~\bibnamefont {Shang}},\ and\ \bibinfo {author} {\bibfnamefont {H.}~\bibnamefont {Zhang}},\ }\bibfield  {title} {\bibinfo {title} {Gatemon qubit based on a thin inas-al hybrid nanowire},\ }\href@noop {} {\bibfield  {journal} {\bibinfo  {journal} {Chinese Physics Letters}\ }\textbf {\bibinfo {volume} {40}},\ \bibinfo {pages} {047302} (\bibinfo {year} {2023})}\BibitemShut {NoStop}%
\bibitem [{\citenamefont {Rigetti}\ \emph {et~al.}(2012)\citenamefont {Rigetti}, \citenamefont {Gambetta}, \citenamefont {Poletto}, \citenamefont {Plourde}, \citenamefont {Chow}, \citenamefont {C\'orcoles}, \citenamefont {Smolin}, \citenamefont {Merkel}, \citenamefont {Rozen}, \citenamefont {Keefe}, \citenamefont {Rothwell}, \citenamefont {Ketchen},\ and\ \citenamefont {Steffen}}]{rigetti2012superconducting}%
  \BibitemOpen
  \bibfield  {author} {\bibinfo {author} {\bibfnamefont {C.}~\bibnamefont {Rigetti}}, \bibinfo {author} {\bibfnamefont {J.~M.}\ \bibnamefont {Gambetta}}, \bibinfo {author} {\bibfnamefont {S.}~\bibnamefont {Poletto}}, \bibinfo {author} {\bibfnamefont {B.~L.~T.}\ \bibnamefont {Plourde}}, \bibinfo {author} {\bibfnamefont {J.~M.}\ \bibnamefont {Chow}}, \bibinfo {author} {\bibfnamefont {A.~D.}\ \bibnamefont {C\'orcoles}}, \bibinfo {author} {\bibfnamefont {J.~A.}\ \bibnamefont {Smolin}}, \bibinfo {author} {\bibfnamefont {S.~T.}\ \bibnamefont {Merkel}}, \bibinfo {author} {\bibfnamefont {J.~R.}\ \bibnamefont {Rozen}}, \bibinfo {author} {\bibfnamefont {G.~A.}\ \bibnamefont {Keefe}}, \bibinfo {author} {\bibfnamefont {M.~B.}\ \bibnamefont {Rothwell}}, \bibinfo {author} {\bibfnamefont {M.~B.}\ \bibnamefont {Ketchen}},\ and\ \bibinfo {author} {\bibfnamefont {M.}~\bibnamefont {Steffen}},\ }\bibfield  {title} {\bibinfo {title} {Superconducting qubit in a waveguide cavity with a coherence time approaching 0.1 ms},\ }\href@noop {} {\bibfield  {journal} {\bibinfo  {journal} {Physical Review B}\ }\textbf {\bibinfo {volume} {86}},\ \bibinfo {pages} {100506} (\bibinfo {year} {2012})}\BibitemShut {NoStop}%
\bibitem [{\citenamefont {Chakram}\ \emph {et~al.}(2021)\citenamefont {Chakram}, \citenamefont {Oriani}, \citenamefont {Naik}, \citenamefont {Dixit}, \citenamefont {He}, \citenamefont {Agrawal}, \citenamefont {Kwon},\ and\ \citenamefont {Schuster}}]{chakram2021seamless}%
  \BibitemOpen
  \bibfield  {author} {\bibinfo {author} {\bibfnamefont {S.}~\bibnamefont {Chakram}}, \bibinfo {author} {\bibfnamefont {A.~E.}\ \bibnamefont {Oriani}}, \bibinfo {author} {\bibfnamefont {R.~K.}\ \bibnamefont {Naik}}, \bibinfo {author} {\bibfnamefont {A.~V.}\ \bibnamefont {Dixit}}, \bibinfo {author} {\bibfnamefont {K.}~\bibnamefont {He}}, \bibinfo {author} {\bibfnamefont {A.}~\bibnamefont {Agrawal}}, \bibinfo {author} {\bibfnamefont {H.}~\bibnamefont {Kwon}},\ and\ \bibinfo {author} {\bibfnamefont {D.~I.}\ \bibnamefont {Schuster}},\ }\bibfield  {title} {\bibinfo {title} {Seamless high-q microwave cavities for multimode circuit quantum electrodynamics},\ }\href@noop {} {\bibfield  {journal} {\bibinfo  {journal} {Physical Reiview Letters}\ }\textbf {\bibinfo {volume} {127}},\ \bibinfo {pages} {107701} (\bibinfo {year} {2021})}\BibitemShut {NoStop}%
\bibitem [{\citenamefont {Milul}\ \emph {et~al.}(2023)\citenamefont {Milul}, \citenamefont {Guttel}, \citenamefont {Goldblatt}, \citenamefont {Hazanov}, \citenamefont {Joshi}, \citenamefont {Chausovsky}, \citenamefont {Kahn}, \citenamefont {{\c{C}}ifty{\"u}rek}, \citenamefont {Lafont},\ and\ \citenamefont {Rosenblum}}]{milul2023superconducting}%
  \BibitemOpen
  \bibfield  {author} {\bibinfo {author} {\bibfnamefont {O.}~\bibnamefont {Milul}}, \bibinfo {author} {\bibfnamefont {B.}~\bibnamefont {Guttel}}, \bibinfo {author} {\bibfnamefont {U.}~\bibnamefont {Goldblatt}}, \bibinfo {author} {\bibfnamefont {S.}~\bibnamefont {Hazanov}}, \bibinfo {author} {\bibfnamefont {L.~M.}\ \bibnamefont {Joshi}}, \bibinfo {author} {\bibfnamefont {D.}~\bibnamefont {Chausovsky}}, \bibinfo {author} {\bibfnamefont {N.}~\bibnamefont {Kahn}}, \bibinfo {author} {\bibfnamefont {E.}~\bibnamefont {{\c{C}}ifty{\"u}rek}}, \bibinfo {author} {\bibfnamefont {F.}~\bibnamefont {Lafont}},\ and\ \bibinfo {author} {\bibfnamefont {S.}~\bibnamefont {Rosenblum}},\ }\bibfield  {title} {\bibinfo {title} {Superconducting cavity qubit with tens of milliseconds single-photon coherence time},\ }\href@noop {} {\bibfield  {journal} {\bibinfo  {journal} {PRX Quantum}\ }\textbf {\bibinfo {volume} {4}},\ \bibinfo {pages} {030336} (\bibinfo {year} {2023})}\BibitemShut {NoStop}%
\bibitem [{\citenamefont {Schleier-Smith}\ \emph {et~al.}(2010)\citenamefont {Schleier-Smith}, \citenamefont {Leroux},\ and\ \citenamefont {Vuleti{\'c}}}]{schleier2010squeezing}%
  \BibitemOpen
  \bibfield  {author} {\bibinfo {author} {\bibfnamefont {M.~H.}\ \bibnamefont {Schleier-Smith}}, \bibinfo {author} {\bibfnamefont {I.~D.}\ \bibnamefont {Leroux}},\ and\ \bibinfo {author} {\bibfnamefont {V.}~\bibnamefont {Vuleti{\'c}}},\ }\bibfield  {title} {\bibinfo {title} {Squeezing the collective spin of a dilute atomic ensemble by cavity feedback},\ }\href@noop {} {\bibfield  {journal} {\bibinfo  {journal} {Physical Review A}\ }\textbf {\bibinfo {volume} {81}},\ \bibinfo {pages} {021804} (\bibinfo {year} {2010})}\BibitemShut {NoStop}%
\bibitem [{\citenamefont {Norcia}\ \emph {et~al.}(2018)\citenamefont {Norcia}, \citenamefont {Lewis-Swan}, \citenamefont {Cline}, \citenamefont {Zhu}, \citenamefont {Rey},\ and\ \citenamefont {Thompson}}]{norcia2018cavity}%
  \BibitemOpen
  \bibfield  {author} {\bibinfo {author} {\bibfnamefont {M.~A.}\ \bibnamefont {Norcia}}, \bibinfo {author} {\bibfnamefont {R.~J.}\ \bibnamefont {Lewis-Swan}}, \bibinfo {author} {\bibfnamefont {J.~R.}\ \bibnamefont {Cline}}, \bibinfo {author} {\bibfnamefont {B.}~\bibnamefont {Zhu}}, \bibinfo {author} {\bibfnamefont {A.~M.}\ \bibnamefont {Rey}},\ and\ \bibinfo {author} {\bibfnamefont {J.~K.}\ \bibnamefont {Thompson}},\ }\bibfield  {title} {\bibinfo {title} {Cavity-mediated collective spin-exchange interactions in a strontium superradiant laser},\ }\href@noop {} {\bibfield  {journal} {\bibinfo  {journal} {Science}\ }\textbf {\bibinfo {volume} {361}},\ \bibinfo {pages} {259} (\bibinfo {year} {2018})}\BibitemShut {NoStop}%
\bibitem [{\citenamefont {Borregaard}\ \emph {et~al.}(2017)\citenamefont {Borregaard}, \citenamefont {Davis}, \citenamefont {Bentsen}, \citenamefont {Schleier-Smith},\ and\ \citenamefont {S{\o}rensen}}]{borregaard2017one}%
  \BibitemOpen
  \bibfield  {author} {\bibinfo {author} {\bibfnamefont {J.}~\bibnamefont {Borregaard}}, \bibinfo {author} {\bibfnamefont {E.}~\bibnamefont {Davis}}, \bibinfo {author} {\bibfnamefont {G.~S.}\ \bibnamefont {Bentsen}}, \bibinfo {author} {\bibfnamefont {M.~H.}\ \bibnamefont {Schleier-Smith}},\ and\ \bibinfo {author} {\bibfnamefont {A.~S.}\ \bibnamefont {S{\o}rensen}},\ }\bibfield  {title} {\bibinfo {title} {One-and two-axis squeezing of atomic ensembles in optical cavities},\ }\href@noop {} {\bibfield  {journal} {\bibinfo  {journal} {New Journal of Physics}\ }\textbf {\bibinfo {volume} {19}},\ \bibinfo {pages} {093021} (\bibinfo {year} {2017})}\BibitemShut {NoStop}%
\bibitem [{\citenamefont {Leroux}\ \emph {et~al.}(2010)\citenamefont {Leroux}, \citenamefont {Schleier-Smith},\ and\ \citenamefont {Vuleti{\'c}}}]{leroux2010implementation}%
  \BibitemOpen
  \bibfield  {author} {\bibinfo {author} {\bibfnamefont {I.~D.}\ \bibnamefont {Leroux}}, \bibinfo {author} {\bibfnamefont {M.~H.}\ \bibnamefont {Schleier-Smith}},\ and\ \bibinfo {author} {\bibfnamefont {V.}~\bibnamefont {Vuleti{\'c}}},\ }\bibfield  {title} {\bibinfo {title} {Implementation of cavity squeezing of a collective atomic spin},\ }\href@noop {} {\bibfield  {journal} {\bibinfo  {journal} {Physical Review Letters}\ }\textbf {\bibinfo {volume} {104}},\ \bibinfo {pages} {073602} (\bibinfo {year} {2010})}\BibitemShut {NoStop}%
\bibitem [{\citenamefont {Tanji-Suzuki}\ \emph {et~al.}(2011)\citenamefont {Tanji-Suzuki}, \citenamefont {Leroux}, \citenamefont {Schleier-Smith}, \citenamefont {Cetina}, \citenamefont {Grier}, \citenamefont {Simon},\ and\ \citenamefont {Vuletić}}]{tanji2011advances}%
  \BibitemOpen
  \bibfield  {author} {\bibinfo {author} {\bibfnamefont {H.}~\bibnamefont {Tanji-Suzuki}}, \bibinfo {author} {\bibfnamefont {I.~D.}\ \bibnamefont {Leroux}}, \bibinfo {author} {\bibfnamefont {M.~H.}\ \bibnamefont {Schleier-Smith}}, \bibinfo {author} {\bibfnamefont {M.}~\bibnamefont {Cetina}}, \bibinfo {author} {\bibfnamefont {A.~T.}\ \bibnamefont {Grier}}, \bibinfo {author} {\bibfnamefont {J.}~\bibnamefont {Simon}},\ and\ \bibinfo {author} {\bibfnamefont {V.}~\bibnamefont {Vuletić}},\ }\bibfield  {title} {\bibinfo {title} {Chapter 4 - interaction between atomic ensembles and optical resonators: Classical description}\ }(\bibinfo  {publisher} {Academic Press},\ \bibinfo {year} {2011})\ pp.\ \bibinfo {pages} {201--237}\BibitemShut {NoStop}%
\bibitem [{\citenamefont {Tavis}\ and\ \citenamefont {Cummings}(1968)}]{tavis1968exact}%
  \BibitemOpen
  \bibfield  {author} {\bibinfo {author} {\bibfnamefont {M.}~\bibnamefont {Tavis}}\ and\ \bibinfo {author} {\bibfnamefont {F.~W.}\ \bibnamefont {Cummings}},\ }\bibfield  {title} {\bibinfo {title} {Exact solution for an n-molecule—radiation-field hamiltonian},\ }\href@noop {} {\bibfield  {journal} {\bibinfo  {journal} {Physical Review}\ }\textbf {\bibinfo {volume} {170}},\ \bibinfo {pages} {379} (\bibinfo {year} {1968})}\BibitemShut {NoStop}%
\bibitem [{\citenamefont {Tavis}\ and\ \citenamefont {Cummings}(1969)}]{tavis1969approximate}%
  \BibitemOpen
  \bibfield  {author} {\bibinfo {author} {\bibfnamefont {M.}~\bibnamefont {Tavis}}\ and\ \bibinfo {author} {\bibfnamefont {F.~W.}\ \bibnamefont {Cummings}},\ }\bibfield  {title} {\bibinfo {title} {Approximate solutions for an n-molecule-radiation-field hamiltonian},\ }\href@noop {} {\bibfield  {journal} {\bibinfo  {journal} {Physical Review}\ }\textbf {\bibinfo {volume} {188}},\ \bibinfo {pages} {692} (\bibinfo {year} {1969})}\BibitemShut {NoStop}%
\bibitem [{\citenamefont {Blaha}\ \emph {et~al.}(2022)\citenamefont {Blaha}, \citenamefont {Johnson}, \citenamefont {Rauschenbeutel},\ and\ \citenamefont {Volz}}]{blaha_beyond_2022}%
  \BibitemOpen
  \bibfield  {author} {\bibinfo {author} {\bibfnamefont {M.}~\bibnamefont {Blaha}}, \bibinfo {author} {\bibfnamefont {A.}~\bibnamefont {Johnson}}, \bibinfo {author} {\bibfnamefont {A.}~\bibnamefont {Rauschenbeutel}},\ and\ \bibinfo {author} {\bibfnamefont {J.}~\bibnamefont {Volz}},\ }\bibfield  {title} {\bibinfo {title} {Beyond the {Tavis}-{Cummings} model: {Revisiting} cavity {QED} with ensembles of quantum emitters},\ }\href@noop {} {\bibfield  {journal} {\bibinfo  {journal} {Physical Review A}\ }\textbf {\bibinfo {volume} {105}},\ \bibinfo {pages} {013719} (\bibinfo {year} {2022})}\BibitemShut {NoStop}%
\bibitem [{\citenamefont {Wineland}\ \emph {et~al.}(1992)\citenamefont {Wineland}, \citenamefont {Bollinger}, \citenamefont {Itano}, \citenamefont {Moore},\ and\ \citenamefont {Heinzen}}]{wineland1992spin}%
  \BibitemOpen
  \bibfield  {author} {\bibinfo {author} {\bibfnamefont {D.~J.}\ \bibnamefont {Wineland}}, \bibinfo {author} {\bibfnamefont {J.~J.}\ \bibnamefont {Bollinger}}, \bibinfo {author} {\bibfnamefont {W.~M.}\ \bibnamefont {Itano}}, \bibinfo {author} {\bibfnamefont {F.}~\bibnamefont {Moore}},\ and\ \bibinfo {author} {\bibfnamefont {D.~J.}\ \bibnamefont {Heinzen}},\ }\bibfield  {title} {\bibinfo {title} {Spin squeezing and reduced quantum noise in spectroscopy},\ }\href@noop {} {\bibfield  {journal} {\bibinfo  {journal} {Physical Review A}\ }\textbf {\bibinfo {volume} {46}},\ \bibinfo {pages} {R6797} (\bibinfo {year} {1992})}\BibitemShut {NoStop}%
\bibitem [{\citenamefont {Kitagawa}\ and\ \citenamefont {Ueda}(1993)}]{kitagawa1993squeezed}%
  \BibitemOpen
  \bibfield  {author} {\bibinfo {author} {\bibfnamefont {M.}~\bibnamefont {Kitagawa}}\ and\ \bibinfo {author} {\bibfnamefont {M.}~\bibnamefont {Ueda}},\ }\bibfield  {title} {\bibinfo {title} {Squeezed spin states},\ }\href@noop {} {\bibfield  {journal} {\bibinfo  {journal} {Physical Review A}\ }\textbf {\bibinfo {volume} {47}},\ \bibinfo {pages} {5138} (\bibinfo {year} {1993})}\BibitemShut {NoStop}%
\bibitem [{\citenamefont {Liu}\ \emph {et~al.}(2021)\citenamefont {Liu}, \citenamefont {Sun}, \citenamefont {Fu}, \citenamefont {Xu}, \citenamefont {Wang}, \citenamefont {He}, \citenamefont {Wang},\ and\ \citenamefont {Zhan}}]{liu2021infidelity}%
  \BibitemOpen
  \bibfield  {author} {\bibinfo {author} {\bibfnamefont {Y.}~\bibnamefont {Liu}}, \bibinfo {author} {\bibfnamefont {Y.}~\bibnamefont {Sun}}, \bibinfo {author} {\bibfnamefont {Z.}~\bibnamefont {Fu}}, \bibinfo {author} {\bibfnamefont {P.}~\bibnamefont {Xu}}, \bibinfo {author} {\bibfnamefont {X.}~\bibnamefont {Wang}}, \bibinfo {author} {\bibfnamefont {X.}~\bibnamefont {He}}, \bibinfo {author} {\bibfnamefont {J.}~\bibnamefont {Wang}},\ and\ \bibinfo {author} {\bibfnamefont {M.}~\bibnamefont {Zhan}},\ }\bibfield  {title} {\bibinfo {title} {Infidelity induced by ground-rydberg decoherence of the control qubit in a two-qubit rydberg-blockade gate},\ }\href@noop {} {\bibfield  {journal} {\bibinfo  {journal} {Physical Review Applied}\ }\textbf {\bibinfo {volume} {15}},\ \bibinfo {pages} {054020} (\bibinfo {year} {2021})}\BibitemShut {NoStop}%
\bibitem [{\citenamefont {Sheng}\ \emph {et~al.}(2018)\citenamefont {Sheng}, \citenamefont {He}, \citenamefont {Xu}, \citenamefont {Guo}, \citenamefont {Wang}, \citenamefont {Xiong}, \citenamefont {Liu}, \citenamefont {Wang},\ and\ \citenamefont {Zhan}}]{sheng2018high}%
  \BibitemOpen
  \bibfield  {author} {\bibinfo {author} {\bibfnamefont {C.}~\bibnamefont {Sheng}}, \bibinfo {author} {\bibfnamefont {X.}~\bibnamefont {He}}, \bibinfo {author} {\bibfnamefont {P.}~\bibnamefont {Xu}}, \bibinfo {author} {\bibfnamefont {R.}~\bibnamefont {Guo}}, \bibinfo {author} {\bibfnamefont {K.}~\bibnamefont {Wang}}, \bibinfo {author} {\bibfnamefont {Z.}~\bibnamefont {Xiong}}, \bibinfo {author} {\bibfnamefont {M.}~\bibnamefont {Liu}}, \bibinfo {author} {\bibfnamefont {J.}~\bibnamefont {Wang}},\ and\ \bibinfo {author} {\bibfnamefont {M.}~\bibnamefont {Zhan}},\ }\bibfield  {title} {\bibinfo {title} {High-fidelity single-qubit gates on neutral atoms in a two-dimensional magic-intensity optical dipole trap array},\ }\href@noop {} {\bibfield  {journal} {\bibinfo  {journal} {{Physical Review Letters}}\ }\textbf {\bibinfo {volume} {121}},\ \bibinfo {pages} {240501} (\bibinfo {year} {2018})}\BibitemShut {NoStop}%
\bibitem [{\citenamefont {Brion}\ \emph {et~al.}(2007)\citenamefont {Brion}, \citenamefont {Pedersen},\ and\ \citenamefont {M{\o}lmer}}]{brion2007adiabatic}%
  \BibitemOpen
  \bibfield  {author} {\bibinfo {author} {\bibfnamefont {E.}~\bibnamefont {Brion}}, \bibinfo {author} {\bibfnamefont {L.~H.}\ \bibnamefont {Pedersen}},\ and\ \bibinfo {author} {\bibfnamefont {K.}~\bibnamefont {M{\o}lmer}},\ }\bibfield  {title} {\bibinfo {title} {Adiabatic elimination in a lambda system},\ }\href@noop {} {\bibfield  {journal} {\bibinfo  {journal} {Journal of Physics A: Mathematical and Theoretical}\ }\textbf {\bibinfo {volume} {40}},\ \bibinfo {pages} {1033} (\bibinfo {year} {2007})}\BibitemShut {NoStop}%
\bibitem [{\citenamefont {Li}\ \emph {et~al.}(2022)\citenamefont {Li}, \citenamefont {Braverman}, \citenamefont {Colombo}, \citenamefont {Shu}, \citenamefont {Kawasaki}, \citenamefont {Adiyatullin}, \citenamefont {Pedrozo-Pe{\~n}afiel}, \citenamefont {Mendez},\ and\ \citenamefont {Vuleti{\'c}}}]{li2022collective}%
  \BibitemOpen
  \bibfield  {author} {\bibinfo {author} {\bibfnamefont {Z.}~\bibnamefont {Li}}, \bibinfo {author} {\bibfnamefont {B.}~\bibnamefont {Braverman}}, \bibinfo {author} {\bibfnamefont {S.}~\bibnamefont {Colombo}}, \bibinfo {author} {\bibfnamefont {C.}~\bibnamefont {Shu}}, \bibinfo {author} {\bibfnamefont {A.}~\bibnamefont {Kawasaki}}, \bibinfo {author} {\bibfnamefont {A.~F.}\ \bibnamefont {Adiyatullin}}, \bibinfo {author} {\bibfnamefont {E.}~\bibnamefont {Pedrozo-Pe{\~n}afiel}}, \bibinfo {author} {\bibfnamefont {E.}~\bibnamefont {Mendez}},\ and\ \bibinfo {author} {\bibfnamefont {V.}~\bibnamefont {Vuleti{\'c}}},\ }\bibfield  {title} {\bibinfo {title} {Collective spin-light and light-mediated spin-spin interactions in an optical cavity},\ }\href@noop {} {\bibfield  {journal} {\bibinfo  {journal} {PRX Quantum}\ }\textbf {\bibinfo {volume} {3}},\ \bibinfo {pages} {020308} (\bibinfo {year} {2022})}\BibitemShut {NoStop}%
\bibitem [{\citenamefont {Schirmer}\ \emph {et~al.}(2001)\citenamefont {Schirmer}, \citenamefont {Fu},\ and\ \citenamefont {Solomon}}]{schirmer_complete_2001}%
  \BibitemOpen
  \bibfield  {author} {\bibinfo {author} {\bibfnamefont {S.~G.}\ \bibnamefont {Schirmer}}, \bibinfo {author} {\bibfnamefont {H.}~\bibnamefont {Fu}},\ and\ \bibinfo {author} {\bibfnamefont {A.~I.}\ \bibnamefont {Solomon}},\ }\bibfield  {title} {\bibinfo {title} {Complete controllability of quantum systems},\ }\href@noop {} {\bibfield  {journal} {\bibinfo  {journal} {{Physical Review A}}\ }\textbf {\bibinfo {volume} {63}},\ \bibinfo {pages} {063410} (\bibinfo {year} {2001})}\BibitemShut {NoStop}%
\bibitem [{\citenamefont {Lloyd}(1995)}]{lloyd1995almost}%
  \BibitemOpen
  \bibfield  {author} {\bibinfo {author} {\bibfnamefont {S.}~\bibnamefont {Lloyd}},\ }\bibfield  {title} {\bibinfo {title} {Almost any quantum logic gate is universal},\ }\href@noop {} {\bibfield  {journal} {\bibinfo  {journal} {Physical Reiview Letters}\ }\textbf {\bibinfo {volume} {75}},\ \bibinfo {pages} {346} (\bibinfo {year} {1995})}\BibitemShut {NoStop}%
\bibitem [{\citenamefont {Kimura}(2003)}]{kimura2003bloch}%
  \BibitemOpen
  \bibfield  {author} {\bibinfo {author} {\bibfnamefont {G.}~\bibnamefont {Kimura}},\ }\bibfield  {title} {\bibinfo {title} {The bloch vector for n-level systems},\ }\href@noop {} {\bibfield  {journal} {\bibinfo  {journal} {Physics Letters A}\ }\textbf {\bibinfo {volume} {314}},\ \bibinfo {pages} {339} (\bibinfo {year} {2003})}\BibitemShut {NoStop}%
\bibitem [{\citenamefont {Wallraff}\ \emph {et~al.}(2004)\citenamefont {Wallraff}, \citenamefont {Schuster}, \citenamefont {Blais}, \citenamefont {Frunzio}, \citenamefont {Huang}, \citenamefont {Majer}, \citenamefont {Kumar}, \citenamefont {Girvin},\ and\ \citenamefont {Schoelkopf}}]{wallraff2004strong}%
  \BibitemOpen
  \bibfield  {author} {\bibinfo {author} {\bibfnamefont {A.}~\bibnamefont {Wallraff}}, \bibinfo {author} {\bibfnamefont {D.~I.}\ \bibnamefont {Schuster}}, \bibinfo {author} {\bibfnamefont {A.}~\bibnamefont {Blais}}, \bibinfo {author} {\bibfnamefont {L.}~\bibnamefont {Frunzio}}, \bibinfo {author} {\bibfnamefont {R.-S.}\ \bibnamefont {Huang}}, \bibinfo {author} {\bibfnamefont {J.}~\bibnamefont {Majer}}, \bibinfo {author} {\bibfnamefont {S.}~\bibnamefont {Kumar}}, \bibinfo {author} {\bibfnamefont {S.~M.}\ \bibnamefont {Girvin}},\ and\ \bibinfo {author} {\bibfnamefont {R.~J.}\ \bibnamefont {Schoelkopf}},\ }\bibfield  {title} {\bibinfo {title} {Strong coupling of a single photon to a superconducting qubit using circuit quantum electrodynamics},\ }\href@noop {} {\bibfield  {journal} {\bibinfo  {journal} {Nature}\ }\textbf {\bibinfo {volume} {431}},\ \bibinfo {pages} {162} (\bibinfo {year} {2004})}\BibitemShut {NoStop}%
\bibitem [{\citenamefont {Helmer}\ and\ \citenamefont {Marquardt}(2009)}]{helmer2009measurement}%
  \BibitemOpen
  \bibfield  {author} {\bibinfo {author} {\bibfnamefont {F.}~\bibnamefont {Helmer}}\ and\ \bibinfo {author} {\bibfnamefont {F.}~\bibnamefont {Marquardt}},\ }\bibfield  {title} {\bibinfo {title} {Measurement-based synthesis of multiqubit entangled states in superconducting cavity qed},\ }\href@noop {} {\bibfield  {journal} {\bibinfo  {journal} {Physical Review. A}\ }\textbf {\bibinfo {volume} {79}},\ \bibinfo {pages} {052328} (\bibinfo {year} {2009})}\BibitemShut {NoStop}%
\bibitem [{\citenamefont {Blais}\ \emph {et~al.}(2004)\citenamefont {Blais}, \citenamefont {Huang}, \citenamefont {Wallraff}, \citenamefont {Girvin},\ and\ \citenamefont {Schoelkopf}}]{blais2004cavity}%
  \BibitemOpen
  \bibfield  {author} {\bibinfo {author} {\bibfnamefont {A.}~\bibnamefont {Blais}}, \bibinfo {author} {\bibfnamefont {R.-S.}\ \bibnamefont {Huang}}, \bibinfo {author} {\bibfnamefont {A.}~\bibnamefont {Wallraff}}, \bibinfo {author} {\bibfnamefont {S.~M.}\ \bibnamefont {Girvin}},\ and\ \bibinfo {author} {\bibfnamefont {R.~J.}\ \bibnamefont {Schoelkopf}},\ }\bibfield  {title} {\bibinfo {title} {Cavity quantum electrodynamics for superconducting electrical circuits: An architecture for quantum computation},\ }\href@noop {} {\bibfield  {journal} {\bibinfo  {journal} {Physical Review A}\ }\textbf {\bibinfo {volume} {69}},\ \bibinfo {pages} {062320} (\bibinfo {year} {2004})}\BibitemShut {NoStop}%
\bibitem [{\citenamefont {Sackett}\ \emph {et~al.}(2000)\citenamefont {Sackett}, \citenamefont {Kielpinski}, \citenamefont {King}, \citenamefont {Langer}, \citenamefont {Meyer}, \citenamefont {Myatt}, \citenamefont {Rowe}, \citenamefont {Turchette}, \citenamefont {Itano}, \citenamefont {Wineland},\ and\ \citenamefont {Monroe}}]{sackett2000experimental}%
  \BibitemOpen
  \bibfield  {author} {\bibinfo {author} {\bibfnamefont {C.~A.}\ \bibnamefont {Sackett}}, \bibinfo {author} {\bibfnamefont {D.}~\bibnamefont {Kielpinski}}, \bibinfo {author} {\bibfnamefont {B.~E.}\ \bibnamefont {King}}, \bibinfo {author} {\bibfnamefont {C.}~\bibnamefont {Langer}}, \bibinfo {author} {\bibfnamefont {V.}~\bibnamefont {Meyer}}, \bibinfo {author} {\bibfnamefont {C.~J.}\ \bibnamefont {Myatt}}, \bibinfo {author} {\bibfnamefont {M.}~\bibnamefont {Rowe}}, \bibinfo {author} {\bibfnamefont {Q.}~\bibnamefont {Turchette}}, \bibinfo {author} {\bibfnamefont {W.~M.}\ \bibnamefont {Itano}}, \bibinfo {author} {\bibfnamefont {D.~J.}\ \bibnamefont {Wineland}},\ and\ \bibinfo {author} {\bibfnamefont {C.}~\bibnamefont {Monroe}},\ }\bibfield  {title} {\bibinfo {title} {Experimental entanglement of four particles},\ }\href@noop {} {\bibfield  {journal} {\bibinfo  {journal} {Nature}\ }\textbf {\bibinfo {volume} {404}},\ \bibinfo {pages} {256} (\bibinfo {year} {2000})}\BibitemShut {NoStop}%
\bibitem [{git()}]{gitlink}%
  \BibitemOpen
  \href@noop {} {\bibinfo {title} {Optimized data and necessary code}},\ \bibinfo {howpublished} {\url{https://github.com/ZhangTao1999/Entanglement_generation_via_single_qubit_operation_in_a_teared_Hilbert_space}}\BibitemShut {NoStop}%
\end{thebibliography}%

\end{document}